\documentclass[aps,prx,twocolumn,superscriptaddress,floatfix]{revtex4-2}
\usepackage{amssymb}
\usepackage{amsmath}
\usepackage[dvipsnames]{xcolor}
\usepackage{color}         
\usepackage{graphics}      
\usepackage[pdftex]{graphicx}      
\usepackage{longtable}     
\usepackage{epsf}          
\usepackage{bm}            
\usepackage{asymptote}     
\usepackage{thumbpdf}
\usepackage{verbatim}
\usepackage{url}
\usepackage{MnSymbol}
\usepackage{physics}
\usepackage{graphicx}
\usepackage[colorlinks=true, hyperindex, breaklinks, linkcolor=blue, urlcolor=blue, citecolor=blue]{hyperref} 
\usepackage{enumitem}	

\usepackage{float}
\usepackage{tipa, extraipa}
\usepackage{color}
\usepackage{verbatim}
\usepackage{soul,xcolor}
\setstcolor{red}
\usepackage{colortbl}
\usepackage{amsthm}
\usepackage[linesnumbered, ruled, vlined]{algorithm2e}
\usepackage{makecell}
\usepackage{stmaryrd}
\usepackage{tikz-cd}
\usepackage{dsfont}
\usepackage{array} 
\usepackage{blkarray}
\newcolumntype{C}[1]{>{\centering\arraybackslash}p{#1}}

\newtheorem{theorem}{Theorem}
\newtheorem{lemma}[theorem]{Lemma}
\newtheorem{corollary}[theorem]{Corollary}
\newtheorem{definition}[theorem]{Definition}
\newtheorem{proposition}[theorem]{Proposition}

\newcommand{\mc}{\mathcal}
\newcommand{\mb}{\mathbf}
\newcommand{\mbb}{\mathbb}
\newcommand{\mr}{\mathrm}
\newcommand{\bs}{\boldsymbol}
\renewcommand{\ker}[1]{\mathrm{ker}(#1)}
\newcommand{\im}[1]{\mathrm{im}(#1)}

\newcommand{\qx}[1]{{\color{black}#1}}

\usepackage{tabularx}

\usepackage[normalem]{ulem}

\SetKwInOut{Input}{Input}
\SetKwInOut{Output}{Output}
\SetKwFunction{FMain}{Rearrange}
\SetKwFunction{Fappend}{append}
\SetKwProg{Fn}{Function}{:}{}
\RestyleAlgo{ruled}

\begin{document}

\title{Batched high-rate logical operations for quantum LDPC codes}

\author{Qian Xu}
\thanks{These authors contributed equally}
\email{qianxu@caltech.edu}
\affiliation{Institute for Quantum Information and Matter, Caltech, Pasadena, CA 91125, USA}
\affiliation{Walter Burke Institute for Theoretical Physics, Caltech, Pasadena, CA 91125, USA}

\author{Hengyun Zhou}
\thanks{These authors contributed equally.}
\affiliation{QuEra Computing Inc., 1284 Soldiers Field Road, Boston, MA 02135, USA}

\author{Dolev Bluvstein}
\affiliation{Department of Physics, Harvard University, Cambridge, MA 02138, USA}

\author{Madelyn Cain}
\affiliation{Department of Physics, Harvard University, Cambridge, MA 02138, USA}

\author{Marcin Kalinowski}
\affiliation{Department of Physics, Harvard University, Cambridge, MA 02138, USA}

\author{John Preskill}
\email{preskill@caltech.edu}
\affiliation{Institute for Quantum Information and Matter, Caltech, Pasadena, CA 91125, USA}
\affiliation{AWS Center for Quantum Computing, Pasadena, CA 91125, USA}

\author{Mikhail D. Lukin}
\email{lukin@physics.harvard.edu}
\affiliation{Department of Physics, Harvard University, Cambridge, MA 02138, USA}

\author{Nishad Maskara}
\email{nishadma@mit.edu}
\affiliation{Department of Physics, Harvard University, Cambridge, MA 02138, USA}
\affiliation{Massachusetts Institute of Technology, Cambridge, Massachusetts 02139, USA}

\begin{abstract}
High-rate quantum LDPC (qLDPC) codes reduce memory overhead by densely packing many logical qubits into a single block of physical qubits. Here we extend this concept to high-rate computation by constructing \emph{batched} fault-tolerant operations that apply the same logical gate across many code blocks in parallel. 
By leveraging shared physical resources to execute many logical operations in parallel, these operations realize high rates in space-time and significantly reduce computational costs.
For \emph{arbitrary} CSS qLDPC codes, we build batched gadgets with \emph{constant space-time overhead} (assuming fast classical computation) for (i) single-shot error correction, state preparation, and code surgeries (ii) code switching, and (iii) addressable Clifford gates. Using these batched gadgets we also construct parallel non-Clifford gates with low space-time cost. We outline principles for designing parallel quantum algorithms optimized for a batched architecture, and show in particular how lattice Hamiltonian dynamical simulations can be compiled efficiently. We also propose a near-term implementation using new self-dual Bivariate-Bicycle codes with high encoding rates ($\sim 1/10$), transversal Clifford gates, and global $T$ gates via parallel magic state cultivation, enabling Hamiltonian simulations with a lower space-time cost than analogous surface-code protocols and low-rate qLDPC protocols.
These results open new paths toward scalable quantum computation via co-design of parallel quantum algorithms and high-rate fault-tolerant protocols.
\end{abstract}

\maketitle

\section{Introduction}
High-rate quantum low-density parity-check (qLDPC) codes enable the efficient encoding of many logical qubits within a single code block~\cite{gottesman2013fault, panteleev2022quantum, breuckmann2020balanced, panteleev2022asymptotically, leverrier2022quantum, gu2022efficient, dinur2022good, lin2022good}, making them compelling candidates for constructing low-space-overhead quantum memories~\cite{tremblay2022constant, xu2024constant, bravyi2024high}.
However, the advantages of qLDPC codes at the memory level do not straightforwardly extend to computation. In high-rate codes, the physical qubits in a single block are densely shared among many logical qubits, reducing space overhead. Yet, existing schemes for implementing addressable logical operations—such as generalized lattice surgery~\cite{cohen2021quantum, cross2024improved, williamson2024low}, homomorphic measurements~\cite{huang2022homomorphic, xu2024fast}, or gate teleportation~\cite{gottesman2013fault, nguyen_qldpc_2024}—typically isolate each logical operation into a separated space-time volume. As a result, these approaches incur increasing space-time costs per logical gate as the problem size increases.

\begin{figure*}
    \centering
    \includegraphics[width=1\linewidth]{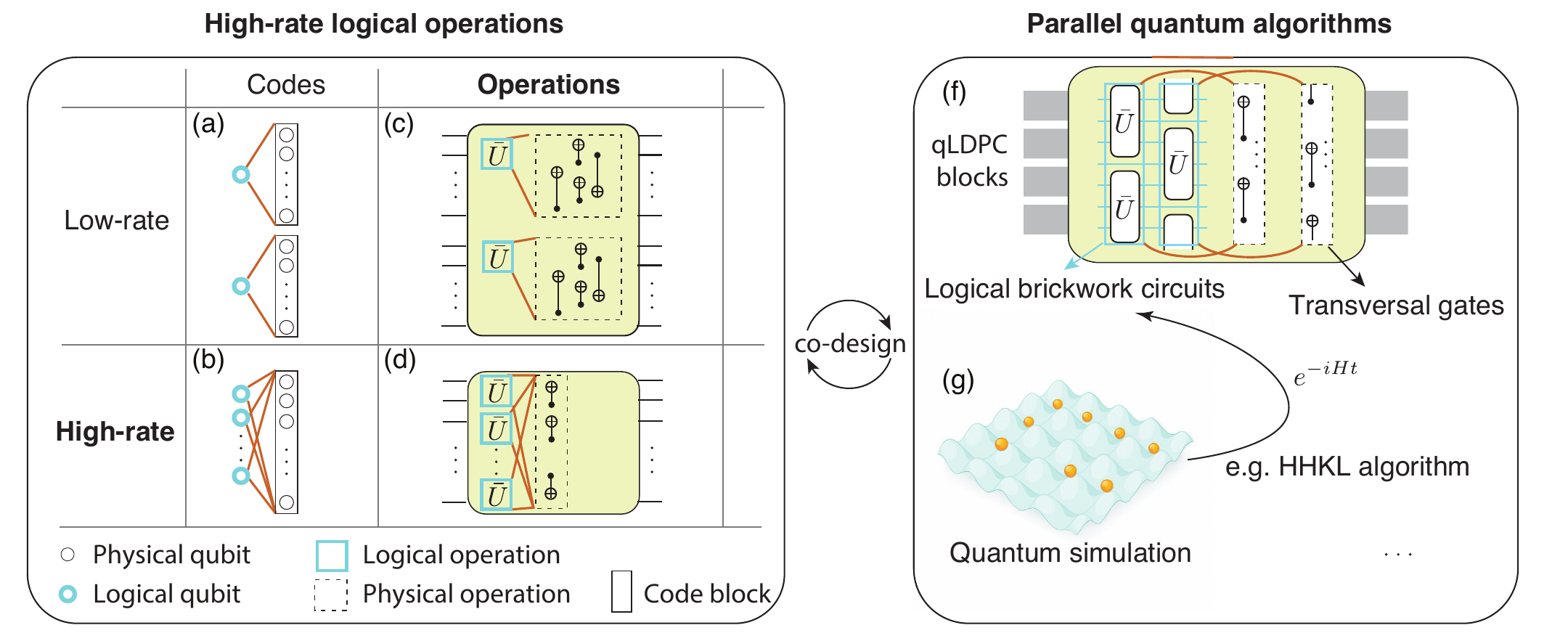}
    \caption{
    \textbf{Overview of new high-rate logical operations and efficiently implementable parallel quantum algorithms.} 
    High-rate qLDPC codes (b) save space costs over low-rate codes (a) by sharing physical qubits within a code block among the many encoded logical qubits. Analogously, high-rate logical operations (d) save space-time costs over low-rate operations (c) by sharing physical gates among the many encoded logical operations. In this work, we introduce new classes of high-rate logical operations by operating on multiple high-rate qLDPC blocks in batches. 
    These batched operations support space-time-efficient implementations of quantum algorithms with a parallel structure. For instance, a brickwork circuit (f) can be implemented using batched logical gates that involve mostly high-rate transversal gates on batched high-rate qLDPC blocks.
    This also enables simulating quantum dynamics of lattice Hamiltonians (g) using algorithms that involve brickwork circuits, e.g. the HHKL algorithm~\cite{haah2021quantum}.
    }
    \label{fig:overview}
\end{figure*}

To reduce space-time costs with qLDPC codes, it would be desirable to develop \emph{high-rate logical operations}: schemes that collectively implement $n$ nontrivial logical gates using $\mc{O}(n)$ shared physical gates, analogous to how high-rate qLDPC codes collectively encode $k$ logical qubits using $\mc{O}(k)$ shared physical qubits (see Fig.~\ref{fig:overview}). A prototypical example is the transversal CNOT between two high-rate CSS codes, which applies $n$ logical CNOTs using only $\Theta(n)$ physical CNOTs. However, such transversal gates act globally on all logical qubits, offering limited programmability and addressability~\cite{breuckmann2022fold, guyot2025addressability}.

In this work, we propose a paradigm for fault-tolerant quantum computation with qLDPC codes, built primarily on high-rate logical operations~\footnote{Although the presented scheme for implementing non-Clifford gates (Theorem~\ref{theorem:batched_global_magic}) incurs a slightly non-constant space-time overhead, we expect this can be improved by using high-rate qLDPC codes that support transversal non-Clifford gates~\cite{zhu2025topological, lin2024transversal, golowich2025quantum}}. We begin by introducing a batched syndrome extraction (BSE) gadget that performs fault-tolerant stabilizer measurements for any CSS qLDPC code with constant space and time overhead (Fig.~\ref{fig:BSE_main}). This enables code-independent single-shot quantum error correction and state preparation~\cite{bombin2015single, kubica2022single, quintavalle2020single, gu2023single} of $n$ (arbitrary) qLDPC code blocks using $\mc{O}(n)$ ancillary code blocks and $\mc{O}(n)$ transversal gates. It also enables batched surgery operations~\cite{horsman2012surface, cohen2022low, cross2024improved} in single shot.

We then significantly broaden the available set of high-rate logical gates by combining transversal operations with a new code-switching gadget that applies to any pair of CSS qLDPC codes, with constant space-time overhead. Crucially, we show that transversal gates—which typically act globally—can be transformed into local, addressable operations on a target code when mediated through this switching protocol (Fig.~\ref{fig:BCS_BAC_main}). 
As a result, we demonstrate how to implement a universal Clifford gate set, with each operation having constant space-time overhead.

Finally, we extend our high-rate gadgets to universal quantum computation by incorporating global non-Clifford gates. This is achieved by temporarily switching to auxiliary codes that natively support transversal non-Clifford gates or transversal Clifford measurements~\cite{gidney2024magic} (Fig.~\ref{fig:BB_cultivation}).

While these gadgets generate a universal gate set, they impose a batching constraint: each gadget must apply the same logical gates to multiple blocks of the same code in parallel. 
The minimal batch size depends on the desired degree of error suppression (e.g. the effective distance). This constraint is illustrated in Fig.~\ref{fig:BCS_BAC_main}. 
Motivated by this structure, we introduce design principles for compiling quantum circuits into parallelizable layers that can be executed with near-constant space-time overhead. In particular, circuits with spatially repeated structures can be compiled efficiently into this architecture. As a concrete example, we show that quantum dynamics simulations on a lattice, such as those in Ref.~\cite{haah2021quantum}, naturally conform to this parallel structure and can be implemented with our high-rate logical operations.

In addition to the asymptotic gadgets, which apply for general qLDPC codes, we also introduce a near-term-friendly scheme based on new instances of finite-size, self-dual Bivariate-Bicycle (BB) codes~\cite{bravyi2024high}. 
These codes offer high encoding rates $(\sim1/10)$, transversal global Clifford gates, translation automorphisms~\cite{sayginel2024fault}, and structured logical operators. 
Leveraging these features, we devise a tailored magic-state cultivation procedure~\cite{gidney2024magic}, which directly transfers magic states from multiple small topological codes into one higher-distance qLDPC code \emph{in parallel}, resulting in effectively high-rate global $T$ gates.
By combining the cultivated $T$ gates with native transversal Clifford gates and matching the symmetry of the codes with that of the circuit, we demonstrate that early fault-tolerant quantum simulations of lattice Hamiltonians can be performed at a logical scale about an order of magnitude larger than unencoded physical circuits (assuming a physical gate error rate of $10^{-3}$), while requiring about $4\times$ lower space–time overhead than analogous surface-code protocols achieving similar logical error rates and employing transversal Clifford gates and parallel magic-state cultivations (Fig.~\ref{fig:quantum_simulation}).
The protocol’s performance is currently limited by the parallel magic-state cultivation scheme, which involves small codes (distance-$3$ color codes and distance-$8$ BB codes) and a relatively large ancilla overhead for their coupling. 
Using larger-distance codes and more efficient ancilla constructions should yield further improvements and greater advantages from qLDPC codes.

These estimates suggest that our high-rate protocols offer a promising path toward low-overhead ``megaquop'' quantum processors capable of demonstrating quantum utility~\cite{preskill2025beyond}.
For lower error-rates and higher-distances, the advantages of high-rate qLDPC codes and operations are expected to grow even further. 

\section{Batched high-rate logical operations for qLDPC codes \label{sec:logical_gates}}
Fault-tolerant quantum computation with low space-time costs necessitates not only high-rate quantum codes but also high-rate quantum operations.
For instance, state-of-the-art surgery schemes for performing logic with high-rate qLDPC codes perform $m$ logical Pauli-product measurements using $m$ \textit{separated} ancilla systems.
Since the operations are implemented using disjoint space-time volume (see Fig.~\ref{fig:overview}c), this limits the efficiency of the computation, and indeed the procedure generally requires an ancilla of size $\tilde{\mathcal{O}}(d)$ and $d$ code cycles, resulting in $\tilde{\mathcal{O}}(d^2)$ space-time cost per measurements. 
In contrast, a transversal CNOT between two high-rate $[[n, k=\Theta(n), d]]$ qLDPC codes directly acts on encoded information, and hence manipulates all logical operators using the same set of $n$ physical CNOTs (see Fig.~\ref{fig:overview}d), 
resulting in $\mc{O}(1)$ space-time cost per gate. 

Here, we expand the set of high-rate logical operations beyond global transversal gates by constructing new batched gadgets \emph{for any CSS qLDPC code} that can efficiently implement (i) fault-tolerant syndrome extraction and state preparation and (ii) universal addressable Clifford gates, with \emph{constant space-time overhead}.
We further show how to implement (iii) global non-Clifford gates with an overhead only slowly growing with $k$. 
Note that, to prove the fault tolerance of these gadgets, we will assume a minimum-weight decoder~\cite{gottesman2013fault} and fast classical computation.

\subsection{Batched single-shot syndrome extraction}
\begin{figure*}
    \centering\includegraphics[width=1\linewidth]{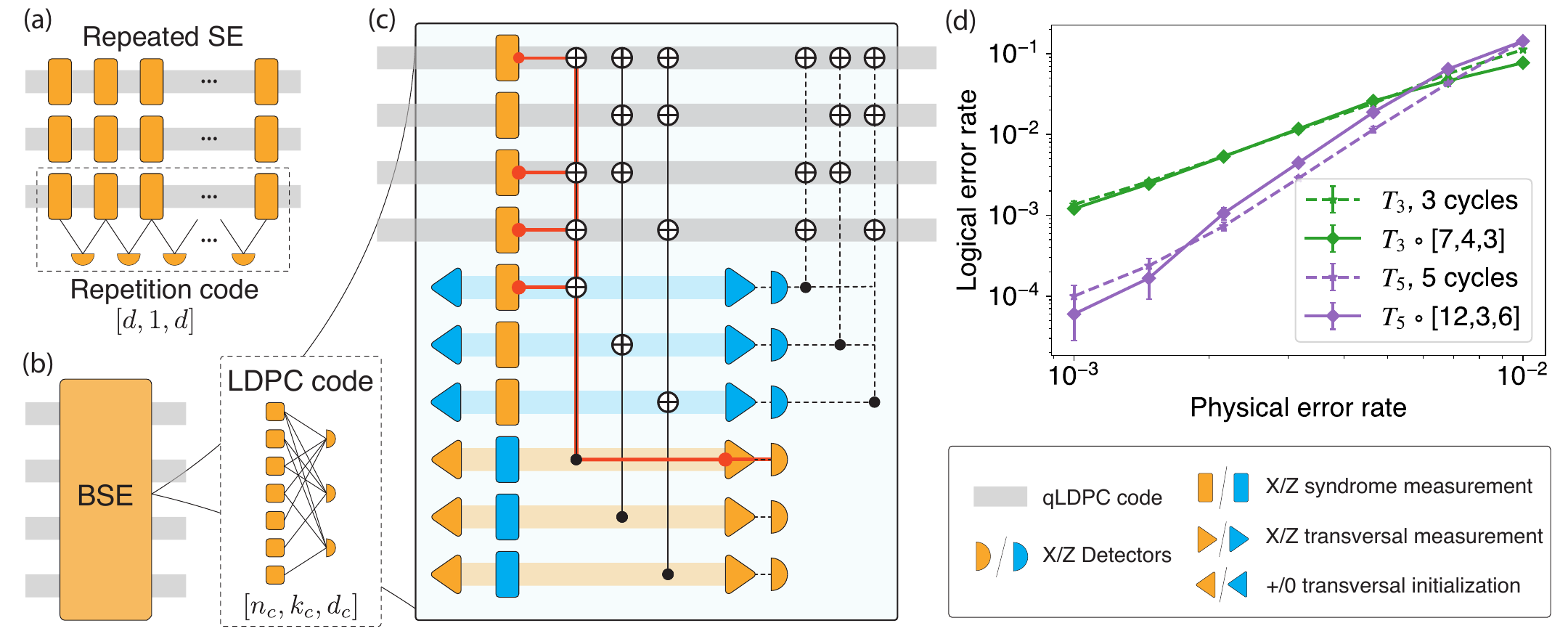}
    \caption{\textbf{Batched syndrome extraction}. 
    Compared to the standard protocol (a), where each quantum code’s syndromes are protected by disjoint, low-rate repetition codes along the time direction---requiring $\mathcal{O}(d_C)$ code cycles---the BSE gadget (b) protects syndromes with a joint, high-rate $[n_C, k_C, d_C]$ classical code spanning multiple quantum codes in the spatial direction, reducing the time cost to a constant depth.
    (c) We illustrate a concrete implementation of BSE with the $[7,4,3]$ Hamming code $\mathcal{C}$, whose Tanner graph is shown in (b).
    By applying transversal CNOTs, information about syndrome errors are transferred to each orange ancilla code block, based on the structure of $\mathcal{C}$.
    The circuit detectors are obtained by comparing the transversal $X$ measurements with the round of noisy $X$ syndromes, and effectively form metachecks that ensure no low-weight syndrome error goes undetected.
    (d) Numerical simulations of BSE are performed using a distance-$d$ toric code $T_d$ for the qLDPC code, and either a $[7,4,3]$ Hamming code~\cite{hamming1950error} or a $[12,3,6]$ quasi-cyclic code~\cite{xu2024fast} for the classical BSE code.
    We simulate the circuit in (c) for preparing a batch of logical $\ket{0}$ states in single shot under standard circuit-level depolarizing noise, excluding idling errors.
    Note, the logical error rate includes so-called ``time-like logical errors'', which are sensitive to syndrome errors.
    The results are compared against standard syndrome extraction repeated $d$ times (dashed lines), and shows that BSE achieves similar performance with only $O(1)$ spacetime overhead.
    }
    \label{fig:BSE_main}
\end{figure*}

Fault-tolerant syndrome extraction is a central subroutine in error-corrected quantum computation.
In particular, as each operation in a stabilizer measurement circuit is noisy, the measured values may themselves be erroneous, making it difficult to extract a reliable syndrome.
As shown in Fig.~\ref{fig:BSE_main}(a), the standard approach for running multiple copies of a $[[n,k,d]]$ qLDPC code $\mathcal{Q}$ is to repeat the syndrome measurements $d_C = \Theta(d)$ times for each block. 
This effectively protects each syndrome with an independent classical repetition code of parameters $[n_C = d_C, 1, d_C = \Theta(d)]$~\cite{bergamaschi2024fault,raussendorf2005quantum}.
Because these repetition codes are embedded in disjoint space–time volumes, they correspond to low-rate operations and incur a $\Theta(d)$ time overhead.

To realize high-rate logical operations, we introduce batched syndrome extraction (BSE), which replaces disjoint repetition codes with a high-rate $[n_C = \mathcal{O}(k_C), k_C, d_C = \Theta(d)]$ classical LDPC code $\mc{C}$, such that the check matrix $H_C \in \mathbb{F}_2^{r_C \times n_C}$ is embedded across multiple $\mathcal{Q}$ blocks in the spatial direction (see Fig.~\ref{fig:BSE_main}(b)).
As shown in Fig.~\ref{fig:BSE_main}(c), a BSE gadget starts by performing a single round of syndrome measurements on $k_C$ data $\mathcal{Q}$ blocks and $n_C - k_C$ ancillary blocks, together corresponding to the $n_C$ bits of $\mathcal{C}$. 
The checks of $\mathcal{C}$ are then measured using $r_C$ additional ancillary $\mathcal{Q}$ blocks: transversal CNOTs are applied between these ``check" blocks and the $n_C$ ``bit" blocks according to $H_C$, followed by transversal measurements.
Crucially, the circuit redundancy introduced by the transversal CNOTs allows the final measurements, together with the initial syndromes, to form detectors (metachecks) that correct initial syndrome errors.
Thus, the BSE circuit achieves fault tolerance with only one round of parallel syndrome extraction and, for constant-rate $\mathcal{Q}$ and $\mathcal{C}$, constant space–time overhead.

We formally prove the fault tolerance and threshold of the procedure in Supplement Note I (using a family of good classical LDPC codes from randomized constructions~\cite{richardson2008modern}), by showing with high-probability that the circuit is equivalent to a round of \textit{noise-free} syndrome extraction on $k_C$ blocks, sandwiched by local physical noise.

\begin{theorem}[Batched syndrome extraction]
    Given a $[n_C, k_C, d_C]$ classical LDPC code with sufficient expansion and $k_C$ blocks of any $[[n, k, d]]$ qLDPC CSS code $\mc{Q}$ with $d_C \geq d$, we can perform a round of stabilizer measurement for $Q^{\otimes k_C}$ fault-tolerantly using $2(n_C-k_C)$ ancillary $\mc{Q}$ blocks and a constant-depth physical circuit.
    \label{theorem:batched_SE}
\end{theorem}

Since Theorem~\ref{theorem:batched_SE} shows that BSE is equivalent to noise-free syndrome extraction followed by local physical errors, it enables not only single-shot QEC, but also single-shot state preparation and code deformation, which are typically more demanding. 
For example, 2D hypergraph product (HGP) codes constructed from expanding classical codes natively support single-shot QEC~\cite{leverrier2015quantum, tremblay2022constant} but not single-shot state preparation~\cite{quintavalle2020single, xu2024fast} due to the lack of soundness~\cite{campbell2019theory}.
We further show that BSE can be applied to achieve single-shot (generalized) lattice surgery~\cite{horsman2012surface, cohen2022low, cross2024improved} in Supplement Note I.

We numerically study the performance of BSE for preparing the computational-basis logical states of the $d=3$ and $d=5$ toric codes
in Fig.~\ref{fig:BSE_main}(d), 
using two simple and practically relevant high-rate classical codes --- the $[7,4,3]$ Hamming code~\cite{hamming1950error} and the $[12,3,6]$ quasi-cyclic code~\cite{xu2024fast}. 
To probe if the syndrome extraction is fault-tolerant, we measure both the traditional space-like logical operators of the quantum codes, as well as time-like operators that are sensitive to residual syndrome errors~\cite{Gidney_stability_2022} (see Supplement Note I for the simulation details).
Comparing repeated rounds of syndrome measurement (dotted lines) with BSE (solid lines), we observe that BSE does not significantly impact the (pseudo-)threshold and logical performance of the underlying quantum code.
This demonstrates the potential of BSE to significantly reduce the space-time costs of quantum fault tolerance in practice with $n_C/k_C \lesssim d_C$.

\subsection{Batched addressable Clifford gates \label{sec:batched_Cliffords}}
\begin{figure}
    \centering
    \includegraphics[width=1\linewidth]{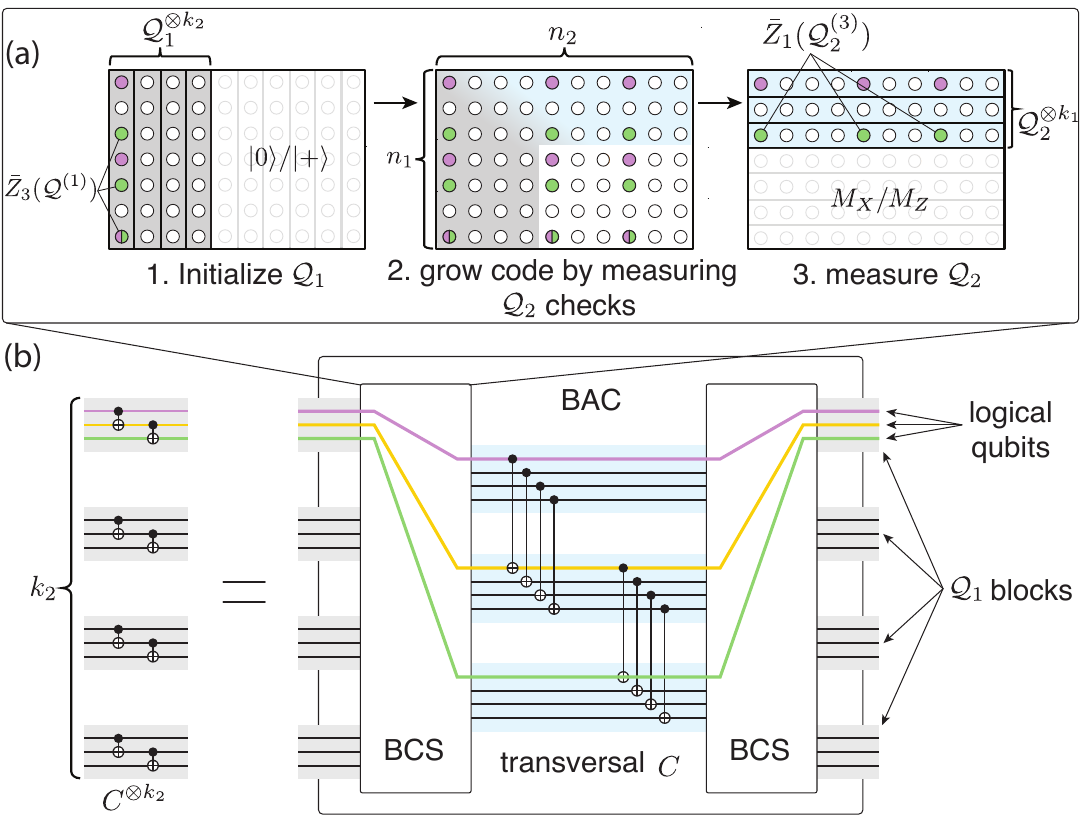}
    \caption{\textbf{Batched code switching and addressable Clifford gates}. 
    (a): A batched code-switching (BCS) gadget that converts between $k_2$ blocks of a $[[n_1, k_1, d_1]]$ CSS qLDPC code $\mathcal{Q}_1$ (a column colored in gray) and $k_1$ blocks of another $[[n_2, k_2, d_2]]$ CSS qLDPC code $\mathcal{Q}_2$ (a row colored in blue), during which the $i$-th logical qubit of the $j$-th $\mathcal{Q}_1$ block, $\mathcal{Q}_1^{(j)}$, is routed to the $j$-th logical qubit of the $i$-th $\mathcal{Q}_2$ block, $\mathcal{Q}_2^{(i)}$ (see the first and third logical operators colored in pink and green, respectively, of $\mathcal{Q}_1^{(1)}$). The physical implementation can be visualized on a 2D grid: $k_2$ columns of $\mathcal{Q}_1$, along with $n_2 - k_2$ ancilla $\mathcal{Q}_1$ initialized in logical $\ket{0}$ or $\ket{+}$, are first entangled into a joint $n_1 n_2$-qubit intermediate code via measuring the $\mc{Q}_2$ checks, where the logical operators are delocalized across the grid, and then disentangled into $k_1$ rows of $\mathcal{Q}_2$ via transversal measurements on the complementary rows.
    (b) A batched addressable Clifford (BAC) gadget that implements an in-block addressable Clifford gate $C$ on a batch of $k_2$ $\mc{Q}_1$ blocks by BCS to $k_1$ blocks of $\mc{Q}_2$ and implementing $C$ transversally. 
    }
    \label{fig:BCS_BAC_main}
\end{figure}

Transversal gates, e.g. transversal CNOTs between two high-rate CSS codes, are high-rate logical operations, but with limited addressability~\cite{guyot2025addressability}.
Here, we present a new fault-tolerant gadget for implementing high-rate and \textit{addressable} Clifford gates for any CSS qLDPC codes.The core of our construction is a new code-switching gadget~\cite{anderson2014fault, bombin2016dimensional} between \emph{any pair of} CSS qLDPC codes:

\begin{theorem}[Batched code switching]
\label{theorem:batched_code_switching}
    Given two CSS qLDPC codes $\mc{Q}_1$ and $\mc{Q}_2$, with parameters $[[n_1, k_1, d_1]]$ and $[[n_2, k_2, d_2]]$ respectively,
    and a set of $K = k_1 k_2$ logical qubits encoded into $k_2$ blocks of $\mc{Q}_1$, there is a fault-tolerant procedure with constant space-time overhead to switch the encoding into $k_1$ blocks of $\mc{Q}_2$. The $i$-th logical qubit in the $j$-th $\mc{Q}_1$ block becomes the $j$-th logical qubit in the $i$-th $\mc{Q}_2$ block after code switching.
\end{theorem}
As illustrated in Fig.~\ref{fig:BCS_BAC_main}(a), a BCS gadget can be implemented on a 2D grid of qubits: $k_2$ columns of $\mathcal{Q}_1$ code blocks, augmented with $n_2 - k_2$ ancillary $\mathcal{Q}_1$ blocks, are first entangled into a joint $n_1 n_2$-qubit (subsystem) intermediate code by measuring the $\mc{Q}_2$ checks and then disentangled into $k_1$ rows of $\mathcal{Q}_2$ code blocks via transversal measurements on the $n_1 - k_1$ complementary rows.
Conceptually, the intermediate code resembles a concatenated $\mathcal{Q}_1$–$\mathcal{Q}_2$ code~\cite{yamasaki2024time, nguyen_qldpc_2024}, with logical operators delocalized across the 2D grid in a tensor-product structure. The protocol thus parallels one in which information is first encoded into the concatenated code and then decoded to $\mathcal{Q}_2^{\otimes k_1}$~\cite{nguyen_qldpc_2024}.
However, it achieves the same effect with a significantly simpler implementation—requiring only single-shot measurements of the stabilizers of $\mc{Q}_1^{\otimes k_2}$ and $\mc{Q}_2^{\otimes k_1}$ using the BSE gadget.
See Sec.~\ref{sec:app_BCS} and Supplement Note II for its detailed construction and proof of fault tolerance.
Notably, the BCS gadget provides a versatile way of interfacing two quantum codes, independent of their code structure. 

As shown in Fig.~\ref{fig:BCS_BAC_main}(a), the BCS gadget routes logical qubits within the same code block in the $\mc{Q}_1$ encoding into different blocks in the $\mc{Q}_2$ encoding. 
Using this, we can now turn global (across-block) logical gates on the $\mc{Q}_2$ codes into locally-addressable (in-block) logical gates on the $\mc{Q}_1$ codes.
As illustrated in Fig.~\ref{fig:BCS_BAC_main}(b), a local pattern of $k_1$-qubit in-block CNOTs can be applied to $\mc{Q}_1^{\otimes k_2}$ by switching to the $\mc{Q}_2$ encoding using the BCS gadget, applying transversal inter-block CNOTs between the $k_1$ $\mc{Q}_2$ blocks, and switching back to the $\mc{Q}_1$ encoding.

By choosing $\mc{Q}_2$ to be a self-dual code supporting global transversal Hadamard and $S$ gates, e.g. using the self-dual BB codes in Sec.~\ref{sec:magic_state_cultivation}, we can also implement addressable Hadamard gates and $S$ gates for $\mc{Q}_1$ in batches, completing the full Clifford group. To guarantee these batched Clifford gates are of constant rate asymptotically, we could choose $\mc{Q}_2$ as the family of hypergraph product codes~\cite{tillich2014quantum} with $[[n_2 = \Theta(d_2^2), k_2 = \Theta(d_2^2), d_2 = \Omega(d_1)]]$ and implement the global Hadamard and $S$ gates by carefully leveraging their native fold-transversal gates~\cite{quintavalle2022partitioning}. 
This gives us a construction of the batched addressable Cliffords (BAC) gadget with asymptotically constant space-time overhead:

\begin{theorem}[Batched addressable Cliffords] 
    Given $N_B = \Omega(d_1^2)$ blocks of any 
    $[[n_1, k_1, d_1]]$ CSS qLDPC code $\mc{Q}_1$ and any target $k_1$-qubit Clifford circuit $C$ with a phyiscal depth $D_P$, we can implement a batched Clifford circuit $C^{\otimes k^{\prime}}$ on any 
    subset of $k' \leq N_B$ $\mc{Q}_1$ code blocks fault-tolerantly using $\mc{O}(N_B)$ ancillary $\mc{Q}_1$ blocks and in a depth $\mc{O}(D_P)$. 
    \label{theorem:batched_Cliffords}
\end{theorem}
The batch size $\Omega(d_1^2)$ arises from our choice of the intermediate codes $\mc{Q}_2$—hypergraph product codes with logical dimension $k_2 = \Theta(d_2^2) = \Omega(d_1^2)$. The targeted Clifford circuit can then be applied to any subset of the $\Omega(d_1^2)$ $\mc{Q}_1$ blocks by replacing the inactive blocks with dummy ones. 
See Sec.~\ref{sec:app_BAC} and Supplement Note III for the proof of Theorem~\ref{theorem:batched_Cliffords} as well as the detailed construction of the BAC gadget. When using constant-rate 
$\mc{Q}_1$, Theorem~\ref{theorem:batched_Cliffords} implies constant space-time overhead for implementing $C^{\otimes k^{\prime}}$.

While the space-time overhead is asymptotically constant (i.e. independent of $d_1$), we can examine more closely what contributes to the constant factor. 
If $\mathcal{Q}_1$ and $\mathcal{Q}_2$ have rates $r_1 = k_1/n_1$ and $r_2 = k_2/n_2$, and if we use a classical code $\mathcal{C}$ with rate $r_c$, then the spacetime overhead of the general BCS protocol scales multiplicatively as $\sim 1/(r_1 r_2 r_c)$.
In certain settings, this encoding rate can be further improved, by e.g. leveraging additional properties of the codes $\mathcal{Q}_1$ and $\mathcal{Q}_2$. For example, in Supplement Note II, we show that for hypergraph-product codes, we can exploit their tensor product structure and use homomorphic logical measurements~\cite{xu2024fast} (now implemented fault-tolerantly in constant-time via BSE gadgets), to perform an equivalent BCS operation with \textit{additive} spacetime overhead $\sim (1/r_1 + 1/r_2)/r_c$, a substantial reduction for small constants $r_1$, $r_2$, and $r_c$.

\subsection{Global non-Clifford gates} 
\begin{figure*}
    \centering
    \includegraphics[width=1\linewidth]{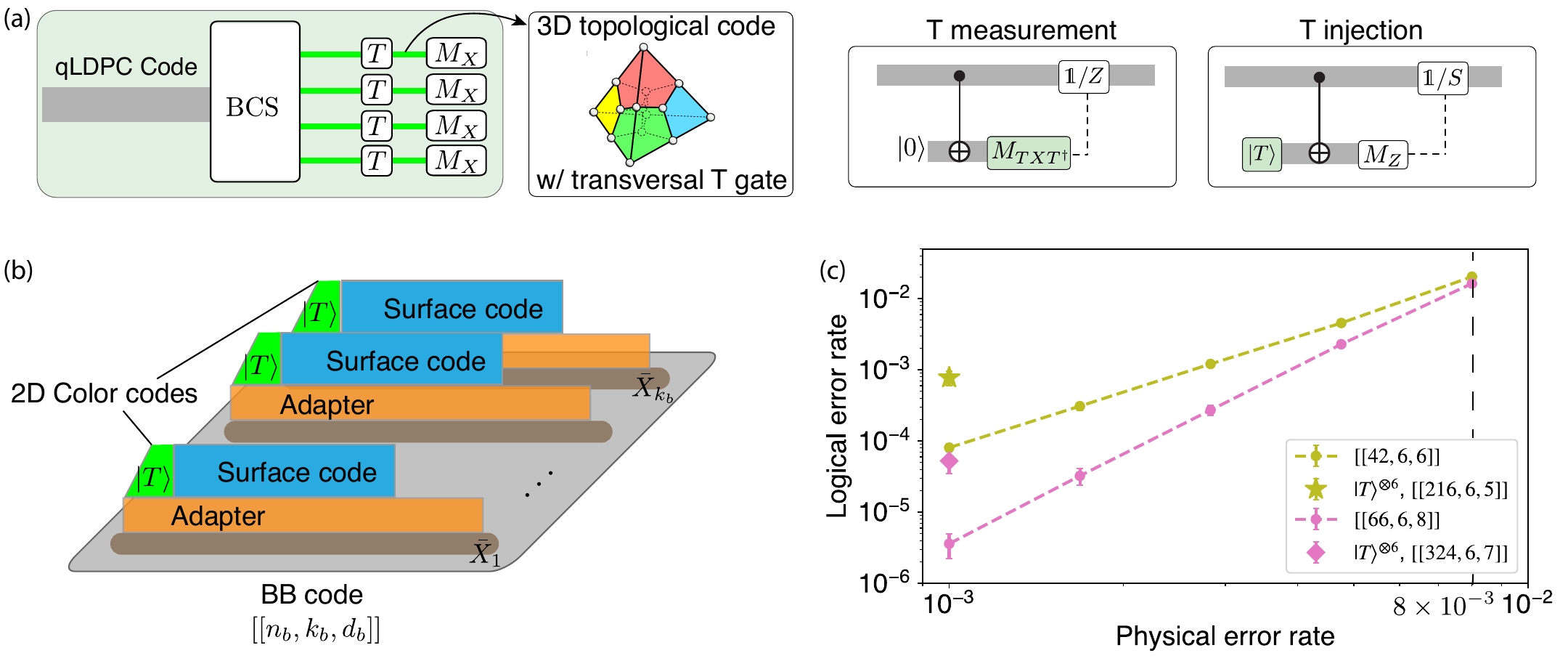}
    \caption{\textbf{Parallel non-Clifford gates.} 
    (a) Batched code switching enables parallel non-Clifford gates in a $[[n_1, k_1, d_1]]$ CSS qLDPC code $\mc{Q}_1$ by switching to/from $k_1$ copies of a $[[n_3,1,d_3=\mc{O}(\log n_1)]]$ code $\mc{Q}_3$ with transversal $T$ gates (e.g., 3D color codes). Global logical $T$ gates on a target $\mc{Q}_1$ block can be implemented via either $T$ measurements (left) or $|T\rangle$-state injection (right) on an ancillary $\mc{Q}_1$ block using the BCS gadget. 
    The $T$ measurement scheme requires only Pauli feedforward on the target code, as opposed to local Clifford gates with higher compilation cost.
    (b) Parallel magic state cultivation for a $[[n_b,k_b,d_b]]$ BB code $\mc{Q}_{\mr{BB}}$ with disjoint logical operators $\{\bar{X}_i\}_{i \in [k_b]}$. 
    The protocol transfers $k_b$ high-fidelity $\ket{T}$ states cultivated in small 2D color codes~\cite{gidney2024magic} into $\mc{Q}_{\mr{BB}}$ in parallel, each via an interface of a thin surface code and modified adapter systems~\cite{swaroop2024universal}. 
    The final $\ket{T}^{k_b}$ states are stored in the merged $[[n>n_b,k=k_b,d\sim d_b]]$ color-surface-BB code $\mc{Q}_{\mr{CSBB}}$ shown in (b).
    (c) Logical error rates (per logical qubit) of cultivated $\ket{T}^{\otimes 6}$ states stored in $[[216,6,5]]$ (green star) and $[[324,6,7]]$ (pink diamond) $\mc{Q}_{\mr{CSBB}}$, starting from $[[42,6,6]]$ and $[[66,6,8]]$ $\mc{Q}_{\mr{BB}}$, compared to $\mc{Q}_{\mr{BB}}$ memory logical error rates (per logical qubit, per code cycle) (dashed).
    }
    \label{fig:BB_cultivation}
\end{figure*}
With addressable Clifford gates provided in Sec.~\ref{sec:batched_Cliffords}, only global non-Clifford gates are needed to complete a universal fault-tolerant gate set (see Sec.~\ref{sec:app_addressable_T}). 
To this end, we use code switching to small topological codes that natively support non-Clifford operations, and show how to build scalable and efficient implementation of global non-Clifford gates for generic qLDPC codes. We further present an optimized, finite-size construction, based on new self-dual bivariate bicycle codes.

\subsubsection{Parallel magic state injection and distillation \label{sec:magic_state_injection}}
Our first strategy implements magic state injection and distillation protocols for general high-rate qLDPC codes. 
The goal is to inject initial magic states---noisy but with bounded error rates---into the qLDPC code blocks, from which distillation can be carried out using their transversal Clifford gates, primarily CNOTs~\cite{xu2024fast}. To supply the initial $\ket{T} := T|+\rangle$ ($\ket{\mr{CCZ}}:= \mr{CCZ}\ket{+}^{\otimes 3}$) states, we leverage small topological codes that support transversal non-Clifford gates (e.g., 3D color codes~\cite{bombin2007exact, bombin2016dimensional} and 3D surface codes~\cite{vasmer2019three}), and perform the injection using the BCS gadget.

A central challenge arises from the overlapping support of logical qubits in high-rate qLDPC codes, which can cause noise to spread in a correlated way. To enable scalable and effective distillation, it is crucial that the \emph{marginal} logical error rate---the probability that any individual logical magic state has an error---remains bounded and does not grow with the code block size.

A naïve scheme injects $k_1$ copies of physical noisy magic states into a $[[n_1, k_1, d_1]]$ data code $\mc{Q}_1$ by code switching from $k_1$ copies of a trivial $[[1,1,1]]$ code 
using the BCS gadget. However, as analyzed in Sec.~\ref{sec:app_magic_injection}, this approach leads to an unbounded marginal error rate $p_L^{\text{marginal}} = \Omega(p d_1)$, which grows with the code distance and hence the block size.

To address this, we propose code-switching from a nontrivial $[[n_3, 1, d_3]]$ auxiliary code $\mc{Q}_3$ with a distance that slowly scales with $n_1$. 
For instance, as shown in Fig.~\ref{fig:BB_cultivation}(a), using small topological codes such as 3D color codes---which support transversal $T$ gates---enables preparation of higher-fidelity initial $\ket{T}$ states. 
As detailed in Sec.~\ref{sec:app_magic_injection}, by choosing $d_3 = \tilde{\mc{O}}(\log n_1)$, we obtain an injection scheme for \emph{any CSS qLDPC codes} in which the marginal logical error rate remains bounded. 
We can similarly inject $\ket{\mr{CCZ}} := \mr{CCZ}|+\rangle^{\otimes 3}$ by code switching from 3D surface codes~\cite{vasmer2019three}.

This protocol for constant-error rate magic state injection is, to the best of our knowledge, state-of-the-art. 
The slightly non-constant overhead is bottlenecked by the use of 3D topological codes, leading to space-time overhead $\tilde{O}(d_3^3) = \tilde{\mc{O}}(\log^3(n_1))$.
This outperforms the $\mr{Poly}(n_1)$ overhead of surgery based approaches~\cite{zhang2025constant}, and the $\mr{Polylog}(n_1)$ overhead of concatenated schemes which typically contain an exponent $\gg 3$~\footnote{the component depends on the rate of the base code and its depth of implementing a target logical circuit}.

Building from the injection protocol for a code $\mc{Q}_1$, we can further reduce the error rate of the magic states by running a logical distillation circuit on $\mc{Q}_1$ code blocks using transversal gates~\cite{xu2024fast} based on existing magic state distillation protocols~\cite{bravyi2005universal}. 
For example, a standard 15-to-1 distillation protocol consumes 15 $\mc{Q}_1$ blocks of injected magic states to produce one $\mc{Q}_1$ block of higher-fidelity magic states. 
To avoid the costly random Clifford corrections required when teleporting a $T$ gate using a $\ket{T}$ state—which can lead to significant compilation overhead (see Sec.~\ref{sec:algorithm_structure})—we instead implement $T$ gates via faulty $T$ measurements~\cite{litinski2019magic}, which can be implemented by code switching from $\mc{Q}_1$ back to $\mc{Q}_3^{\otimes k_1}$, effectively reversing the state-injection protocol (see Fig.~\ref{fig:BB_cultivation}(a)). As further detailed in Supplement Note IV, high-fidelity $T$ measurements on $\mc{Q}_1$ can be ``distilled" from many faulty $T$ measurements by further encoding (using Def.~\ref{def:encoding_decoding_protocol}) into a distillation code $\mc{Q}_D$ supporting transversal $T$ gates via, e.g. the Bravyi and Haah distillation protocol~\cite{bravyi2012magic}.  We formalize these results in the following theorem:

\begin{theorem}[Parallel non-Clifford gates]
    Given $N_B = \Omega(d_1)$ blocks of any $[[n_1, k_1, d_1]]$ CSS qLDPC code $\mc{Q}_1$, we can prepare global $\ket{T}$ ($\ket{\mr{CCZ}}$) states or perform global $T$ ($\mr{CCZ}$) measurements on $\mc{Q}_1^{\otimes N_B}$ with a marginal error rate $\mc{O}(1)$ and a space-time overhead $\tilde{\mc{O}}(\log^3 n_1)$.  
    Furthermore, the faulty $T$ ($\mr{CCZ}$) measurement protocol can be combined with any magic-state distillation protocol for implementing lower-error-rate global $T$ ($CCZ$) gates.  \label{theorem:batched_global_magic}
\end{theorem}
Note that Theorem~\ref{theorem:batched_global_magic} assumes a batch size $\mc{\Omega}(d_1)$ due to the implicit use of the BSE gadget (involving a good classical LDPC code with logical dimension $k_C = \Omega(d_1)$) for measuring the stabilizers of $\mc{Q}_1^{\otimes k_C}$ in single shot when code switching from the 3D topological codes $\mc{Q}_3^{\otimes k_C k_1}$. 
Such a batch size $k_C$ can be relaxed in practice when we do not strictly need single-shot syndrome extraction.
In principle, the non-Clifford gates need not be global. For example, addressable $T$ gates on a single $\mc{Q}_1$ block can be realized by code switching from $\mc{Q}_3^{\otimes k_1}$, where only a subset encodes $\ket{T}$ states. Here, however, we focus on global non-Clifford gates, as they allow us to compress the overall space–time overhead per non-Clifford operation.

Finally, we note that, 
using the recently-developed high-rate magic state distillation schemes~\cite{wills2024constant, golowich2025asymptotically, nguyen_qldpc_2024}, we can implement global non-Clifford gates in parallel on any CSS qLDPC code with almost a constant asymptotic space-time overhead.
For instance, using the distillation scheme of Ref.~\cite{nguyen_qldpc_2024}, which utilizes high-rate Reed-Solomon codes as distillation codes, exponentially suppresses the error rates in $k_1$ with a space-time overhead $\mc{O}(k_1^{o(1)})$, where $o(1) \sim 1/\log \log k_1$~\cite{nguyen_qldpc_2024}.  

\subsubsection{Parallel magic state cultivation \label{sec:magic_state_cultivation}}
Recent advances show that high-fidelity magic states can be prepared using small topological codes $\mc{Q}_3$, such as distance-3 2D color codes, operated in error-detection mode by measuring Clifford operators~\cite{chamberland2020very, gidney2024magic}. Final error-corrected states are then obtained by code-switching (``escaping'') into a large-distance surface code~\cite{gidney2024magic}. Since $\mc{Q}_3$---in error-detection mode---has relatively high encoding rate, escaping directly into high-rate qLDPC codes rather than low-rate surface codes would yield an effectively high-rate magic-state-preparation protocol. While this could be done in principle via the generic BCS protocol (Fig.~\ref{fig:BB_cultivation}(a)), preserving low error rates requires substantial postselection before reaching large qLDPC distance, so applying BCS directly could generate prohibitively low success probability.
As such, we further optimize the code-switching procedure to leverage code-specific structures.

In particular, we propose a more practical scheme, suitable for near-term experiments, by coupling $\mc{Q}_3$ with a qLDPC code $\mc{Q}_1$ using surgery~\cite{cohen2022low, cross2024improved, swaroop2024universal} (Fig.~\ref{fig:BB_cultivation}(b)). To ensure a high-rate protocol, multiple $\mc{Q}_3$ must couple simultaneously to $\mc{Q}_1$ without excessive overhead, enabling parallel magic-state preparation with low space-time cost. This is difficult for generic qLDPC codes with highly overlapping logical operators (e.g., the Gross code~\cite{bravyi2024high}). We resolve this by introducing a new set of bivariate bicycle (BB) codes, which feature disjoint logical operators that enable parallel magic-state cultivation. These codes also exhibit self-duality for transversal Clifford gates, and support useful automorphism gates via simple qubit permutations, which we leverage in Sec.~\ref{sec:quantum_simulation}.
 
Each self-dual BB code is defined on two $l \times m$ blocks of physical qubits with overlapped $X$ and $Z$ checks that are specified by a bivariate polynomial $c \in \mbb{F}_2[x, y]/(x^l - 1, y^m - 1)$.
See Sec.~\ref{sec:app_magic_cultivation} for the detailed construction of these codes. 
We show three examples of these codes in Table~\ref{tab:QC_code_params}, where we fix a choice of $l$ and the check polynomial $c$ and increase the distance of the code up to $10$ by increasing $m$. These codes have parameters $[[n_b = 2lm, k_b = 2l, d_b \leq m]]$ and high pseudo-thresholds $\gtrsim 0.8\%$ (under circuit-level depolarizing noise, excluding idling errors, see Fig.~\ref{fig:BB_cultivation}(c)).
Unlike the BB codes in Ref.~\cite{bravyi2024high}, they have a disjoint logical operator basis 
$\{(\bar{X}_i, \bar{Z}_i)\}_{i \in [k_b]}$, 
where the logical $X$- and $Z$- operators $\bar{X}_{i}$ and $\bar{Z}_i$ of the $i$-th logical qubit overlap and different logical operators have disjoint support. 

\begin{table*}
    \centering
    \begin{tabular}{C{4cm}|C{2cm}|C{3cm}|C{2cm}}
    \hline
    \hline
          Code parameters & $l, m$ & $c$ & $kd^2/n$\\
          \hline
          $[[42, 6, 6]]$ & $3, 7$ & $1 + y^3 + xy^2 + xy^4$ & $\approx 5$ \\
          \hline
          $[[66, 6, 8]]$ & $3, 11$ & $1 + y^3 + xy^2 + xy^4$ & $\approx 6$ \\
          \hline
          $[[78, 6, 10]]$ & $3, 13$ & $1 + y^3 + xy^2 + xy^4$ & $\approx 8$\\
    \hline
    \hline
    \end{tabular}
    \caption{\textbf{List of self-dual BB codes with translational automorphisms and a disjoint logical basis.} For a $\mr{BB}(c, c^T; R_{l, m})$ code defined in Sec.~\ref{def:BB_code}, we list its code parameter $[[n, k, d]]$, choice of $l$ and $m$ as well as the check polynomial $c \in R_{l, m}$, and the qubit saving $kd^2/n$ against the $[[kd^2, k, d]]$ (rotated) surface codes with the same logical dimension and distances.}
    \label{tab:QC_code_params}
\end{table*}

The disjoint logical operator basis of the codes enables us to perform addressable surgery operations~\cite{cross2024improved, williamson2024low} on different logical qubits \emph{in parallel} using separated ancilla systems. In particular, we can transfer $k_b$ high-fidelity logical $\ket{T}$ states hosted in $k_b$ copies of small 2D color codes $\mc{Q}_C$ with parameters $[[n_c \ll n_b, 1, d_c \ll d_b]]$ into the $k_b$ logical qubits of a BB code $\mc{Q}_{\mr{BB}}$ in parallel by simultaneously coupling each copy of $\mc{Q}_C$ to a logical qubit in $\mc{Q}_{\mr{BB}}$ using a separate interface consisting of a thin surface code and an adapter~\cite{swaroop2024universal} (see Fig.~\ref{fig:BB_cultivation}(b)).

The merged color-surface-BB code $\mc{Q}_{\mr{CSBB}}$ has a distance comparable to $d_b$ (see Supplement Note V) and encodes $k_b$ logical qubits each hosting a $\ket{T}$ state inherited from a color code. Analogous to the code-growth procedure for surface codes~\cite{li2015magic}, the code-merging protocol prepares the data qubits of the surface code, the BB code, and the adapters in either $\ket{0}$ or $\ket{+}$ states and measures the stabilizers of the merged code in a certain order, effectively growing the logical operators of the color codes into those of the merged code. To avoid ruining the logical error rates of the magic states, we postselect on detectors at space-time locations with effectively low distances ($\sim d_c$). 
Crucially, since the merged code has a large distance $\sim d_b$, the system grows to the full distance immediately and we thus only need to postselect on a small space-time volume, leading to a large success rate even though the final merged code could be large. See Sec.~\ref{sec:app_magic_cultivation} for the detailed code merging protocol as well as how the postselection is performed.

Finally, the merged code $\mc{Q}_{\mr{CSBB}}$ can be coupled fault-tolerantly to another $[[n_b, k_b, d_b]]$ data BB code $\mc{Q}_{\mr{BB}}^{\prime}$ by transversally measuring out the complement of the BB component in $\mc{Q}_{\mr{CSBB}}$ and applying transversal CNOTs between its remaining BB component and the data BB code $\mc{Q}_{\mr{BB}}^{\prime}$ (see Supplement Note V).
This can be used to perform teleported $T^{\otimes k_b}$ gates on $\mc{Q}_{\mr{BB}}^{\prime}$ in parallel (see Supplement Note V).

We numerically evaluate the performance of the above parallel cultivation protocol for the first two BB codes in Table~\ref{tab:QC_code_params} and plot it in Fig.~\ref{fig:BB_cultivation}(c).
Under a circuit-level noise model excluding idling errors~\cite{xu2024constant} with a $10^{-3}$ physical error rate, using a BP+LSD decoder~\cite{hillmann2025localized}, we estimate that we can obtain cultivated $\ket{T}^{\otimes 6}$ states with a logical error rate (summing over logical $X$ and $Z$ error rates) about $5\times 10^{-5}$ (resp. $8 \times 10^{-4}$) per logical qubit stored in the merged $[[324, 6, 7]]$- (resp. $[[216, 6, 5]]$-) $\mc{Q}_{\mr{CSBB}}$ code, starting from the $[[66, 6, 8]]$- ($[[42, 6, 6]]$-) BB code, with an overall success rate about $35\%$. See Supplement Note V for details. 
We note that, although the color codes are small compared to the BB code, the ancilla systems linking them are much larger: the thin surface codes and adapters (blue and orange in Fig.~\ref{fig:BB_cultivation}(b)) each occupy about twice the space of $\mc{Q}_{\mr{BB}}$. As a result, the merged code is roughly five times larger than the bare BB code, which is the main bottleneck of the protocol, limiting both the space–time cost (including success probability) and the logical error rate. More efficient ancilla constructions will therefore be crucial for substantial performance improvements.

To conclude, by leveraging the disjoint logical basis structure of the presented BB codes, we can implement parallel magic state injection by code switching from small 2D color codes via surgery techniques---a practical alternative to the BCS gadget. Crucially, by carefully applying post-selection and cultivation techniques, we can prepare low-error-rate magic states for all logical qubits in a large high-rate qLDPC code while only using small topological codes, effectively rendering the protocol high rate.

\section{Structured Quantum Algorithms \label{sec:algorithm_structure}}

\begin{figure}
    \centering
    \includegraphics[width=0.5\textwidth]{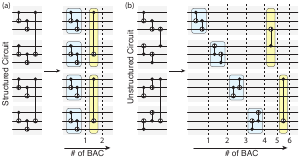}    \caption{\textbf{Fault-tolerant compilation with batched logical gates}. 
    Structured circuits (a) with parallel gates can be compiled more efficiently than unstructured circuits (b) using batched gadgets. We illustrate fault-tolerant implementations of a structured (a) and an unstructured (b) logical CNOT circuit with the BAC gadgets (Theorem~\ref{theorem:batched_Cliffords}) on a batch of qLDPC-encoded sectors (gray blocks). In (a), the circuit decomposes into two layers of parallel gates with identical distributions across $\mc{Q}_1$ blocks, requiring only two BAC gadgets. In contrast, (b) lacks parallel structure, forcing serial execution with the number of BAC gadgets scaling with the logical block size of the qLDPC code. 
    }
    \label{fig:alg_structure}
\end{figure}

The high-rate logical operations discussed above form a flexible, universal gate set for implementing fault-tolerant algorithms with low error-correction overhead.
However in practice, the \textit{compilation cost} of the algorithm is also a significant source of overhead. 
Here, we analyze the structure of circuits which can be efficiently compiled into high-rate logical operations, and develop a set of design principles for co-designing quantum algorithms with the batched gadgets.

We sketch the main idea here and formalize the compilation problem in Sec.~\ref{sec:app_addressable_T}.
Our high-rate gate set $\mb{B}_L$ discussed in Sec.~\ref{sec:logical_gates} is comprised of \emph{batched}, addressable Clifford gates, and global non-Clifford (T) gates. This is a subset of the typical Clifford + T gate set $\mb{B}_P$ used for fault-tolerant compiling of quantum algorithms, since $\mb{B}_L$ is subject to an additional batching constraint---the gates being applied to a batch of qLDPC blocks have to be identical (or identity).
Since $\mb{B}_L$ is more restricted than  $\mb{B}_P$, there will be an extra overhead of further compiling (re-synthesizing) an \textit{arbitrary} Clifford + T circuit into $\mb{B}_L$.
As detailed in Sec.~\ref{sec:app_compilation}, we show that such a worst-case extra compilation overhead scales linearly with the logical size of each qLDPC block.

Nevertheless, we show that a broad class of structured Clifford + T circuits --- those that consist of layers of parallel gates divisible into small sectors with identical patterns --- can be efficiently compiled into $\mb{B}_L$ with only a \emph{constant} overhead. 
This motivates us to design quantum algorithms and circuits with such a parallel structure that can be implemented fault-tolerantly with an extremely low overhead.

More concretely, we illustrate the essential compilation ideas in Fig.~\ref{fig:alg_structure}.
We first divide the logical circuit into small spatial sectors each encoded in a qLDPC block. 
The compilation cost then depends on the distribution of the circuit across its small circuit sectors. 
If a circuit has identical distributions, i.e. different sectors implement the same operations, it can be implemented efficiently with a low compilation overhead. 
As an example, each layer of the brickwork circuit in Fig.~\ref{fig:alg_structure}(a) can be partitioned into qLDPC sectors with identical distributions, and are thus efficiently implementable with a constant compilation overhead with the batched gadgets.
In contrast, circuits without such a parallel structure are costly to implement. As an example, the circuit in Fig.~\ref{fig:alg_structure}(b) has to be compiled sequentially using many batched gadgets, incurring a compilation overhead proportional to the qLDPC logical block size.
See Sec.~\ref{sec:app_compilation} for more details on a concrete compilation scheme and an analysis of its overhead.
We emphasize once more the key algorithmic and circuit feature that enables efficient compilation into our batched gadgets in general: the circuit’s parallel structure, realized as layers of parallel gates with a repeated pattern across multiple small-qubit sectors.

\section{Application to Quantum Simulation \label{sec:quantum_simulation}}

To best leverage high-rate logical operations, our results show that it is important to find and design quantum algorithms with parallel structure built in.
Quantum simulation turns out to be a particularly promising application for these techniques, as we discuss in this section.
The key feature is that the time-evolution unitary for a lattice Hamiltonian with translational symmetry involves parallel, multi-Pauli rotations, which can be readily compiled into high-rate logical operations with only a constant compilation overhead.

\begin{figure*}
    \centering
\includegraphics[width=1.0\linewidth]{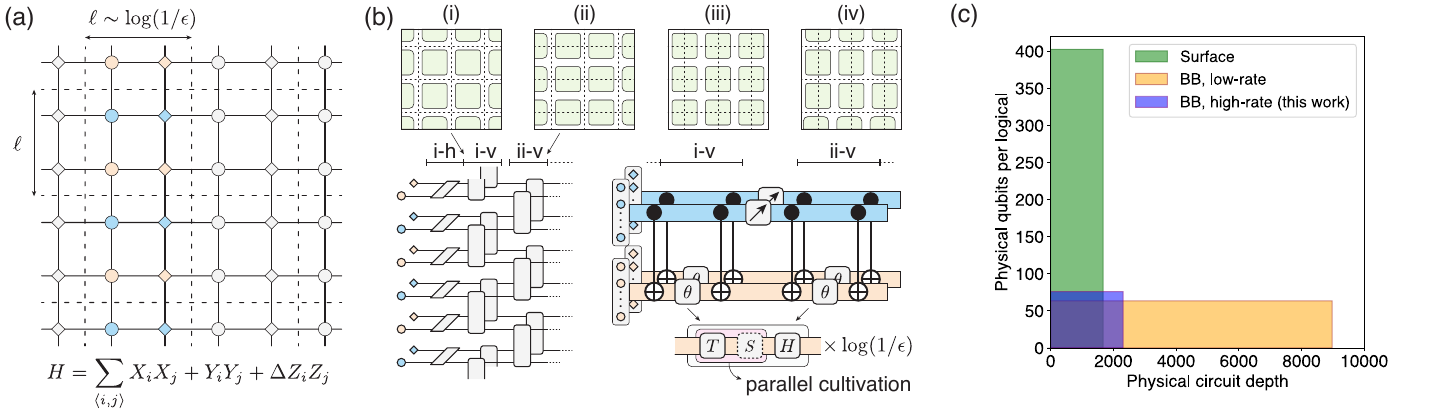}
    \caption{\textbf{Efficient fault-tolerant quantum simulation of lattice Hamiltonian}. 
    (a) The 2D XXZ model on an $L\times L$ periodic lattice.
    (b) HHKL-style brickwork decomposition where each green brick acts on an $\ell\times\ell$ sublattice; translational invariance makes bricks identical and therefore amenable to batched qLDPC gadgets. 
    Bottom-left: alternative Trotterization circuit for $\ell=2$ restricted to two colored columns; vertical interactions factor into two parallel layers of two-qubit rotations (labeled ``i-v'' and ``ii-v''). 
    Bottom-right: column encoding into two qLDPC codes $\mathcal{Q}^{(1)}$ (blue) and $\mathcal{Q}^{(2)}$ (yellow) by grouping the $\alpha$-th qubit of each sublattice into $\mathcal{Q}^{(\alpha)}$. 
    In-sublattice rotations (``i-v") become transversal cross-block $ZZ$ rotations (and, by conjugation, $XX$/$YY$); 
    the offset cross-sublattice rotations (``ii-v'') are realized by utilizing a logical translation gadget that cyclically shifts logical qubits (an arrowed box). Parallel transversal $ZZ$ rotations are compiled into transversal Clifford layers plus global $T$ gates implemented via parallel magic-state cultivation (Sec.~\ref{sec:magic_state_cultivation}; Supplement Note V). 
    (c) Space–time cost per logical qubit (area $=$ space $\times$ time) to implement a layer of small-angle parallel two-qubit rotations (example $\theta=\pi/64$). Blue: high-rate protocol (this work) using the self-dual BB code $[[66,6,8]]$ (Table~\ref{tab:QC_code_params}) and the parallel magic state cultivation; green: leading surface-code protocols using transversal Clifford gates and parallel surface-code cultivation~\cite{gidney2024magic}; yellow: low-rate protocol using the same BB code but sequentially coupling to cultivated surface-code factories~\cite{yoder2025tour}.
    }
    \label{fig:quantum_simulation}
\end{figure*}

\subsection{Simulating local Lattice Hamiltonians \label{sec:simulation_batched_gates}}

We start by presenting a general technique for simulating time-evolution $e^{-i H t}$ under a geometrically local Hamiltonian $H = \sum h_X$ composed of $K$ sites, where $h_X$ are terms that act non-trivially only on qubits in a local region $X$.

To achieve high-accuracy simulation, we start by using the HHKL decomposition~\cite{haah2021quantum} to break time-evolution over the entire lattice $\Lambda$ into smaller blocks, while introducing an exponentially small error in block size.
For instance, for a 2D $L$ by $L$ lattice $\Lambda$, we divide it into $\ell \times \ell$ sized blocks $\{\Lambda_b \subset \Lambda\}_b$.
Then, we define four translations of the blocks $T_s(\Lambda_b)$, for $s = $ (i), (ii), (iii), and (iv), as illustrated in Fig.~\ref{fig:quantum_simulation}(b). 
The corresponding HHKL decomposition generates a brickwork circuit with four layers, written formally as:
\begin{equation}
    e^{-i H t} \approx \prod_{s} U^{s} = \prod_{s} \bigotimes_{b} U^{s}_{T_s(\Lambda_b)},
    \label{eq:HHKL_decomp}
\end{equation}
where $U^{s}_{Y \subseteq \Lambda}$ acts nontrivially only on the block $Y$.
Note that the backward time evolution invoked by the HHKL algorithm is implicitly absorbed into the unitaries in Eq.~\eqref{eq:HHKL_decomp}, yielding distinct $U^s$ for each layer $s$.
The small unitaries $\{U^s_{T_s(\Lambda_b)}\}$ on each block are further decomposed into circuits with single- and two-qubit gates using approximate Hamiltonian simulation algorithms such as Trotter formulas~\cite{trotter1959product}, linear combination of unitaries~\cite{childs2012hamiltonian}, and qubitization~\cite{low2019hamiltonian}. 
Exact unitary synthesis can also be used when $\ell$ is sufficiently small.

If $H$ is translationally invariant under $\{T_s\}_s$, then the different bricks $\{U^{s}_{T_s(\Lambda_b)}\}_b$ of each layer of the circuit are identical (regardless of the specific Hamiltonian simulation algorithms), thereby exhibiting a parallel structure in Sec.~\ref{sec:algorithm_structure} that can be exploited by our batched qLDPC gadgets with a constant compilation overhead. 
For example, in the 2D setting, we can encode the qubits of each first-layer blocks $\Lambda_b$ into a $[[n_1, k_1 = \ell^2, d_1]]$ qLDPC code $\mc{Q}_1$. 
The first-layer circuit $U^{(i)}$ can be implemented using the batched single-block gates on the $\mc{Q}_1$ blocks. For each layer of (ii), (iii), and (iv), where the bricks are offset from the first-layer bricks, which necessitates inter-block CNOTs for the $\mc{Q}_1$ blocks, we can implement it by grouping every pair of blocks into a larger code block and applying the batched gadgets on these bigger blocks.
Crucially, the inherent parallel structure of the HHKL circuit decomposition ensures that the batched implementation (see Sec.~\ref{sec:algorithm_structure}) incurs only a constant compilation overhead beyond the standard Clifford+T compilation cost.

The choice of the qLDPC code $\mc{Q}_1$ for each block is flexible, provided they satisfy the following constraints to fault-tolerantly simulate a $K$-qubit logical Hamiltonian for time $T$ with accuracy $\epsilon$ using the logical HHKL algorithm:
\begin{enumerate}
\item To ensure the approximation error from decomposing $e^{-iHt}$ into products of unitaries on $\ell \times \ell$ blocks remains small, the block size must satisfy $k_1 = \ell^2= \Omega(\log^2(KT/\epsilon))$, as implied by the Lieb–Robinson bounds in Ref.~\cite{haah2021quantum}.
\item To control the Hamiltonian simulation error within each block, and assuming that the chosen simulation algorithm (e.g., LCU~\cite{childs2012hamiltonian}) has polynomial cost in system size, we require an upper bound on the block size $k_1 = O(\mathrm{polylog}(KT/\epsilon))$.
\item To suppress logical error rates to $\epsilon$, the code distance must satisfy $d_1 = \Omega(\mathrm{polylog}(KT/\epsilon))$.
\item To enable the use of batched logical gadgets with batch size $\Omega(d_1^2)$ (see Theorem~\ref{theorem:batched_Cliffords}), we need $k_1 = O(K/d_1^2)$.
\end{enumerate}

One concrete choice of the $\mc{Q}_1$ parameters satisfying these constraints is $[[n_1 = \Theta(k_1), k_1 = \Theta(\mathrm{polylog}(KT/\epsilon)), d_1 = \Theta(\mathrm{poly}(k_1))]]$. 
The resulting total space–time overhead for the simulation is $\tilde{O}(\mathrm{polylog}(KT/\epsilon))$, primarily from the costs of the Hamiltonian-simulation algorithm for each of the $k_1$-qubit block. Importantly, the additional overhead from fault tolerance is asymptotically low due to three factors: the use of constant-rate codes, batched logical gates with constant compilation overhead, and high-rate magic-state distillation with only a $k_1^{o(1)}$ overhead~\cite{nguyen_qldpc_2024}. The parallel structure of the HHKL algorithm is essential here: without it, batching constraints would incur an $\Omega(k_1)$ compilation overhead (see Sec.~\ref{sec:algorithm_structure}), significantly increasing the total space–time cost.
Finally, note that one could choose larger qLDPC block sizes (e.g., polynomial in $K$) if the Hamiltonian simulation algorithm on small blocks has polylogarithmic cost in system size—for instance, by applying the HHKL algorithm~\cite{haah2021quantum} iteratively within each small block. 

\subsection{Co-design with code  automorphisms\label{sec:BB_simulation}}
Although the HHKL protocol above can be implemented with any qLDPC code by matching the universal batched gadgets to the algorithm’s parallel structure, it demands large batch sizes for adequate error suppression. 
Here,we show that co-designing codes with native symmetries relaxes this batching constraint and makes simulations of smaller lattices feasible on near-term devices. As a concrete example we consider the XXZ model on a 2D $L\times L$ lattice with periodic boundary conditions:
\begin{equation}
H=\sum_{\langle i,j\rangle} X_i X_j + Y_i Y_j + \Delta Z_i Z_j.
\end{equation}

Using first-order Trotterization~\cite{trotter1959product} we implement $e^{-iHt}$ as $\big(e^{-iH\delta t}\big)^{t/\delta t}$ and split each Trotter step into vertical and horizontal layers,
\begin{equation}
e^{-iH\delta t}\approx e^{-iH_v\delta t}\,e^{-iH_h\delta t},
\end{equation}
where $H_v$ (resp. $H_h$) contains only vertical (resp. horizontal) interactions. Dividing the qubits into $2 \times 2$ sublattices turns each Trotter step again into a sequence of brickwork circuits ---  with connectivity analogous to the HHKL circuit with $\ell = 2$. Rather than implementing these bricks with batched gadgets and code switching, we can implement them natively by choosing a qLDPC code that admits (i) global transversal-Cliffords and low-overhead $T$ gates and (ii) a logical translation gadget that cyclically shifts logical qubits under periodic boundary conditions.

Focusing on the vertical interactions, $e^{-iH_v\delta t}=\prod_{c=1}^L e^{-iH_{v,c}\delta t}$ factors into column unitaries that can be executed in parallel. For a given column $c$ subdivide the qubits into two sublattices and apply
\begin{equation}
e^{-iH_{v,c}\delta t}=e^{-iH_{v,c}^{(ii)}\delta t}\,e^{-iH_{v,c}^{(i)}\delta t},
\end{equation}
where $H_{v,c}^{(i)}$ acts within each brick and $H_{v,c}^{(ii)}$ couples bricks (see the ``i-v" and ``ii-v" gate layers in Fig.~\ref{fig:quantum_simulation}(b)). 

More generally, for a brickwork circuit with $\ell \times \ell$ sized blocks, our approach involves encoding a column into $\ell$ qLDPC codes $\{\mathcal{Q}^{(\alpha)}\}_{\alpha\in[\ell]}$, each storing $L/\ell$ logical qubits. 
Then the operations within a sub-block can be implemented transversely across the $\ell$ codes.
For instance, in the XXZ model, each Hamiltonian term decomposes into three layers of $XX$, $YY$, $ZZ$ rotations between two codes, where each rotation is realized by conjugating a global $Z(\theta)$ rotation on one code with transversal CNOT layers.
See the bottom-right panel of Fig.~\ref{fig:quantum_simulation}(b).

Terms between sub-blocks, like $e^{-iH_{v,c}^{(ii)}\delta t}$ in Fig.~\ref{fig:quantum_simulation}(b), can be efficiently implemented by applying the translation auto-morphism to shift one pair of each interaction, such that the transversal interaction couples adjacent sub-blocks. While sites in the same sublattice cannot directly interact using this circuit gadget, this constraint can be readily circumvented by appropriate compilation. 
Horizontal interactions $e^{-iH_h \delta t}$ can be implemented analogously, by applying transversal gates and translation automorphisms to interact qLDPC blocks in different columns.

In what follows, we specify a concrete code choice using the self-dual BB codes in Table~\ref{tab:QC_code_params} and analyze the performance of corresponding circuits.
Because these codes are self-dual, all global Clifford gates are transversal.
As detailed in Supplement Note V, each BB code also admits a low-overhead logical translation gadget via its automorphism gates (see Sec.~\ref{sec:app_magic_cultivation}).
Global $T$ gates are implemented by teleporting a block of $\ket{T}$ states prepared through the parallel cultivation scheme in Sec.~\ref{sec:magic_state_cultivation}, with the subsequent $S$-gate corrections handled as in Supplement Note V.
Finally, the global $Z(\theta)$ rotations in Fig.~\ref{fig:quantum_simulation}(b) can be synthesized from ${H, S, T}$ using standard techniques such as Solovay--Kitaev~\cite{dawson2005solovay} and modern improvements~\cite{kliuchnikov2013fastefficientexactsynthesis,ross2016optimalancillafreecliffordtapproximation}.
As a concrete example, using the $[[66,k_b=6,8]]$ code with $\ell=2$ allows simulation of a lattice of size $L=\ell k_b=12$. 

We now estimate the space–time cost and logical error rate of a layer of parallel $ZZ(\theta)$ rotations based on the $[[66,6,8]]$ BB code, assuming physical error rate $10^{-3}$. 
Each column is encoded in two BB codes of footprint $N_{\mathrm{BB}}=132$, and one of them couples to a merged color–surface–BB code $\mathcal{Q}_{\mathrm{CSBB}}$ of footprint $N_{\mathrm{CSBB}}=648$ for magic-state cultivation. This yields an effective footprint per logical lattice qubit of $(2N_{\mathrm{BB}}+N_{\mathrm{CSBB}})/2k_b \approx 76$.
For a small angle such as $\theta=\pi/64$, a synthesis using $n_T=11$ $T$ gates achieves error $<10^{-4}$. Each global $T$ layer (including the Clifford fixups) requires about $15$ $\mathcal{Q}_{\mathrm{CSBB}}$ code cycles of physical depth $14$, so the total is $
t_{\mathrm{BB}}=n_T\times 15\approx 165$ code cycles, or $ 
T_{\mathrm{BB}}=t_{\mathrm{BB}}\times14\approx 2300$ physical depth.
With logical error per cycle $p_{L,0}\approx4\times10^{-6}$, the estimated error per logical qubit for one layer of parallel rotations is $
p_L \approx t_{\mathrm{BB}}\,p_{L,0}\approx 7\times10^{-4}$.

Assuming the same circuit for parallel two-qubit rotations, an analogous surface-code protocol---i.e., one achieving similar logical error rates, using transversal Clifford gates~\cite{zhou2024algorithmic} and parallel, cultivated surface-code magic-state factories~\cite{gidney2019efficient}—requires about $5\times$ more qubits (green rectangle in Fig.~\ref{fig:quantum_simulation}(c)).
In contrast, a low-rate BB-code protocol that uses the same BB codes but couples sequentially to the surface-code magic-state factory~\cite{gidney2019efficient, yoder2025tour} incurs $\sim 7\times$ more code cycles due to the serialization of global $T$ gates. 
Relative to the parallel surface-code protocol, this design merely trades time for space, leaving the overall space–time overhead similar (orange rectangle in Fig.~\ref{fig:quantum_simulation}(c)).
The longer runtime also raises logical error rates to $\approx 4\times 10^{-3}$.
Therefore, only the parallel BB protocol, which sustains a high rate in the space–time picture, offers genuine savings---reducing the space–time overhead by about $4\times$ compared with the surface-code protocol or the low-rate BB protocol (blue rectangle in Fig.~\ref{fig:quantum_simulation}(c)). 
Detailed estimates appear in Supplementary Note V.

The logical error rate per qubit per two-qubit rotation layer using our high-rate BB protocol ($\sim 7\times 10^{-4}$) is about an order of magnitude lower than in unencoded implementations ($\sim 7\times 10^{-3}$) --- specifically for parallel $YY$ rotations implemented with seven gate layers (parallel $Z$ rotations sandwiched between two layers each of CNOT, $H$, and $S$) --- even after accounting for the $T$-synthesis overhead. 
This reduction enables simulations of larger quantum systems, supporting either bigger lattice sizes or longer evolution times.
We further note that the Clifford parts of the logical circuits are implemented very efficiently by matching the symmetry of the BB codes with that of the simulation circuit. 
The main bottleneck lies in the relatively high cost and error rate of the parallel magic-state cultivation protocol. With small codes (distance-$3$ color codes and distance-$d_b = 8$ BB codes), the logical error rate per code cycle is limited to the $10^{-6}$ level, accumulating to $>10^{-5}$ for $T$ gates that require $\sim d_b$ cycles. Employing larger codes together with more efficient ancilla systems could reduce the logical error rate to the level required for a megaquop quantum machine~\cite{preskill2025beyond}. 
Moreover, because the cultivation protocol relies heavily on postselection, even a modest improvement in the physical error rate would yield an exponential reduction in logical cost and a substantial boost in performance~\cite{gidney2024magic}.
Finally, decoder choice strongly impacts the cultivated magic-state error rates, suggesting room for further improvement with better, tailored decoders.

\section{Conclusion and discussion}

In this work, we introduce a complete, high-rate toolkit of logical operations for generic qLDPC codes—encompassing syndrome extraction, addressable Clifford gates, and global non-Clifford gates—--implemented via code-switching and transversal gates.
By executing these operations in batches, we can efficiently embed logical circuits into non-locally connected physical circuits, achieving high logical operation rates with low space–time overhead --- mirroring the efficiency of high-rate codes with low space cost, and enabling genuine reductions in total space–time resources.
While our batched logical operations are asymptotically advantageous, it will be important to further reduce the constant factors, such that resource reductions can be achieved even in near- to medium-term implementations.

The batching structure of our high-rate gadgets suggests new design principles for quantum algorithms and circuits. Beyond the usual focus on $T$-count, the degree of parallel structure in a target algorithm can strongly influence the total space–time cost of its logical implementation.
In both our logical constructions and in leading physicalarchitectures~\cite{bluvstein2023logical, xu2024constant}, a common philosophy emerges: parallel, batched controls can dramatically reduce space–time overhead. This motivates rethinking algorithm design to align with and fully exploit such parallel execution patterns.

Finally, since our batched gadgets are largely built upon transversal gates and parallel stabilizer measurements, they are within the capabilities of reconfigurable atom arrays~\cite{bluvstein2023logical}. 
In addition, BB codes can also be deployed on such platforms through parallel syndrome-extraction circuits~\cite{viszlai2024efficient}. 
Looking beyond, shuttlable trapped-ion architectures~\cite{pino2021demonstration} and modular superconducting systems~\cite{yoder2025tour} also present compelling routes, each leveraging their strengths in connectivity, control precision, and modular scalability to realize high-rate quantum processors.

The magic-state cultivation protocol for BB codes is currently limited by the large space overhead required for ancillary systems that mediate between the BB codes and the cultivated color codes. While preparing this manuscript, we became aware of several new optimized cultivation protocols implemented directly on surface codes, without relying on color codes~\cite{chen2025efficient, vaknin2025magic, sahay2025fold, claes2025cultivating},
which achieve further reductions in space–time volume. These approaches may also be compatible with the BB protocol introduced here. Moreover, we anticipate that the space cost for preparing low-error-rate magic states can be further reduced by developing more efficient injection and cultivation protocols tailored directly to BB codes~\cite{zhou2025resource}.

\section{Acknowledgement}
We thank Pablo Bonilla, Hossein Dehghani, Roland Farrell, Jeongwon Haah, Zhiyang He, Liang Jiang, Christian Kokail, Milan Kornjaca, Gideon Lee, Quynh Nguyen, Christopher Pattison, Chen Zhao, Guo Zheng for helpful discussions. We acknowledge financial support from IARPA and the Army Research Office, under the Entangled Logical Qubits program (Cooperative Agreement Number W911NF-23-2-0219), the DARPA MeasQuIT program (HR0011-24-9-0359), the Center for Ultracold Atoms (a NSF Physics Frontiers Center, 
PHY-2317134),
the Institute for Quantum Information and Matter (a NSF Physics Frontiers Center, PHY-2317110), the National Science Foundation (grant number PHY-2012023 and grant number CCF-2313084), the Army Research Office MURI (grant number W911NF-20-1-0082), DOE/LBNL (grant number DE-AC02-05CH11231), DOE Quantum Systems Accelerator Center, contract number 7568717, the Wellcome Leap Quantum for Bio program.
Q.X. is funded in part by the Walter Burke Institute for Theoretical Physics at Caltech.

\clearpage
\newpage
\section{Methods}
\subsection{Batched syndrome extraction \label{sec:app_BSE}}

Here, we provide the construction of the BSE gadget. We present the protocol for measuring $X$ stabilizers and the protocol for measuring $Z$ stabilizers is mirrored. 

Given a $[n_C, k_C, d_C]$ classical LDPC code with a check matrix $H_C \in \mbb{F}_2^{\sigma_C\times n_C}$, where $\sigma_C = n_C - k_C$ (assuming full rank). 
Up to permutation of the bits, we assume that $H_C$ is in its canonical form with a conjugate generator matrix $G_C = (I_{k_C}, P)$, where $P \in \mbb{F}_2^{k_C \times (n_C - k_C)}$, with $H_C G_C^T = 0 \mod 2$. 
We measure the $X$ checks of a $[[n, k, d]]$ CSS qLDPC code $\mc{Q}$ using the following protocol:
\begin{enumerate}
    \item Input: $k_C$ $\mc{Q}$ blocks. Denote this set of blocks $\bs{B}_0$. Start by measuring the $X$ checks of $\mc{Q}^{\otimes k_C}$ in $\bs{B}_0$ once.
    \item Initialize $\sigma_C$ ancillary $\mc{Q}$ blocks in $\vert 0 \rangle^{\otimes n}$, and  measure the $X$ checks of $\mc{Q}^{\otimes \sigma_C}$ once. Denote this set of blocks $\bs{B}_1$.
    \item Initialize $\sigma_C$ ancilla blocks in $\vert + \rangle^{\otimes n}$, and measure the $Z$ checks of $\mc{Q}^{\otimes \sigma_C}$. Denote this set of blocks $\bs{A}$.
    \item Apply $\bs{A}$-controlled transversal CNOTs targeting $\bs{B}_0$ and $\bs{B}_1$ based on $H_C$: Let the blocks of $\bs{B}_0 \cup \bs{B}_1$ be indexed by $[n_C]$ and the blocks of $\bs{A}$ indexed by $[\sigma_C]$. Apply a transversal CNOT between the $i$-th block of $\bs{A}$ and the $j$-th block of $\bs{B}_0 \cup \bs{B}_1$ if $H_C[i,j] = 1$. 
    \item Perform transversal measurements of the blocks in $\bs{B}_1$ in the $Z$ basis and the blocks in $\bs{A}$ in the $X$-basis. 
    \item Perform error correction using detectors in the circuits and apply a feedback Pauli $X$ correction on $\bs{B}_0$ based on the $Z$ measurement outcomes of $\bs{B}_1$. See Supplement Note I for details of the detectors and the Pauli feedback.
\end{enumerate}

\begin{figure}
    \centering
    \includegraphics[width=1\linewidth]{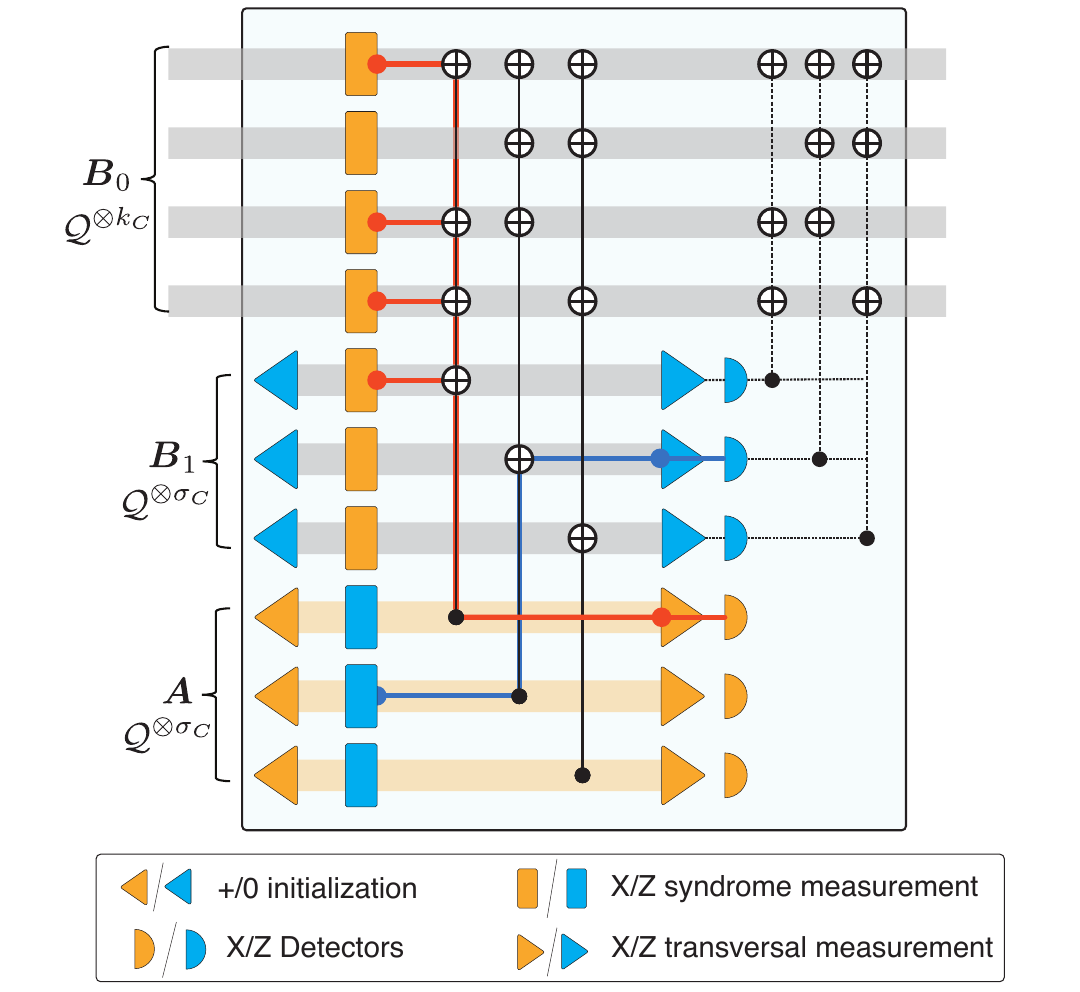}
    \caption{\textbf{Illustration of the batched syndrome extraction protocol.} 
    Each blue (orange) rectangle represents a round of $X$ ($Z$) syndrome measurement for the underlying quantum code. Each layer of CNOT gates represent a set of transversal CNOTs controlled by the $B_2$ blocks. Each $X$ ($Z$) detector consists of a stabilizer measurement outcome $s_X$ ($s_Z$) reconstructed from the transversal $X$ ($Z$) measurements of the $B_2$ ($B_1$) block as well as the initial $X$ ($Z$) syndrome on the $ B_0\cup B_1$ ($B_2$) block that corresponds to the back propagation of $s_X$ ($s_Z$) through the transversal CNOTs. }
    \label{fig:BSE}
\end{figure}

See Fig.~\ref{fig:BSE} for an illustration of the above protocol. Error correction is performed using the $X$ (resp. $Z$) detectors of the circuit: each $X$ (resp. $Z$) detector consists of the measurement outcome of a $X$ (resp. $Z$) stabilizer $s_X$ (resp. $s_Z$) constructed from the transversal $X$ (resp. $Z$) measurements of the $\bs{A}$ (resp. $\bs{B}_1$) blocks as well as the initial $X$ (resp. $Z$) syndrome on the $\bs{B}_0\cup \bs{B}_1$ (resp. $\bs{A}$) block that corresponds to the back propagation of $s_X$ (resp. $s_Z$) through the transversal CNOTs. In particular, the $X$ (resp. $Z$) detectors can be used to correct the $X$ (resp. $Z$) syndrome errors on the $\bs{B}_0 \cup \bs{B}_1$ (resp. $\bs{A}$) blocks, guaranteeing the robustness of the BSE protocol as a syndrome extraction gadget. 

According to Theorem~\ref{theorem:batched_SE}, when using a sufficiently expanding classical code, a noisy BSE gadget is equivalent to a perfect SE gadget sandwiched by two layers of locally stochastic data noise (See Supplement Note I for the proof). 
In the setting of a memory experiment, where the BSE gadget is repeatedly applied, the standard analysis for single-shot QEC~\cite{bombin2015single, kubica2022single, quintavalle2020single, gu2023single} implies that the residual error is under control and there is a sustainable single-shot QEC threshold.
More nontrivially, the BSE gadget also enables single-shot code deformations, particularly single-shot state preparation in the computational basis.

\begin{corollary}[Batched state preparation]
    Given a $[n_C, k_C, d_C]$ classical LDPC code with sufficient expansion and $k_C$ blocks of any $[[n, k, d]]$ CSS qLDPC code $\mc{Q}$ with $d_C \geq d$, we can perform a single-shot state preparation in either the $Z$ or $X$ basis fault-tolerantly for $Q^{\otimes k_C}$ using $2(n_C - k_C)$ ancillary $\mc{Q}$ blocks.
    \label{corollary:batched_state_prep}
\end{corollary}
\begin{proof}
    We prepare the logical states $\overline{\ket{0}}^{\otimes k}$ (resp. $\overline{\ket{+}}^{\otimes k}$) of $\mc{Q}$ by initializing the qubits in $\ket{0}^{\otimes n}$ (resp. $\ket{+}^{\otimes n}$), 
    measuring the $X$ (resp. $Z$) checks using a BSE gadget, and applying a Pauli $Z$ (resp. $X$) feedback based on the (corrected) stabilizer measurement outcomes from BSE. 
    According to Theorem~\ref{theorem:batched_SE}, a noisy BSE for measuring $X$ (resp. $Z$) checks is equivalent to a noiseless round of $X$ (resp. $Z$) stabilizer measurements for $\mc{Q}$ sandwiched by two layers of locally stochastic data noise, where the first-layer noise before the BSE can be absorbed into the initial state preparation.   
    Then, it follows that the state preparation gadget prepares the ideal logical states up to a locally stochastic residual data noise, thereby being fault tolerant.
\end{proof}

Our BSE gadget can be readily extended to other tasks involving state preparation and code deformations.
For example, in Steane-type logical measurements, homomorphic measurements~\cite{huang2022homomorphic, xu2024fast}, and teleportation-based gates~\cite{knill2004quantum}—all of which rely on prepared ancilla states—our method enables single-shot execution when performed in batches.
We also show its applications to (generalized) lattice surgeries~\cite{horsman2012surface, cohen2022low, cross2024improved}, where we can measure the stabilizers of the merged codes just once using the BSE gadget and still obtain the correct logical measurement with high probability.

\subsection{Batched code switching \label{sec:app_BCS}}
Here, we provide the construction of the BCS gadget. The BCS gadget can be viewed as a generalization of an encoding/decoding protocol for any CSS quantum code in its canonical logical basis. 

\begin{definition}[Canonical basis of CSS codes~\cite{gottesman1997stabilizer}]
Given any $[[n, k, d]]$ CSS code $\mc{Q}$ with $Z$ and $X$ check matrices with rank $r_Z$ and $r_X$, respectively, we can, up to qubit permutation, write its logical operators in the following canonical form~\cite{gottesman1997stabilizer}:
\begin{equation}
L_Z = 
\begin{blockarray}{ccccc}
 & L & Z & X \\
\begin{block}{c(ccc)c}
 & I_k & A & 0 & \\
\end{block}
\end{blockarray}, \quad 
L_X = 
\begin{blockarray}{ccccc}
 & L & Z & X \\
\begin{block}{c(ccc)c}
 & I_k & 0 & B & \\
\end{block}
\end{blockarray},
\label{eq:canonical_basis}
\end{equation}
    where $A \in \mbb{F}_2^{k \times r_X}$ and $B \in \mbb{F}_2^{k \times r_Z}$. The qubits are partitioned into three zones labeled by $L$, $Z$, and $X$, which support the intersections of $Z$- and $X$-logical operators, $Z$-logical operators, and $X$-logical operators, respectively.
    \label{def:canonical_basis}
\end{definition}

\begin{definition}[Measurement-based encoding/decoding protocol for any CSS code]
    Given any $[[n, k, d]]$ CSS code $\mc{Q}$, we can encode $k$ physical qubits associated with the $L$ zone of $\mc{Q}$ into the $k$ logical qubits of $\mc{Q}$ in the canonical basis by initializing $r_X$ (resp. $r_Z$) extra physical qubits associated with the $Z$ (resp. $X$) zone in $\ket{0}$ (resp. $\ket{+}$) states, measuring the stabilizers of $\mc{Q}$, and fixing the Pauli frame. 
    Conversely, we can decode from the $k$ logical qubits of $\mc{Q}$ to the $k$ physical qubits of zone $L$ by measuring the $r_X$ (resp. $r_Z$) physical qubits of zone $Z$ (resp. $X$) in the $Z$ (resp. $X$) basis and applying a feedback Pauli correction on the output $k$ physical qubits. 
    \label{def:encoding_decoding_protocol}
\end{definition}
Note that the encoding protocol in Def.~\ref{def:encoding_decoding_protocol} can be viewed as growing the logical operators from $Z_i$ (resp. $X_i$) to $\bar{Z}_i$ (resp. $\bar{X}_i$), where $Z_i$ (resp. $X_i$) denotes the single-qubit $Z$ (resp. $X$) Pauli operator on the $i$-th physical qubit of zone $L$ and $\bar{Z}_i$ (resp. $\bar{X}_i$) the logical operator associated with the $i$-th row of $L_Z$ (resp. $L_X$) in Eq.~\eqref{eq:canonical_basis}).
Correspondingly, the decoding protocol can be viewed as shrinking the canonical logical operators to single-qubit operators supported on zone $L$. 
\begin{figure*}
    \centering
    \includegraphics[width=1.0\linewidth]{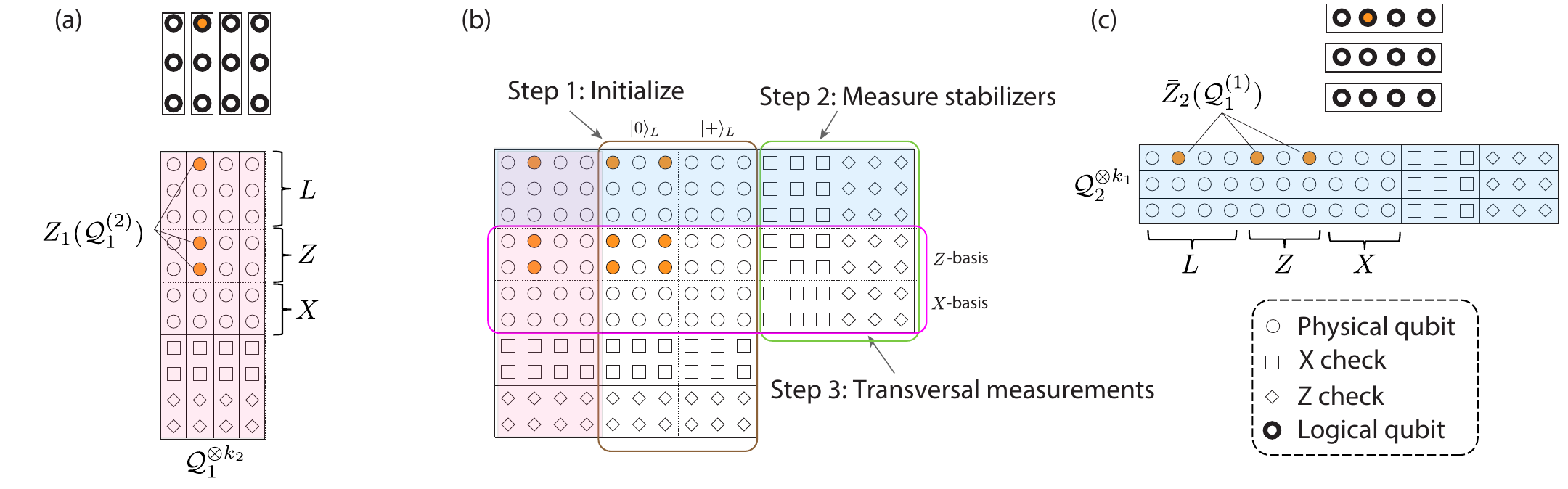}
    \caption{\textbf{Illustration of the batched code switching gadget}. Given $k_2$ blocks of $[[n_1, k_1, d_1]]$ codes $\mc{Q}_1$ (a) and $k_1$ blocks of $[[n_2, k_2, d_2]]$ codes $\mc{Q}_2$ (c), we can efficiently switch from the $\mc{Q}_1$ encoding to the $\mc{Q}_2$ encoding following the three steps illustrated in (b). 
    }
    \label{fig:batched_switching}
\end{figure*}

Equivalently, the encoding (resp. decoding) protocol of Def.~\ref{def:encoding_decoding_protocol} can be viewed as a code-switching protocol from $k$ copies of the $[[1, 1, 1]]$ trivial code $\mc{Q}_{\mr{trivial}}$ (resp. $\mc{Q}$) to $\mc{Q}$ (resp. $\mc{Q}_{\mr{trivial}}^{\otimes k}$). Now, we can generalize this code switching protocol by modifying $\mc{Q}_{\mr{trivial}}$ to another nontrivial CSS qLDPC code. 
Let $\mc{Q}_1$ and $\mc{Q}_2$ be any two qLDPC CSS codes with parameters $[[n_1, k_1, d_1]]$ and $[[n_2, k_2, d_2]]$ as well as $Z$ (resp. $X$) check rank $r_{Z, 1}$ (resp. $r_{X, 1}$) and $r_{Z, 2}$ (resp. $r_{X, 2}$), respectively.
Given $k_2$ blocks of $\mc{Q}_1$ and $k_1$ blocks of $\mc{Q}_2$, we denote the canonical logical operators of $\mc{Q}_1^{\otimes k_2}$ and $\mc{Q}_{2}^{\otimes k_1}$ as $\{\bar{P}_i(\mc{Q}_1^{(j)})\}_{i \in [k_1], j \in [k_2]}$ and $\{\bar{P}_i(\mc{Q}_2^{(j)})\}_{i \in [k_2], j \in [k_1]}$ (for $P \in \{X, Z\}$), respectively (see Fig.~\ref{fig:batched_switching}(a) and (c)).
We can efficiently switch from $\mc{Q}_1^{\otimes k_2}$ to $\mc{Q}_2^{\otimes k_1}$ in their canonical basis, i.e. $\bar{P}_i(\mc{Q}_1^{(j)}) \rightarrow \bar{P}_j(\mc{Q}_2^{(i)})$, following the three steps illustrated in Fig.~\ref{fig:batched_switching}(b):
\begin{enumerate}
    \item Prepare $r_{X, 2}$ (resp. $r_{Z, 2}$) blocks of $\mc{Q}_1$ associated with the $Z$ (resp. $X$) zone of $\mc{Q}_2$ in logical $\overline{\ket{0}}^{\otimes k_1}$ (resp. $\overline{\ket{+}}^{\otimes k_1}$) (see the brown box). Using the BSE gadget, we can perform these state preparations in single shot.
    \item Measure the stabilizers of $\mc{Q}_2^{\otimes n_1}$ (see the green box) using the BSE gadget.
    \item Measure transversally the $r_{X, 1}$ (resp. $r_{Z, 1}$) blocks of $\mc{Q}_2$ associated with the $Z$ (resp. $X$) zone of $\mc{Q}_1$ in the $Z$ (rep. $X$) basis (see the pink box) and apply Pauli feedback on the output $\mc{Q}_2^{\otimes k_1}$ block (see Supplement Note II for details of the feedback).

\end{enumerate}
Note that step 2 grows the logical operators of $\mc{Q}_1$ into a tensor product form, i.e. $\bar{P}_i(\mc{Q}_1^{(j)}) \rightarrow \bar{P}_i(\mc{Q}_1^{(j)})\otimes \bar{P}_j(\mc{Q}_2^{(i)})$ (see the highlighted operator in Fig.~\ref{fig:batched_switching} for an illustration), by running the measurement-based encoding protocol described above for $\mc{Q}_2$. 
The final step (3) then shrinks the tensor-product logical operators to those of $\mc{Q}_2$, i.e. $\bar{P}_i(\mc{Q}_1^{(j)})\otimes \bar{P}_j(\mc{Q}_2^{(i)}) \rightarrow \bar{P}_j(\mc{Q}_2^{(i)})$. 

It is easy to check that the above BCS protocol reduces to the encoding/decoding protocol for $\mc{Q}_1$ (resp. $\mc{Q}_2$) if we choose $\mc{Q}_2$ (resp. $\mc{Q}_1$) as the trivial code $\mc{Q}_{\mr{trivial}}$. 
The fault tolerance of the BCS protocol is essentially guaranteed by: (i) the logical $\overline{\ket{0}}$ and $\overline{\ket{+}}$ states of the $\mc{Q}_1$ blocks in step 1 as well as the stabilizers of the $\mc{Q}_2$ blocks in step (2) are fault-tolerantly prepared/measured in a constant depth using the BSE gadget (see Theorem~\ref{theorem:batched_Cliffords} and Corollary~\ref{corollary:batched_state_prep}) up to locally stochastic errors on the data qubits 
(ii) the parent subsystem code associated with this code-deformation protocol~\cite{vuillot2019code} is LDPC and has a sufficiently large distance $\geq \min\{d_1, d_2\}$, guaranteeing fault tolerance against the data errors. See Supplement Note II for details of the proof.

\subsection{Batched addressable Cliffords\label{sec:app_BAC}}
Here, we sketch the construction of the batched Clifford gates (BAC) gadget that has a constant space-time overhead (see Theorem~\ref{theorem:batched_Cliffords}) and provide the details in Supplement Note III.

Generally, a BAC gadget for implements a $k_1$-qubit, depth-$D_P$ Clifford circuit $C_{k_1}$ for any $[[n_1, k_1, d_1]]$ CSS qLDPC code $\mc{Q}_1$ in a batch size $\mc{O}(k_2)$ following the three steps below
\begin{enumerate}
    \item Code switch to $\mc{Q}_2^{\otimes k_1}$, for some $[[n_2, k_2, d_2]]$ CSS qLDPC code $\mc{Q}_2$, using the BCS gadget. 
    \item Perform global fault-tolerant Clifford gates on $\mc{Q}_2^{\otimes k_1}$ according to $C_{k_1}$ in depth $\mc{O}(D_P)$. See below for the detailed implementation of these $\mc{Q}_2$-encoded logical gates.
    \item Code switch back to $\mc{Q}_1^{\otimes k_2}$ using the BCS gadget.
\end{enumerate}

By choosing a $[[n_2 = \Theta(k_2), k_2, d_2 \geq d_1]]$ CSS qLDPC code $\mc{Q}_2$, we can implement any $k_1$-qubit CNOT circuit $C_{\mr{CNOT}}$ on $\mc{Q}_1^{\otimes k_2}$ by code switching to $\mc{Q}_2^{\otimes k_1}$ using the BCS gadget, applying transversal CNOTs on the $k_1$ $\mc{Q}_2$ blocks according to $C_{\mr{CNOT}}$ in a depth $\mc{O}(D_P(C_{\mr{CNOT}}))$ (see Supplement Note III for details of these transversal CNOTs), and switching back to $\mc{Q}_1^{\otimes k_2}$ with another BCS gadget.   

To implement a $k_1$-qubit $H$ (or $S$) circuit $C_H$ ($C_S$) on $k_2$ blocks of $\mc{Q}_1$, we apply a similar three-step protocol -- code switching to $\mc{Q}_2^{\otimes k_1}$ with a BCS gadget, apply global $H$ or $S$ gates on a subset of the $\mc{Q}_2$ blocks according to $C_H$ ($C_S$), and then switching back to $\mc{Q}_1^{\otimes k_2}$ with another BCS gadget. We could have implemented the global $H$ and $S$ gates on $\mc{Q}_2$ easily with a constant space-time overhead if $\mc{Q}_2$ supports transversal $H$ and $S$ gates (e.g. for the self-dual BB codes in Table~\ref{tab:QC_code_params}). However, since we do not know such family of codes that scale asymptotically, we turn to codes that do not admit transversal $H$ or $S$ gates, yet support global fault-tolerant $H$ and $S$ gates with a low space-time overhead. 
One such choice is a symmetric HGP code (i.e. the two base classical codes of $\mc{Q}_2$ are the same) with parameters $[[n_2 = \Theta(k_2), k_2, d_2 = \Theta(\sqrt{k_2}) \geq d_1]]$. As detailed in Supplement Note III, we can implement global $H$ and $S$ gates on such a code with a constant space-time overhead by leveraging its fold transversal gates~\cite{quintavalle2022partitioning} and homomorphic gates~\cite{xu2024fast}. 
The choice of such constant-rate, square-root distance HGP codes determined the batch size $k_2 = \Omega(d_1^2)$ in Theorem~\ref{theorem:batched_Cliffords}.

We note that the BCS and the BAC gadgets are also analogous to the Gottesman-style gate teleportation protocol in Refs.~\cite{gottesman2013fault, nguyen_qldpc_2024, tamiya2024polylog} that implements addressable Clifford operations using (distilled) Clifford resource states.
In particular, the scheme in Ref.~\cite{nguyen_qldpc_2024} implements addressable Clifford gates on a high-rate qLDPC code $\mc{Q}_1$ in batches by first injecting noisy resource states into $\mc{Q}_1$ using concatenated schemes~\cite{gottesman2013fault} (which incurs a non-constant space-time overhead) and then distilling them using another high-rate quantum code $\mc{Q}_2$.
Although the gadgets constructed in \cite{nguyen_qldpc_2024} have similar logical action, our gadgets circumvent the need for the complex injection scheme using concatenated codes, enabling implementations that not only have low asymptotic space-time overhead but are also potentially relevant for early fault-tolerant devices. 

\subsection{Universal gate set with global non-Clifford gates and addressable Clifford gates \label{sec:app_addressable_T}}
Here, we show that we can obtain a universal gate set using addressable Clifford gates (giving the full $k$-qubit Clifford group $\mc{C}(k)$) and global $T$ gates $T^{\otimes k}$ with a constant overhead for a $[[n, k, d]]$ code $\mc{Q}$. It suffices to show that any layer of addressable $T$ gates can be implemented with a layer of global $T^{\otimes k}$ and a constant number of layers of addressable gates in $\mc{C}(k)$ (also using a constant number of ancilla codes). 

\begin{figure}[h!]
    \centering
    \includegraphics[width=1\linewidth]{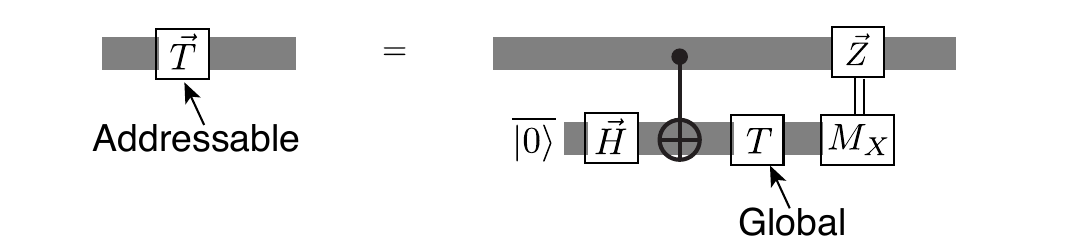}
    \caption{\textbf{Circuit for implementing addressable $T$ gates using global $T$ gates and addressable Clifford gates.}}
    \label{fig:addressable_T}
\end{figure}

As shown in Fig.~\ref{fig:addressable_T}, we can implement the target addressable $T$ gates, denoted by $\vec{T}$, on a data $\mc{Q}$ code by (1) initializing an ancilla code $\mc{Q}^{\prime}$ in logical $\overline{\ket{0}}$,  (2) applying a layer of addressable $H$ gates on logical qubits of $\mc{Q}^{\prime}$ on which $\vec{T}$ is \emph{not} supported, a layer of transversal CNOT between $\mc{Q}$ and $\mc{Q}^{\prime}$, and a layer of global $T$ gates on $\mc{Q}^{\prime}$, (3) transversally measuring $\mc{Q}^{\prime}$ in the $X$ basis, and (4) applying addressable Pauli $Z$ corrections on $\mc{Q}$. The key ingredient is that the addressable $H$ gates are applied in a way that, after which, a subset of ancilla logical qubits are not entangled with the data logical qubits, giving rise to the addressable $T$ gates. 

\subsection{Parallel magic state injection and distillation \label{sec:app_magic_injection}}
Here, we provide a detailed analysis of the magic state injection protocol described in the main text that utilizes small 3D topological codes and the BCS gadget. In particular, we show that parallel injection into an arbitrary $[[n_1, k_1, d_1]]$ CSS qLDPC code with a bounded marginal error rates can be achieved with a $\tilde{O}(\log^3 n_1)$ space-time overhead (see Theorem~\ref{theorem:batched_global_magic}).

We first analyze the error rate of the simplest injection scheme by code switching from $k_1$ copies of physical $\ket{T}$ states in the trivial $[[1,1,1]]$ codes to a 
$[[n_1, k_1, d_1]]$ qLDPC code using the BCS gadget.
Such an injection protocol simply reduces to the encoding protocol into $\mc{Q}_1$ in Def.~\ref{def:encoding_decoding_protocol}.
This protocol is noisy and, in particular, it suffers from the state preparation errors on the physical qubits in the $Z$ and $X$ zone. Concretely, for the $i$-th logical qubit of $\mc{Q}_1$ with logical operators $\bar{Z}_i$ and $\bar{X}_i$, each state preparation error on the support of $\bar{Z}_i$ (rep. $\bar{X}_i$) on zone $Z$ (resp. $X$) will cause an undetectable logical $X$ (resp. $Z$) error.
As such, the marginal logical error rate of each injected magic state will grow with $d_1$.

Now, we show that the marginal error rates can be reduced to a constant by switching from 3D color codes with slowly increasing distances $d_3 = \tilde{\mc{O}}(\log n_1)$, which essentially proves Theorem~\ref{theorem:batched_global_magic}. 

The marginal logical error rate of the injected magic states/faulty non-Clifford measurements is upper bounded by the total logical error rate of the protocol, which is dominated by that of the BCS gadget that code switches between $\mc{Q}_1$ and $\mc{Q}_3^{\otimes k_1}$ in a batch size $\mc{O}(d_1)$:
\begin{equation}
    P_{L}^{\mr{marginal}} \leq P_{L} \leq \mc{O}\left(d_1 n_3 n_1(p/p_{\mr{th}}\right)^{\Omega(d_3)}),
\end{equation}
provided that the physical error rate $p$ is below some constant threshold $p_{\mr{th}}$. See Supplement Note III for the bound on the BCS error rate used here. As such, choosing choosing $d_3 = \tilde{\mc{O}}(\log n_1)$ suffices for $P_{L}^{\mr{marginal}} =\mc{O}(1)$. 

\subsection{Self-dual BB codes and magic state cultivation protocols\label{sec:app_magic_cultivation}}
\begin{figure}
    \centering
    \includegraphics[width=1\linewidth]{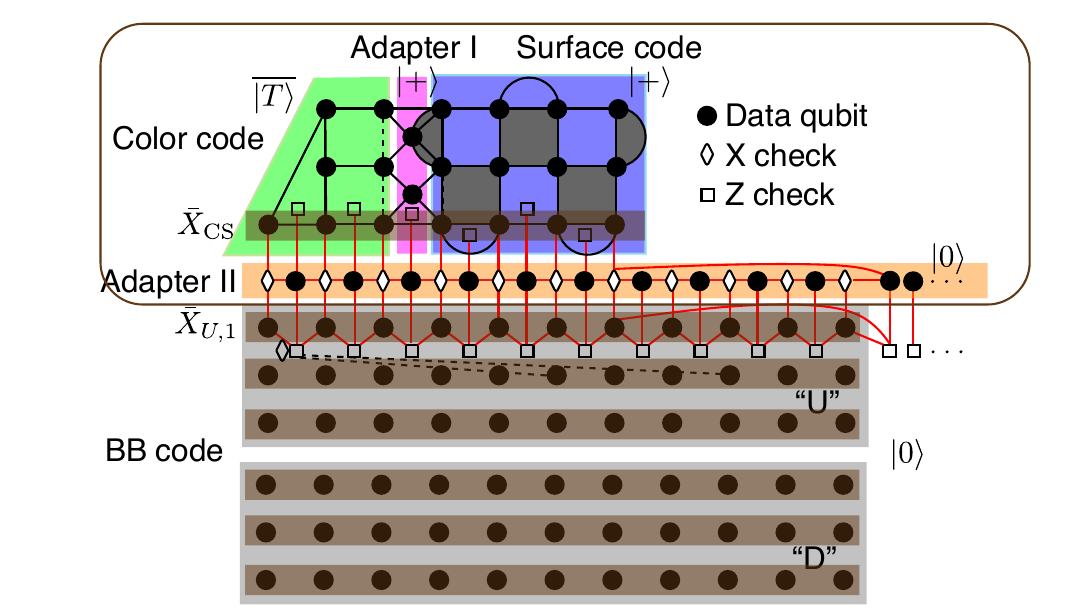}
    \caption{\textbf{Illustration of a self-dual BB code (gray) and the parallel magic-state cultivation protocol}. A self-dual $\mr{BB}(c, c^T; R_{l,m})$ code is defined on two blocks (``U" and ``D") of $l\times m$ qubits labeled by the monomials $\{x^i y^j\}_{i \in [l], j \in [m]}$. 
    An example pair of $Z$ and $X$ checks for $c = 1 + y + xy^5 + xy^8$ are illustrated by the square and the diamond, respectively, and all other checks are obtained by translating the example checks on the torus. This code is equivalent to the $[[66, 6, 8]]$ code in Table~\ref{tab:QC_code_params}. 
    The $2l$ pairs of logical operators in Eq.~\eqref{eq:disjoint_logical_basis}, each supporting on a row of the ``U" or the ``D" block, are illustrated by the brown strips.
    An adapter that couples the first logical qubit of the ``U" BB block with a distacne-3 color code (green) and an ancillary thin surface code (blue) is illustrated by the orange strip.
    }
    \label{fig:BB_illustration}
\end{figure}
Here, we provide the detailed construction of the self-dual BB codes listed in Table~\ref{tab:QC_code_params} as well as the parallel magic state cultivation protocol for these codes.

Given a commutative quotient polynomial ring $R_{l, m} := \mbb{F}_2[x, y]/(x^l - 1, y^m - 1)$ and two polynomials $c, d \in R_{l, m}$, a Bivariate Bicycle (BB) code $\mr{BB}(c, d; R_{l,m})$ is defined as a CSS code with check matrices~\cite{bravyi2024high}
\begin{equation}
    H_X = (c, d); \quad H_Z = (d^T, c^T),
    \label{eq:BB_checks}
\end{equation}
\label{def:BB_code}
where $f^T$ for any polynomial $f \in R_{l,m}$ is defined by taking the inverse of each of its monomial terms.

As shown in Fig.~\ref{fig:BB_illustration}(a), the physical qubits of a $\mr{BB}(c, d; R_{l,m})$ code can be arranged on two $l \times m$ blocks $U$ and $D$, where each qubit $q(U, \alpha)$ or $q(R, \alpha)$ within the block $U$ or $D$ is labeled by the monomial $\alpha \in \mc{M} = \{x^i y^j\}_{i \in [l], j \in [m]}$. 
Note that $R_{l, m}$ is equivalent to a vector space $\mbb{F}_2^{\mc{M}}$ with basis elements in $\mc{M}$. Given a $f \in \mbb{F}_2^{\mc{M}}$, it can also be interpreted as a set in the powerset of $\mc{M}$. 
Let $X(P, Q)$ (resp. $Z(P, Q))$, where $P, Q \in \mbb{F}_2^{\mc{M}}$, denote a $X$ (resp. $Z$) Pauli operator supported on $\{q(U, \alpha)\}_{\alpha \in P} \cup \{q(D, \beta)\}_{\beta \in Q}$.

In this work, we will focus on self-dual BB codes, which support transversal Hadamard and $S$ gates, by choosing odd $l$ and $m$ as well as polynomials that satisfy $d = c^T$. 
In addition, by carefully choosing the polynomial $c$, we identify codes with disjoint logical operators that can be translationally shifted on a toric layout by performing physical permutations. These choices result in the codes in Table~\ref{tab:QC_code_params} with weight-$8$ stabilizers and code parameters $[[n_b=2lm, k_b = 2l, d_b \leq m]]$. Below, we summarize the nontrivial features of these BB codes compared to the original ones in Ref.~\cite{bravyi2024high}. 
\begin{enumerate}
\item \textbf{Transversal Clifford gates:} The codes are self-dual and support transversal Hadamard and $S$ gates, i.e., $\overline{H}^{\otimes k_b}$ and $\overline{S}^{\otimes k_b}$ are implemented by $H^{\otimes n}$ and $S^{\otimes n}$, respectively.
\item \textbf{Disjoint logical support:} We can find the following logical-operator basis:
\begin{equation}
\begin{aligned}
    \bar{X}_{U, i} & = X(x^i \chi, 0), \quad \bar{Z}_{U, i} = Z(x^i \chi, 0), \\
    \bar{X}_{D, i} & = X(0, x^i \chi), \quad \bar{Z}_{D, i} = Z(0, x^i \chi),
\end{aligned}
\label{eq:disjoint_logical_basis}
\end{equation}
where $\chi := \sum_{j = 0}^{m - 1}y^j$, and $i$ iterates from $0$ to $l - 1$. As shown in Fig.~\ref{fig:BB_illustration}, each logical qubit indexed by $i$ on the $U$ ($D$) block has overlapped logical $X$ and $Z$ operators supported on the $i$-th row of the $U$ ($D$) block of the physical qubits. 

\item \textbf{Cyclic automorphisms:} Logical qubits within each block can be cyclically permuted via automorphisms induced by shifts of the physical qubits, i.e., $(\bar{X}_{U,i}, \bar{Z}_{U,i}) \mapsto (\bar{X}_{U,(i+1) \bmod l}, \bar{Z}_{U,(i+1) \bmod l})$ (and similarly for $D$) is implemented by $q(U, \alpha) \mapsto q(U, x\alpha)$ and $q(D, \alpha) \mapsto q(D, x\alpha)$.
\end{enumerate}

To prepare high-fidelity magic states for these BB codes, we first cultivate $\overline{\ket{T}}$ using small 2D color codes following the standard protocol in Ref.~\cite{gidney2024magic} and then transfer them into the BB codes using the following code-merging protocol.

The protocol utilizes the following subroutine for merging two quantum codes: Given two CSS codes $\mc{Q}_A$ and $\mc{Q_B}$ with logical operators $(\bar{X}_A, \bar{Z}_A)$ and $(\bar{X}_B, \bar{Z}_B)$ (we focus on one particular logical qubit for ease of presentation"), respectively, we could use an adapter $\mc{A}_{\mr{A-B}}$ to merge the two codes by performing a joint logical measurement $\bar{X}_A \bar{X}_B$. 
Starting from the code $\mc{Q}_A$ and initializing the data qubits of $\mc{Q}_B$ and $\mc{A}_{\mr{A-B}}$ in the product $\ket{0}$ states, we can also view this procedure as growing the code from $\mc{Q}_A$ to the merged code $\mc{Q}_{AB}$ with a logical $X$ operator $\bar{X}_A$ and a grown logical $Z$ operator $\bar{Z}_A \bar{Z}_B$. Thus, this merging protocol grows the $Z$ distance of the code (rigorously proved in Supplement Note V) utilizing the adapter that enables a joint $\bar{X}_A \bar{X}_B$ logical measurement. 
We refer to the above protocol as an $X$-merging protocol with a $X$-type adapter. We can similarly construct a $Z$-merging protocol with a $Z$-type adapter, which enables a joint $\bar{Z}_A \bar{Z}_B$ measurement, growing the $X$ distance of the code.  

For a BB code encoding $k_b$ logical qubits, starting from $k_b$ copies of $[[n_c, 1, d_c]]$ color codes $\mc{Q}_C^{\otimes k_b}$, we first grow the $X$ distance of the codes by merging $\mc{Q}_C^{\otimes k_b}$ with $k_b$ copies of $[[n_s, 1, d_{s, X}, d_{s, Z} = d_c]]$ thin surface codes $\mc{Q}_S^{\otimes k_b}$ through $k_b$ $Z$-type adapters (see the purple strip in Fig.~\ref{fig:BB_illustration}). The $k_b$ copies of the merged color-surface code $\mc{Q}_{\mr{CS}}^{\otimes k_b}$ then have a $Z$ distance $d_c$ and a $X$ distance $d_c + d_{s, X}$. Next, we grow the $Z$ distance of the code by merging $\mc{Q}_{\mr{CS}}^{\otimes k_b}$ with a $[[n_b, k_b, d_b]]$ BB code through a $X$-type adapter $\mc{A}_{\mr{CS-BB}}$ that consists of $k_b$ disjoint and identical adapters $\{\mc{A}_{\mr{CS-BB}}^{(i)}\}_{i = 1}^{k_b}$, each coupling the $i$-th logical qubit of $\mc{Q}_{\mr{BB}}$ to a copy of $\mc{Q}_{\mr{CS}}$ (see the orange strip in Fig.~\ref{fig:BB_illustration} for $\mc{A}_{\mr{CS-BB}}^{(1)}$). By choosing $d_c + d_{s, X} \sim d_b$, we obtain a merged color-surface-BB code $\mc{Q}_{\mr{CSBB}}$ that encodes $k_b$ logical qubits with a distance $\sim d_b$ (see Supplement Note V). 
The above two-step merging protocol, analogous to the grafting procedure from a color code to a surface code~\cite{gidney2024magic}, then transfers the magic states cultivated in the $k$ color codes to the merged code $\mc{Q}_{\mr{CSBB}}$. 

To maintain the low space overhead of the BB code, we use small color codes with $d_c \ll d_b$.
In this regime, post-selection (error detection) must be applied before the code reaches full distance $d_b$ to avoid degrading the logical error rate, which could potentially reduce the overall success probability. 
Fortunately, aided by the thin surface codes, our protocol rapidly grows the code distance to $d_b$.
Moreover, by restricting post-selection to small spatial regions centered on the color codes, we achieve low logical error rates while maintaining high success probabilities. See Supplementary Note V for details.

As a concrete example, we can merge $6$ copies of $[[7, 1, 3]]$ color codes with a $[[66, 6, 8]]$-BB code with $6$ copies of $[[12, 1, d_{s,X} = 4, d_{s, Z} = 3]]$ thin surface codes, resulting in a $[[324, 6, 7]]$ merged color-surface-BB code. See Supplement Note V for details.

\subsection{Fault-tolerant compilation\label{sec:app_compilation}}
Here, we formalize the fault-tolerant compilation problem using our batched high-rate logical gadgets $\mb{B}_L$ and analyze the compilation overhead in detail.

Considzer a Clifford + $T$ circuit on $K$ qubits, 
$C = \prod_{t = 1}^{D_P} C_t$, where each layer $C_t$ of the circuit is drawn from the gate set $\mb{B}_P$, which consists of $\{$arbitrary $K$-qubit Hadamard gates$\}$ $\cup $ $\{$ arbitrary $K$-qubit $S$ gates $\}$ $\cup$ $\{$ arbitrary \emph{non-overlapping} CNOTs on $K$-qubits $\}$ $\cup$ $\{$ arbitrary $K$-qubit $T$ gates $\}$. Note that $\mb{B}_P$ are essentially all possible parallel gates of the same type that are implementable with physical qubits under long-range connectivity. Note that any conventional Clifford + T circuit can be turned into the above form with a constant overhead~\footnote{According to Ref.~\cite{aaronson2004improved}, any Clifford circuit can be decomposed into a canonical form of 11 blocks, each consisting of only one type of gates from gate set $\{\mr{CNOT}, S, H\}$.  In addition, as detailed in Sec.~\ref{sec:app_addressable_T}, each layer of addressable $T$ gates can be implemented by a layer of global $T$ gates and a constant number of layers of addressable Clifford operations}. 
We aim to analyze the overhead of further compiling each layer $C_t$ into our batched gate set $\mb{B}_L$ in depth $D_{\mb{B}_L}(C_t)$, i.e. $C_t = \prod_{t^{\prime} = 1}^{D_{\mb{B}_L}(C_t)} C_{t, t^{\prime}}$, where each $C_{t, t^{\prime}}$ is drawn from $\mb{B}_L$. We simply refer to the extra overhead for compiling $C$, $\left[ \sum_{t = 1}^{D_P} D_{\mb{B}_L}(C_t) \right]/D_P$, as the compilation overhead.

For concreteness, we specify our batched gate set $\mb{B}_L$ here.
We distribute the total of $K$ logical qubits into $K/k_1$ blocks of $[[n_1, k_1, d_1]]$ $\mc{Q}_1$ codes. 
Here, we assume that $K/k_1$ is larger than the batch size for the BAC gadget, i.e. $K = \Omega(d_1^2)$.
Let $\mb{B}_{\mr{BC}}$ denote the set of depth-$1$ physical Clifford gates that are implementable by a single BAC gadget in Theorem~\ref{theorem:batched_Cliffords}, which takes the following form: take any depth-$1$ Clifford circuit $C_1$ on $k_1$ qubits and assign it to an arbitrary subset of $k^{\prime} \leq K/k_1$ $Q_1$ blocks (the rest being applied with identity)~\footnote{Note that when $K/k_1$ is much lager than the batch size $\mc{O}(d_1^2)$ of a BAC gadget, we can have multiple batches implementing distinct $k_1$-qubit circuits, which are beyond the operations in the current $\mb{B}_{\mr{BC}}$. We neglect such a flexibility for now.}. Let $\mb{B}_{\mr{GT}} = \{\bar{T}^{\otimes K}\}$ be the global $T$ gates. We thus take the logical gadget basis as $\mb{B}_L = \mb{B}_{\mr{BC}} \cup \mb{B}_{\mr{GT}}$. 

We note that the compilation overhead is not equivalent to the time overhead of fault-tolerantly implementing a circuit as the global $T$ gates in $\mb{B}_L$ are not implemented using a constant-depth physical circuit. Rather, it characterizes the extra time overhead that arises from the limited addressability/parallelism of the gadgets in $\mb{B}_L$ (compared to $\mb{B}_P$) on top of the time overhead of the gadgets themselves in $\mb{B}_L$. 
\subsubsection{Worst-case compilation overhead}
We show that the worst-case compilation overhead defined above using our batched gadgets \( \mb{B}_L \) scales at least linearly with \( k_1 \), the number of logical qubits per qLDPC block \( \mc{Q}_1 \).

Specifically, we construct a \( K \)-qubit CNOT circuit \( C_{\mr{CNOT}} \)—comprising only CNOT gates—that can be implemented with physical depth \( \tilde{\mc{O}}(K) \), yet requires logical depth at least \( \tilde{\Omega}(K k_1) \) when compiled using \( \mb{B}_L \).

As shown in Ref.~\cite{jiang2020optimal}, any CNOT circuit \( C_{\mr{CNOT}} \) can be decomposed into layers of non-overlapping CNOT gates within \( \mb{B}_P \), yielding a physical depth
\begin{equation}
    D_{\mb{B}_P}(C_{\mr{CNOT}}) = \mc{O}(K/\log K).
    \label{eq:depth_physical_CNOT}
\end{equation}

To lower bound the corresponding logical depth \( D_{\mb{B}_L}(C_{\mr{CNOT}}) \), we analyze the size of \( \mb{B}_L \) and apply a simple counting argument.

\begin{lemma}[Size of the batched gadgets $\mb{B}_L$]
Given $K$ logical qubits distributed in $K/k_1$ blocks of a $[[n_1, k_1, d_1]]$ CSS qLDPC code $\mc{Q}_1$, where $K = \Omega(k_1^2)$, the batched gadgets have a size
    \begin{equation}
        |\mb{B}_L| = \exp [\tilde{\mc{O}}(K/k_1)].
    \end{equation}
    \label{lemma:size_batched_gadgets}
\end{lemma}
\begin{proof}
    Recall that the batched gadgets consist of batched $H$ gates, batched $S$ gates, batched CNOT gates, and global $T$ gates. The size$|\mb{B}_L|$ is dominated by the number of distinct batched CNOT gates. Let $\mb{S}_2 = \{\{\alpha_i, \beta_i\}\}_i$ be a two-block partition of the indices of the $K/k_1$ blocks, $[K/k_1]$, i.e. $\{\alpha_i, \beta_i\}$ and $\{\alpha_j, \beta_j\}$ are two distinct pairs for $i \neq j$ and $\bigcup_i \{\alpha_i, \beta_i\} = [K/k_1]$. A BAC gadget can apply any $2k_1$-qubit CNOT circuit of a physical depth $1$ to any subset of the pairs of the blocks in $\mb{S}_2$. Denote the set of all two-block partitions of $[K/k_1]$ as $\boldsymbol{\mc{S}}_2(K/k_1)$ and the set of all depth-$1$ $2k_1$-qubit CNOT circuits as $\mb{C}_{\mr{CNOT, P}}(2k_1)$. The size of all distinct batched CNOTs is then given by
    \begin{equation}
        |\boldsymbol{\mc{S}}_2(K/k_1)|\times |\mb{C}_{\mr{CNOT, P}}(2k_1)| \times 2^{K/(2k_1)},
    \end{equation}
    where the contribution $2^{K/(2k_1)}$ comes from the fact that for each of the $K/(2k_1)$ pairs of blocks, we can decide whether or not to apply the selected $2k_1$-qubit CNOT circuit. Following a simple combinatorial analysis, we have $|\boldsymbol{\mc{S}}_2(K/k_1)| = (K/k_1)!/[K/(2k_1)]! = \exp{\tilde{\Theta}(K/k_1)}$. According to Ref.~\cite{jiang2020optimal}, we have $|\mb{C}_{\mr{CNOT}}(2k_1)| = \exp{\tilde{\Theta}(k_1)}$.  In the regime where $K = \Omega(k_1^2)$ (there are more than $\Theta(k_1)$ $\mc{Q}_1$ blocks), we then have $|\mb{B}_L| = \exp{\tilde{\mc{O}}(K/k_1)}$.
\end{proof}

\begin{proposition}[Worst-case compilation overhead of the batched gadgets]
    There exists a $K$-qubit CNOT circuit $C_{\mr{CNOT}}$ with 
    \begin{equation}
        D_{\mb{B}_{L}}(C_{\mr{CNOT}})/D_{\mb{B}_{P}}(C_{\mr{CNOT}}) = \Tilde{\Omega}(k_1),
        \label{eq:worst_case_seri_overhead}
    \end{equation}
    \label{prop:hardness_CNOT}
\end{proposition}
\begin{proof}
    According to Ref.~\cite{jiang2020optimal}, the number of distinct $K$-qubit CNOT circuits is $|\mb{C}_{\mr{CNOT}}(K)| = \exp\{\tilde{\Theta}(K^2)\}$, Whereas, according to Lemma.~\ref{lemma:size_batched_gadgets}, the number of distinct gates in each layer of batched gadgets is $|\mb{B}_L| = \exp{\tilde{O}(K/k_1)}$. Therefore, it takes a minimal logical depth of 
    \begin{equation}
        D^* = \log{|\mb{C}_{\mr{CNOT}}(K)|}/\log{|\mb{B}_L|} = \tilde{\Omega}(K k_1), 
    \end{equation}
    to reach all possible $K$-qubit CNOTs. This indicates that there exists at least one $K$-qubit CNOT $C_{\mr{CNOT}} \in \mb{C}_{\mr{CNOT}}(K)$ such that the logical depth $D_{\mb{B}_L}(C_{\mr{CNOT}}) = \tilde{\Omega}(K k_1)$. In contrast, the physical depth is always $D_{\mb{B}_P}(C_{\mr{CNOT}}) = \tilde{\mc{O}}(K)$~\cite{jiang2020optimal}. This leads to the compilation overhead $\tilde{\Omega}(k_1)$ in Eq.~\eqref{eq:worst_case_seri_overhead}.
\end{proof}

\subsubsection{Batched compilation}
Now, we present a concrete compilation scheme with the batched gadgets $\mb{B}_L$. 

Given \( K \) qubits in $K/k_1$ $k_1$-qubit blocks, we can partition them into multiple sectors, each composed of several \( k_1 \)-qubit blocks. 
Specifically, we define a partition \( \mb{S} = \{S_i\} \) of the set of the block indices \( [K/k_1] \), where each \( S_i \subseteq [K/k_1] \) and \( [K/k_1] = \bigcup_i S_i \).

If a \( K \)-qubit circuit \( C \) can be written as a tensor product over the sectors,
\[
    C = \bigotimes_{S_i \in \mb{S}} C_{S_i},
\]
where each \( C_{S_i} \) is a \( (|S_i| \times k_1) \)-qubit circuit supported only on the blocks indexed by \( S_i \), we denote the sector-wise decomposition of $C$ as
\begin{equation}
    (C)_{\mb{S}} := \{C_{S_i}\}_{S_i \in \mb{S}}.
\end{equation}

As an example, consider a \( K \)-qubit Hadamard circuit \( C_H \). Its decomposition into a collection of \( k_1 \)-qubit Hadamard circuits acting on each $k_1$-qubit block individually corresponds to the partition \( [[K/k_1]] := \{\{i\}\}_{i \in [K/k_1]} \), so we write the decomposition as \( (C_H)_{[[K/k_1]]} \).

Now, we describe a compilation scheme whose overhead for a given circuit $C$ is related to the property of the distribution of the sectors that $C$ is partitioned into. 

We first compile $C$ into layers of physical circuits consisting of only local $H$, $S$, CNOT, or global $T$ gates, i.e. $C = \prod_{t = 1}^{D_{\mb{B}_P}(C)} C_t$, where each $C_t$ consists exclusively of either $H$ gates, $S$ gates, non-overlapping CNOTs, or global $T$ gates $T^{\otimes K}$. Then, we further compile each circuit layer $C_t$ using the batched gadgets. A layer of global $T$ gates can be implemented using simply a batched $T$ gadget. 
It thus remains to compile circuit layers of type $H$, $S$, or CNOT. 

We first consider a circuit layer of $H$ gates, $C_H$, and the compilation for a layer of $S$ gates is analogous.  Choosing a single-block partition $[[K/k_1]]$, $C_H$ can be partitioned into $K/k_1$ $k_1$-qubit sectors $(C_H)_{[[K/k_1]]}$. 
We say that a set of generators $\mb{G} = \{g_j\}_{j = 1}^{|\mb{G}|}$, where $g_i$ is a $k_1$-qubit $H$ circuit, generate $(C_H)_{[[K/k_1]]}$ if 
\begin{equation}
        C_i = \prod_{j = 1}^{|\mb{G}|} g_j^{a_{i,j}},
\end{equation}
for any $C_i \in (C_H)_{[[K/k_1]]}$, where $a_{i,j} \in \{0, 1\}$. 
As an example, $\mb{G}_H := \{H_i\otimes I^{\otimes(k_1 - 1)}\}_{i \in [k_1]}$ with $|\mb{G}_H| = k_1$, where $H_i\otimes I^{\otimes(k_1 - 1)}$ denotes a circuit that applies a Hadamard gate on the $i$-th qubit and identity on the remaining $k_1-1$ qubits, generates any collection of $k_1$-qubit Hadamard circuits. 

Given such a generating set $\mb{G}$, an elementary BAC gadget in Theorem~\ref{theorem:batched_Cliffords} can naturally implement a generator element in $\mb{G}$ on any subset of the $K/k_1$ blocks in one logical step and execute the full circuit layer $C_H$ by implementing $\mb{G}$ sequentially:
\begin{equation}
    C_H = \prod_{j = 1}^{|\mb{G}|} \bigotimes_{i = 1}^{K/k_1} (g_j^{a_{i, j}}).
    \label{eq:compilation_generating_set}
\end{equation}
Choosing a minimal generating set $\mb{G}$ minimizes the compilation depth $|\mb{G}|$ for $C_H$.
Note that here we assume a large batch size of $K/k_1$, i.e. the same gate (or identity) is applied to all the $K/k_1$ blocks at a given time step, which results in the strictly sequential implementation of $\mb{G}$. 
More efficient compilation is possible by using a smaller batch size, as long as it satisfies the batching constraints from the construction of the batched logical gadgets. 

Next, we consider a circuit layer of CNOTs, $C_{\mr{CNOT}}$. 
We first decompose it as an intra-block CNOT circuit $C_{\mr{CNOT}}^{\mr{intra}}$ and an inter-block CNOT circuit $C_{\mr{CNOT}}^{\mr{inter}}$
\begin{equation}
    C_{\mr{CNOT}} = C_{\mr{CNOT}}^{\mr{inter}} C_{\mr{CNOT}}^{\mr{intra}},
\end{equation}
where $C_{\mr{CNOT}}^{\mr{inter}}$ (resp. $C_{\mr{CNOT}}^{\mr{intra}}$) only consists of CNOTs across (resp. within) different $k_1$-qubit blocks. 
To implement $C_{\mr{CNOT}}^{\mr{intra}}$, we can again simply choose the one-block partition $\mb{S} = [[K/k_1]]$ and implement a minimal generating set $\mb{G}$ of $(C_{\mr{CNOT}})_{[[K/k_1]]}$ sequentially, in a depth $|\mb{G}|$ analogous to Eq.~\eqref{eq:compilation_generating_set}. 

To implement $C_{\mr{CNOT}}^{\mr{inter}}$, we partition it into certain two-block circuits, i.e. choosing $\mb{S}$ in the form of $\{\{\alpha_i, \beta_i\}\}_i$, in the following. 
We define a graph $G(\mb{V}, \mb{E})$, where the vertices $\mb{V}$ are associated with the $k_1$ blocks and each physical CNOT in $C_{\mr{CNOT}}^{\mr{inter}}$ is associated with an edge $(\alpha, \beta) \in \mb{E}$ if it is supported on two qubits in the $\alpha$-th and the $\beta$-th block, respectively. An edge-coloration $\mb{E} = \bigcup_{c \in \mb{c}} \mb{E}_c$
gives a decomposition of $C_{\mr{CNOT}}^{\mr{inter}}$ into $|\mb{c}|$ layers of CNOTs,
$C_{\mr{CNOT}}^{\mr{inter}} = \prod_{c \in \bs{\vec{c}}} C_c$,
where $C_c$ has a partition into two-block CNOTs according to $\mb{E}_c$, $(C_c)_{\mb{E}_c}$. 

Then, we implement the $|\mb{c}|$ layers of the two-block CNOTs $\{C_c\}_{c \in \mb{c}}$ sequentially using the batched logical gadgets. For each layer $C_c$, we again implement a minimal generating set $\mb{G}$ of its two-block partition $(C_c)_{\mb{E}_c}$ sequentially, analogous to Eq.~\eqref{eq:compilation_generating_set}. Note that the minimum compilation depth of $C_{\mr{CNOT}}^{\mr{inter}}$ is obtained by minimizing over the choice of the coloration and the corresponding generating set.

\pagebreak

\clearpage
\onecolumngrid


\begin{center}
\textbf{\large Supplementary Information for ``Batched high-rate logical operations for quantum LDPC codes"}
\end{center}

\newcommand{\beginsupplement}{%
        \setcounter{table}{0}
        \renewcommand{\thetable}{S\arabic{table}}%
        \setcounter{figure}{0}
        \renewcommand{\thefigure}{S\arabic{figure}}%
     }
\setcounter{section}{0}
\makeatletter


\renewcommand{\thesection}{S\arabic{section}}
\vspace{0.8 in}
\newcommand{\D}{\Delta}
\newcommand{\tD}{\tilde{\Delta}}
\newcommand{\K}{K_{PP}}
\newcommand{\bn}{\bar{n}_P}
\newcommand{\G}{\Gamma}
\newcommand{\LH}{\underset{L}{H}}
\newcommand{\HL}{\underset{H}{L}}
\vspace{-1in}



\section{Batched Syndrome Extraction}
Here, we present the detailed construction of the batched syndrome extraction (BSE) gadget and prove its fault tolerance (see Theorem~\ref{theorem:formal_BSE}, the formal version of the Theorem 1 in the main text).

We first introduce some helpful notations and definitions in Sec.~\ref{sec:notations_BSE}. 
We then detail the BSE construction in Sec.~\ref{sec:BSE_construction} and put together a formal description of the logical action of the BSE gadget (Prop.~\ref{prop:BSE_action}) as well as its fault tolerance (Theorem~\ref{theorem:formal_BSE}).
Then, in Sec.~\ref{sec:proof_BSE_formal}, we prove the fault tolerance Theorem~\ref{theorem:formal_BSE}. 
In Sec.~\ref{sec:cluster_state}, we draw connections between our BSE gadget and the cluster-state scheme in Ref.~\cite{bergamaschi2024fault}.
Finally, in Sec.~\ref{sec:BSE_numerics}, we provide the details of the numerical simulation of the BSE gadget presented in the main text. 

\subsection{Notations and definitions
\label{sec:notations_BSE}}
Given a vector $f \in \mbb{F}_2^n$, we denote $\mr{supp}(f) \subseteq [n]$ as its support. Given another matrix $M \in \mbb{F}_2^{m \times n}$, we denote $[f]_M := \{f^{\prime} \mid Mf^{\prime} = M f\}$ as the equivalent class of $f$ with respect to $M$. Furthermore, we denote $|f|^r_M := \min_{f^{\prime} \in [f]_M} |f^{\prime}|$ as the reduced weight of $f$ with respect to $M$.

We denote $e^n_i \in \mbb{F}_2^n$ as a unit vector with the $i$-th entry being $1$.  

\begin{definition}[Left-expanding graph]
    Let $G = (A\cup B, E)$ be a bipartite regular graph with degree-$\Delta_A$ vertices $A$ and degree-$\Delta_B$ vertices $B$. 
    We say that $G$ is $(\gamma_A, \delta_A)$-left-expanding if for any $S \subseteq A$ with $|S| \leq \gamma_A |A|$, we have
    \begin{equation}
        |\Gamma(S)| \geq (1 - \delta_A)\Delta_A |S|,
    \end{equation}
    where $\Gamma(S) \subseteq B$ denotes the neighbor of $S$.  
\end{definition}

\begin{definition}[Expander Code]
    A classical code $C$ with check matrix $H_C$ is $(\gamma, \delta)$-expanding if the bipartite Tanner graph $G = (A\cup B, E)$ defined from $H_C$, where $A$ and $B$ represent the bits and the checks of $H_C$, respectively, is $(\gamma, \delta)$-left expanding.
    \label{def:expander_code}
\end{definition}

\begin{definition}[Locally stochastic noise]
A distribution $\mr{Pr}(e)$ of a random vector $e$ is locally stochastic if there exists some constant $p$ such that
\begin{equation}
    \tilde{\mr{Pr}}(e) := \sum_{\mr{supp}(e') \supseteq \mr{supp}(e)}\mr{Pr}(e') \leq p^{|e|}.
\end{equation}
\label{def:locally_stochastic_noise}
\end{definition}

\subsection{Construction of the BSE gadget
\label{sec:BSE_construction}}
\begin{figure}
    \centering
    \includegraphics[width=1\linewidth]{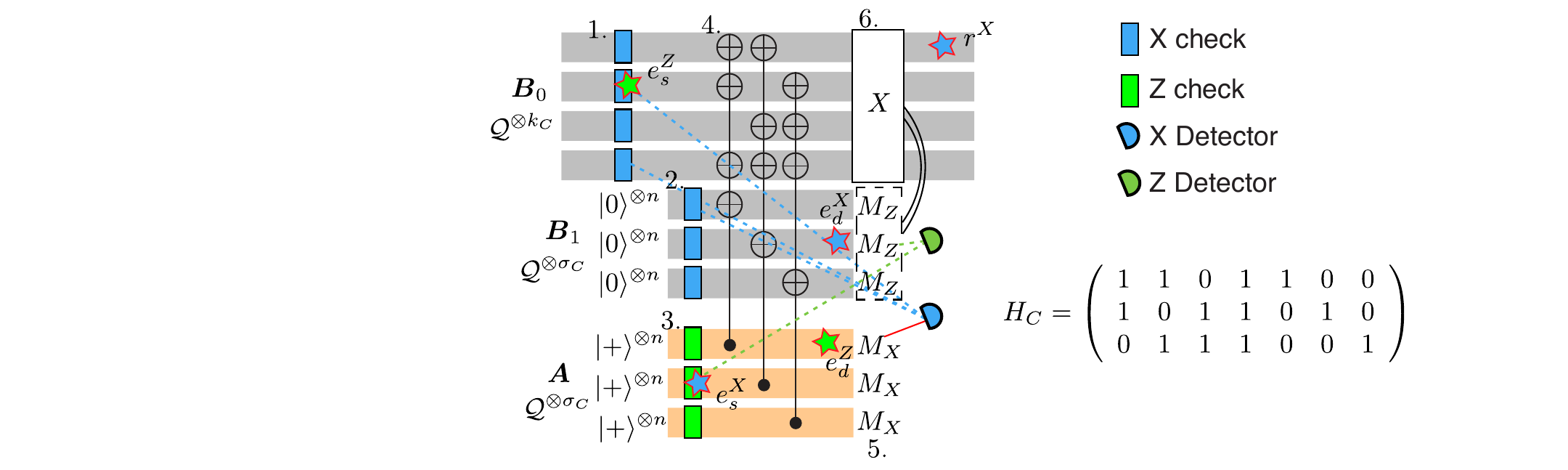}
    \caption{\textbf{Illustration of the proof for BSE.} The stars represent the four types of phenomenological syndrome or data errors that we consider during the protocol, which are detailed in the texts. }
    \label{fig:BSE_proof}
\end{figure}

Here, we present a detailed construction of the BSE gadget.

Given a $[n_C, k_C, d_C]$ classical LDPC code with a check matrix $H_C \in \mbb{F}_2^{\sigma_C\times n_C}$, where $\sigma_C = n_C - k_C$. 
Up to permutation of the bits, we assume that $H_C$ is in its canonical form with a conjugate generator matrix $G_C = (I_{k_C}, P)$, where $P \in \mbb{F}_2^{k_C \times (n_C - k_C)}$, with $H_C G_C^T = 0 \mod 2$. 
We measure the $X$ checks of a $[[n, k, d]]$ CSS qLDPC code $\mc{Q}$ with $X$ checks $H_X\in \mbb{F}_2^{\sigma_X \times n}$ and $Z$ checks $H_Z \in \mbb{F}_2^{\sigma_Z \times n}$ using the following protocol:

\begin{enumerate}
    \item Input: $k_C$ $\mc{Q}$ blocks. Denote this set of blocks $\bs{B}_0$. Start by measuring the $X$ checks of $\mc{Q}^{\otimes k_C}$ in $\bs{B}_0$ once.
    \item Initialize $\sigma_C$ ancilla blocks in $\vert 0 \rangle^{\otimes n}$, and  measure the $X$ checks of $\mc{Q}^{\otimes \sigma_C}$ once. Denote this set of blocks $\bs{B}_1$.
    \item Initialize $\sigma_C$ ancilla blocks in $\vert + \rangle^{\otimes n}$, and measure the $Z$ checks of $\mc{Q}^{\otimes \sigma_C}$ once. Denote this set of blocks $\bs{A}$.
    \item Apply $\bs{A}$-controlled transversal CNOTs targeting $\bs{B}_0$ and $\bs{B}_1$ based on $H_C$: Let the blocks of $\bs{B}_0 \cup \bs{B}_1$ be indexed by $[n_C]$ and the blocks of $\bs{A}$ indexed by $[\sigma_C]$. Apply a transversal CNOT between the $i$-th block of $\bs{A}$ and the $j$-th block of $\bs{B}_0 \cup \bs{B}_1$ if $H_C[i,j] = 1$. 
    \item Measure transversely blocks in $\bs{B}_1$ in the $Z$ basis and blocks in $\bs{A}$ in the $X$-basis. 
    \item Perform error correction using detectors in the circuits (described in the following) and apply a feedback Pauli $X$ correction on $\bs{B}_0$ based on the $Z$ measurement outcomes of $\bs{B}_1$ as follows: apply a classically-controlled transversal CNOT from the $j$-th block of $\bs{B}_1$ to the $i$-th block of $\bs{B}_0$ if $P_{ij}=1$.
\end{enumerate}
See Fig.~\ref{fig:BSE_proof} for an illustration of the protocol. 

We establish the following notation for describing qubits and checks of $m$ blocks of $\mc{Q}$, $\mc{Q}^{\otimes m}$: 
Let $e^m_i\otimes e^n_j$ be a basis vector associated with the $j$-th qubit of the $i$-th $\mc{Q}$ block, and similarly, $e^m_i \otimes e^{\sigma_X}_j$ ($e^m_i \otimes e^{\sigma_Z}_j$) be a basis vector associated with the $j$-th $X$ ($Z$) check of the $i$-th $\mc{Q}$ block. The space of $m n$-qubit Pauli $X$ (or $Z$) operators are then spanned by $\{e^m_i \otimes e^n_j\}_{i \in [m], j \in [n]}$. Correspondingly, the $X$ ($Z$) check matrix of $\mc{Q}^{\otimes m}$ is given by $I_m \otimes H_X$ ($I_m \otimes H_Z$). Similarly, a vector indexing a subset of $X$ ($Z$) checks of $\mc{Q}^{\otimes m}$ is in $\langle e^{m}_i \otimes e^{\sigma_X}_j\rangle_{i \in [m], j \in [\sigma_X]}$ 
($\langle e^{m}_i \otimes e^{\sigma_X}_j\rangle_{i \in [m], j \in [\sigma_Z]}$).

We first show that in the absence of errors within the BSE gadget, the protocol is equivalent to only measuring a round of $X$ checks for $\bs{B}_0$. 
\begin{proposition}[Action of the BSE gadget in the absence of errors]
    In the absence of errors during the BSE gadget, the six-step protocol above is equivalent to measuring the $X$ checks of $\bs{B}_0$ once. 
    \label{prop:BSE_action}
\end{proposition}
\begin{proof}
Measuring the $X$ checks of $\bs{B}_0$ once projects $\bs{B}_0$ into eigenstates of $I_{k_C} \otimes H_X$, which will be superposition of eigenstates of the $Z$ checks and logicals of $\bs{B}_0$, i.e. $\{I_{k_C}\otimes H_Z, I_{k_C}\otimes L_Z\}$, where $L_Z \in \mbb{F}_2^{k \times n}$ denotes the $Z$ logical matrix of $\mc{Q}$. 
Since the circuit of the BSE gadget commutes with $I_{k_C} \otimes H_X$, it remains to prove that any state that is also an eigenstate of $\{I_{k_C}\otimes H_Z, I_{k_C}\otimes L_Z\}$ remains the same after the BSE gadget. It suffices to show this by checking it for the $+1$ eigenstates of $\{I_{k_C}\otimes H_Z, I_{k_C}\otimes L_Z\}$. In this case, $\bs{B}_0$ is initially stabilized by $\{I_{k_C}\otimes H_Z, I_{k_C}\otimes L_Z\}$ and we aim to show that it remains so at the output of the BSE gadget. We prove this by checking the transformation of $\{I_{k_C}\otimes H_Z, I_{k_C}\otimes L_Z\}$ throughout the circuit. 

It is easy to see that after step 2, the blocks $\bs{B}_0$ and $\bs{B}_1$ are stabilized by $\{I_{n_C}\otimes H_Z, I_{n_C}\otimes L_Z\}$. 
Next, the transversal $X$ measurements of $\bs{A}$ contains the measurements of the $X$ operators $\{H_C\otimes H_X, H_C\otimes L_X\}$ on $\bs{B}_0 \cup \bs{B}_1$. This transforms the $Z$-type stabilizers to $\{I_{n_C}\otimes H_Z, G_C\otimes L_Z\}$. Note that the $Z$ logical operators have delocalized across the $n_C$ blocks. The final transversal measurements on $\bs{B}_1$ then projects the system into an eigenstate of $\{I_{k_C}\otimes H_Z, I_{k_C}\otimes L_Z\}$, with the $Z$ checks $I_{k_C}\otimes H_Z$ having eigenvalues $+1$. However, $I_{k_C}\otimes L_Z$ might have random eigenvalues. This is finally fixed by the Pauli $X$ feedback in step 4, which ensures the eigenvalues of $I_{k_C}\otimes L_Z$ get back to $+1$. 
\end{proof}

Now, we analyze the fault tolerance of the BSE gadget by considering a phenomenological noise model that contains the following four sets of errors $\mc{E} = \{e^Z_d, e^Z_s, e^X_d, e^X_s\}$:
\begin{enumerate}
    \item $e^Z_d \in \mbb{F}_2^{\sigma_C}\otimes \mbb{F}_2^{n}$: $Z$ data error on $\bs{A}$ before the final transversal $X$ measurements.
    \item $e^Z_s \in \mbb{F}_2^{n_C} \otimes \mbb{F}_2^{\sigma_{X}}$: syndrome error on the initial measurements of the $X$ checks of $\bs{B}_0 \cup \bs{B}_1$. 
    \item $e^X_d \in \mbb{F}_2^{\sigma_C}\otimes \mbb{F}_2^{n}$: $X$ data error on $\bs{B}_1$ before the final $Z$ transversal measurements. 
    \item $e^X_s \in \mbb{F}_2^{\sigma_C}\otimes \mbb{F}_2^{\sigma_Z}$: syndrome error on the initial measurements of the $Z$ checks of $\bs{A}$.
\end{enumerate}
See the blue and green stars in Fig.~\ref{fig:BSE_proof} for an illustration of these errors. 
We consider that these errors $\mc{E}$ are locally stochastic (see Def.~\ref{def:locally_stochastic_noise}) with the same physical error rate $p$.
The fault tolerance of the BSE gadget can be fully captured by its tolerance against $\mc{E}$ since all locally stochastic errors at other locations are equivalent to the combination of $\mc{E}$ and/or some locally stochastic residual errors at the output. 

Now, we describe the detectors for correcting $\mc{E}$. 
By assumption, $H_C$ is in the following canonical form: $H_C = (H_C^0, H_C^1)$, where $H_C^0 \in \mbb{F}_2^{\sigma_C \times k_C}$ represents the first $k_C$ columns of $H_C$ and $H_C^1 \in \mbb{F}_2^{\sigma_C \times \sigma_C}$, being full rank, represents the last $\sigma_C$ columns of $H_C$.

The $X$ detectors for correcting a $Z$-type error $e^Z = \left(
\begin{array}{c}
     e^Z_d \\
     e^Z_s
\end{array}\right)$ are described by a $X$ detector matrix
\begin{equation}
    D_X = \left( I_{\sigma_C}\otimes H_{X}, H_C \otimes I_{\sigma_{X}}\right),\label{eq:X_detector_mat}
\end{equation}
and the $Z$ detectors for correcting a $X$-type error $e^X = \left(
\begin{array}{c}
     e^X_d \\
     e^X_s
\end{array}\right)$ are given by a $Z$ detector matrix
\begin{equation}
    D_Z = \left(I_{\sigma_C} \otimes H_{Z}, (H_C^{1})^{T} \otimes I_{\sigma_{Z}} \right).
    \label{eq:Z_detector_mat}
\end{equation}
See Fig.~\ref{fig:BSE_proof} for an illustration of these detectors. 

After performing error correction using the $X$ and $Z$ detectors, there will be two types of residual errors on the output $\bs{B}_0$ block:
\begin{enumerate}
    \item $r_s^Z \in \mbb{F}_2^{k_C}\otimes \mbb{F}_2^{\sigma_X}$:residual syndrome error on the initial measurements of the $X$ checks of $\bs{B}_0$. 
    \item $r^X \in \mbb{F}_2^{k_C}\otimes \mbb{F}_2^{n}$: residual $X$ data error on $\mb{B}_0$. 
\end{enumerate}
\qx{Note that the residual $Z$ data errors will simply be the initial $Z$ errors on the $\bs{B}_0$ commuting through the circuit, therefore being locally stochastic automatically}.

The fault tolerance of the BSE gadget in the presence of $\mc{E}$ is formally stated in the following theorem:
\begin{theorem}[Formal statement of Theorem 1 in the main text]
    If $H_C = (H_C^0, H_C^1)$ satisfies the following condition
    \begin{enumerate}
        \item $H_C^1$ is full rank.
        \item $H_C$ is $(\gamma, \delta)$-expanding for some constants $\gamma$ and $\delta < 1/4$ (see Def.~\ref{def:expander_code}). 
        \item $(H_C^{1})^T$ is $(\gamma^{\prime}, \delta^{\prime})$-expanding for some constants $\gamma^{\prime}$ and $\delta^{\prime} < 1/4$. 
        \item $H_C$ is LDPC and $d_C = \Omega(d)$.
    \end{enumerate}
    Then, after applying minimum-weight corrections based on the circuit detectors, in the limit $d \rightarrow \infty$ and provided that the physical error rate $p$ is below some constant threshold,  the residual $X$ syndrome error $r^Z_s$ and the residual $X$ data error $r^X$ satisfy:
    \begin{enumerate}
        \item $r^Z_s$ is the boundary of some locally stochastic $Z$ data error on $\mb{B}_0$, i.e., there exists $h^Z \in \mbb{F}_2^{k_C \otimes n}$ such that
        \begin{equation}
        \mr{Pr}(r^Z_s = (I_{k_C}\otimes H_X)h^Z) \leq \mc{O}(p)^{|h^Z|},
        \label{eq:th_residual_syndrome}
    \end{equation}
    \item $r^X$ is locally stochastic.    \end{enumerate}
    \label{theorem:formal_BSE}
\end{theorem}

\subsection{Proof of BSE Theorem~\ref{theorem:formal_BSE}
\label{sec:proof_BSE_formal}}
We sketch the main ideas behind the proof here. For both the $Z$- and $X$-type errors, we aim to show that some effective residual data error $h$ (e.g. $h^Z$ in Eq.~\eqref{eq:th_residual_syndrome}) is locally stochastic after error correction using the BSE detectors Eqs.~\eqref{eq:X_detector_mat}\eqref{eq:Z_detector_mat}. This residual error $h$ is caused by some locally stochastic circuit fault $f$. Analogous to the proof for single-shot QEC assuming confinement or soundness~\cite{quintavalle2020single, campbell2019theory}, we first show that the weight of the $h$ can be upper-bounded by the weight of $f$, indicating that small faults would only produce small residual errors. This utilizes the structure of the detectors matrices Eqs.~\eqref{eq:X_detector_mat}\eqref{eq:Z_detector_mat} as the check matrices of some HGP code and a ``soundness" lemma (Lemma~\ref{lemma:HGP_soundness}) for a HGP code with one expanding base code. Then, combining with some variants of the cluster-counting techniques from Ref.~\cite{gottesman2013fault} (Lemma.~\ref{lemma:prob_equivalent_classes}), which upper-bound the number of circuit faults that lead to the same residual error $h$, we can prove the local stochastic property of $r^Z_s$ and $r^X$ in Sec.~\ref{sec:r_Z_s} and Sec.~\ref{sec:r_X}, respectively.
Finally, we present a family of classical codes that satisfy the requirements in Theorem~\ref{theorem:formal_BSE} via random constructions~\cite{sipser1996expander, richardson2008modern}. 

\subsubsection{Important mathematical lemmas}
We first establish two key mathematical lemmas that facilitate our proof: Lemma~\ref{lemma:prob_equivalent_classes} that bounds the probability of an equivalent class $[e]_{M}$ with respect to a LDPC matrix $M$ when $e$ is locally stochastic; Lemma~\ref{lemma:HGP_soundness} that relates the syndrome weight of a HGP code with its data error weight when the base classical codes are sufficiently expanding. 

\begin{lemma}[Lemma 2 of Ref.~\cite{gottesman2013fault}]
Let $G = (\mb{V}, \mb{E})$ be a graph with bounded degree $z$, and $\mb{S}\subseteq \mb{V}$ be a subset of vertices. Let $M_z(s, \mb{S})$ be the number of size-$s$ sets containing $\mb{S}$ (i.e., with $s - |\mb{S}|$ extra nodes beyond those in $\mb{S}$), and which are a union of connected clusters, each of which contains a
node in $\mb{S}$. Then 
$$M_z(s, \mb{S}) \leq e^{|\mb{S}|-1} (ze)^{s-|\mb{S}|}.$$
\label{lemma:size_clusters_containing}
\end{lemma}

\begin{corollary}
Let $G = (\mb{V}, \mb{E})$ be a graph with bounded degree $z$, and $\mb{S}\subseteq \mb{V}$ be a subset of vertices. Let $M^{\prime}_z(s, \mb{S})$ be the number of size-$s$ sets that are a union of connected clusters each containing at least one vertex in $\mb{S}$. 
Then 
$$M^{\prime}_z(s, \mb{S}) \leq \frac{(ze)^{s}}{e}[(1 + \frac{1}{z})^{|\mb{S}|} - 1].$$
\label{corollary:size_clusters_touching}
\end{corollary}
\begin{proof}
Let $\mb{M}(s, \mb{S})$ be the collection of size-$s$ sets that contain $\mb{S}$ and are a union of connected clusters each containing at least one vertex in $\mc{S}$, $\mb{M}^{\prime}(s, \mb{S})$ be the collection of size-$s$ sets that are a union of connected clusters each containing at least one vertex in $S$. Then,
\begin{equation}
\mb{M}^{\prime}(s, \mb{S}) \subseteq \bigcup_{\mb{S}_0 \subseteq \mb{S}} \mb{M}(s, \mb{S}_0). 
\end{equation}
Therefore,
\begin{equation}
M^{\prime}_z(s, S) = |\mb{M}^{\prime}(s, \mb{S})| \leq \sum_{\mb{S}_0 \subseteq \mb{S}} |\mb{M}(s, \mb{S}_0)| = \sum_{\mb{S}_0 \subseteq \mb{S}} M_z(s, \mb{S}_0).
\end{equation}
According to Lemma~\ref{lemma:size_clusters_containing}, $M_z(s, \mb{S}_0) \leq e^{|\mb{S}_0| - 1}(ze)^{s - |\mb{S}_0|}$. Then, we have
\begin{equation}
    M^{\prime}_z(s, \mb{S}) \leq \sum_{\mb{S}_0 \subseteq \mb{S}}e^{|\mb{S}_0| - 1}(ze)^{s - |\mb{S}_0|} = \frac{(ze)^{s}}{e}\sum_{t = 1}^{|\mb{S}|} \left( \begin{array}{c}
             |\mb{S}|\\
             t \\
        \end{array}\right) z^{-t} = \frac{(ze)^{s}}{e}[\frac{(1 + z)^{|\mb{S}|}}{z} - 1].
\end{equation}
\end{proof}

\begin{lemma}[Probability of equivalent error classes]
Let $H \in \mbb{F}_2^{\sigma\times n}$ be a $(r,c)$-LDPC matrix, if the error $e \in \mbb{F}_2^n$ is subject to a locally stochastic noise with a noise rate $p$, then the probability of the equivalent classes $[e]_{H}$ satisfies:
\begin{equation}
    \mr{Pr}([e]_{H}) := \sum_{e^{\prime} \in [e]_H}\mr{Pr}(e^{\prime}) \leq \mc{O}((p/p_z)^{|e|_H^r})
\end{equation}
if $p < p_z$, where $p_z$ is a constant depending on $z:=(r-1)c$.
\label{lemma:prob_equivalent_classes}
\end{lemma}
\begin{proof}
    \begin{equation}
        \mr{Pr}([e]_{H}) = \sum_{e^{\prime} \in [e]_{H}} \mr{Pr}(e) = \sum_{e_c \in \ker{H}}\mr{Pr}(e_m + e_c),
        \label{eq:sum1}
    \end{equation}
    where $e_m$ is a minimum-weight representation in $[e]_{H}$.
    
    Following Ref.~\cite{gottesman2013fault}, we define a syndrome adjacency graph $G$ where we associate each of the $n$ bits with a vertex and two bits $i$ and $j$ are connected by an edge if there is a check in $H$ that acts nontrivially on both $i$ and $j$. $G$ has a bounded vertex degree $z = (r-1)c$. Then, any element $e_c \in \ker{H}$ has the following structure on $G$: $e_c$ is always a union of connected clusters, i.e. $e_c = \sum_i e_c^{(i)}$ with $\{e_c^{(i)}\}$ being disjoint connected clusters on $G$, where each connected cluster $e_c^{(i)}$ is itself in the kernel of $H$. 
    
    Now, let $\mb{J}_{e_m}$ be the union of $\{0\}$ and the set of $e_c$ such that it only consists of a union of connected clusters each intersecting nontrivially with $e_m$. Then we have 
    \begin{equation}
        \mr{Pr}([e]_{H}) \leq \sum_{e_c \in \mb{J}_{e_m}} (\sum_{e' \supseteq e_m + e_c}\mr{Pr}(e')) = \sum_{e_c \in \mb{J}_{e_m}} \tilde{\mr{Pr}}(e_m + e_c),
        \label{eq:sum2}
    \end{equation}
    as for any $e_c^{\prime} \notin \mb{J}_{e_m}$, there exists a $e_c \in \mb{J}_{e_m}$ such that $e_c^{\prime} + e_m \supset e_c + e_m$, and thus the set that is summed over in Eq.~\eqref{eq:sum2} is strictly larger than that in Eq.~\eqref{eq:sum1}. Since the noise is locally stochastic, we have $\tilde{\mr{Pr}}(e_m + e_c) \leq p^{|e_m + e_c|}$. Furthermore, since $e_m$ is a minimum-weight representation, $|e_m + e_c| \geq \max\{|e_m|, |e_c| - |e_m|\}$. Therefore, 
    \begin{equation}
    \begin{aligned}
        \mr{Pr}([e]_H) & \leq \sum_{e_c \in \mb{J}_{e_m}, |e_c| < 2|e_m|} p^{|e_m|} + \sum_{e_c \in \mb{J}_{e_m}, |e_c| \geq 2|e_m|} p^{|e_c| - |e_m|} \\
        & = (\sum_{w = 0}^{2|e_m| - 1} n_w)p^{|e_m|} + (\sum_{w = 2|e_m|}^{\infty} n_w p^w)/p^{|e_m|},
    \end{aligned}
    \end{equation}
    where $n_w$ denotes the number of $e_c \in \mb{J}_{e_m}$ with weight $w$. According to Corollary~\ref{corollary:size_clusters_touching}, we have $n_w = M^{\prime}_z(w, e_m) \leq \frac{(ze)^{w}}{e}[(1 + \frac{1}{z})^{|e_m|} - 1]$. Therefore,
    \begin{equation}
        \mr{Pr}([e]_H) \leq \mc{O}((p/p_z)^{|e_m|}), 
    \end{equation}
    where $p_z = \frac{1}{(1 + z)ze^2}$, for $p <1/(ze)$.
\end{proof}

\begin{lemma}[Locally stochastic residual error under minimum-weight decoder]
Let $e \in \mbb{F}_2^n$ be some locally stochastic error with an error rate $p$ and a decoder returns a minimum-weight correction $e^{\prime}$ with respect to some check matrix $H \in \mbb{F}_2^{r \times n}$, i.e. $H(e + e^{\prime}) = 0$ and $e^{\prime}$ is of minimum weight among all valid corrections, then the residual error $f = e + e^{\prime}$ is locally stochastic with an error rate $\leq 2\sqrt{p}$.
\label{lemma:residual_locally_stochastic}
\end{lemma}
\begin{proof}
    Given a residual error $f$, it is easy to see that $e$ must have a lafger intersection with $f$: $|e \cap f| > |f|/2$ since otherwise $e^{\prime}$ would not be a minimum-weight correction. Then, we have
    \begin{equation}
        \mr{Pr}(f) \leq \sum_{f_0 \subseteq f, |f_0| > |f|/2}\mr{Pr}(e \supseteq f_0) \leq \sum_{f_0 \subseteq f, |f_0| > |f|/2} p^{|f_0|} = \sum_{t = |f|/2}^{|f|} \left( \begin{array}{c}
             |f| \\
             t
        \end{array}\right) p^t \leq (2\sqrt{p})^{|f|}.
    \end{equation}
\end{proof}

\begin{lemma}[Unique-neighbor expansion~\cite{sipser1996expander}]\label{lemma:single_site_flip}
    Let $G = (A\cup B, E)$ be a $(\gamma_A, \delta_A)$-left-expanding graph with $\delta_A < 1/2$ and left and right degree $\Delta_A$ and $\Delta_B$, respectively. Let $S \subseteq A$ be any subset of $A$, and $\Gamma_u (S)$ be the set of unique neighbors of $S$, i.e. the set of vertices $v \in B$ such that $v$ has degree $1$ in the graph $S \cup \Gamma(S)$ induced by $S$. Then, for $|S| \leq \gamma_A |A|$, we have
    \begin{equation}
        \left|\Gamma_u\left(S\right)\right| \geq\left(1-2 \delta_A\right) \Delta_A\left|S\right|.
    \end{equation}
\end{lemma}

\begin{corollary}[majority-unique neighbor condition] Let $G = (A\cup B, E)$ be a $(\gamma_A, \delta_A)$-left-expanding graph with $\delta_A < 1/4$. Then, for any subset $S \subseteq A$, where $|S| \leq \gamma_A |A|$, with neighbors $\Gamma(S)$ and unique neighbors $\Gamma_u(S)$ forming subsets in $B$, there is at least one element $s \in S$ with majority unique neighbors, i.e. $|\Gamma_u(s,S)| > \frac{1}{2} |\Gamma(s,S)|$, where $\Gamma_u(s,S),\Gamma(s,S)$ are the subsets of elements in $\Gamma_u(S),\Gamma(S)$ respectively with an edge connected to $s$.
\label{corollary:maj_unique_neighbors}
\end{corollary}

\begin{lemma}[``Soundness" of HGP code]
Let $\tilde{D}_X$ and $\tilde{D}_Z$ be the $X$ and $Z$ check matrix of a HGP code $\mr{HGP}(H_1, H_2)$ with two base matrices $H_1 \in \mbb{F}_2^{r_1 \times n_1}$ and $H_2 \in \mbb{F}_2^{r_2 \times n_2}$:
\begin{equation}
    \begin{aligned}
        \tilde{D}_X & = (I_{n_1}\otimes H_2, H_1^T \otimes I_{r_2}), \\
        \tilde{D}_Z & = (H_1 \otimes I_{n_2}, I_{r_1}\otimes H_2^T).
    \end{aligned}
\end{equation}
If $H_1^T$ is $(\gamma, \delta)$-expanding with $\delta < 1/4$, then we have the following ``linear soundness" relation: for any $f = \left(\begin{array}{c}
            f_{a}\\
            f_{b}
        \end{array}\right) = \tilde{D}_Z^T h \in \im{\tilde{D}_Z^T}$, 
        we have
        \begin{equation}
            |f| \geq \min\{|h|^r_{I_{r_1} \otimes H_2}, d(H_1^T), \gamma r_1\},
            \label{eq:HGP_soundness}
        \end{equation}
        where $d(H_1^T)$ denotes the distance of the classical code with check matrix $H_1^T$.

\label{eq:linear_soundness_HGP}
\label{lemma:HGP_soundness}
\end{lemma}
\begin{proof}
    Let $\mb{B}_i = [n_i]$ and $\mb{C}_i = [n_1 + r_1]\backslash [n_1]$ denote the indices of the bits and checks of $H_i$, respectively, for $i = 1, 2$. $f_a$, $f_b$, and $h$ are supported on the 2D grids $\mb{B}_1 \times \mb{B}_2$, $\mb{C}_1\times \mb{C}_2$, and $\mb{C}_1 \times \mb{B}_2$, respectively. 
    We refer to $r \times (\mb{B}_2 \cup \mb{C}_2)$, where $r \in {\mb{B}_1 \cup \mb{C}_1}$ as a row indexed by $r$. Given any vector $v$, we denote $v|_{r}$ as its restriction on row $r$, and $|v|_{\mr{RW}}$ (row weight of $v$) as the number of nonzero rows that $v$ is supported on.
    Given any subset $\mb{R} \subseteq \mb{C}_1$ and a row $r \in \mb{C}_1$, denote $\Gamma(\mb{R}) \subseteq \mb{B}_1$ as the neighbors of $\mb{R}$ and $\Gamma_u(r, \mb{R}) \subseteq \Gamma(\mb{R})$ the unique neighbor of $r$ (see Corollary.~\ref{corollary:maj_unique_neighbors}) according to the Tanner graph defined by $H_1^T$. 

    Observe that Eq.~\eqref{eq:HGP_soundness} is equivalent to proving that if $|f| \leq \min\{d(H_1^T), \gamma r_1\}$, we have $|f| \geq |h|^r_{I_{r_1}\otimes H_2}$. So now we assume that $|f| \leq \min\{d(H_1^T), \gamma r_1\}$. 
    
    Let $\mb{R} \subseteq \mb{C}_1$  with $|\mb{R}| = q $ be the set of rows that $f_b$ has nontrivial supported on. By assumption, we have that $q \leq |f_b| \leq |f| \leq \gamma r_1$.
    \qx{We now use an iterative proof procedure similar to that for small-set-flip decoders~\cite{fawzi2017efficient} by}
    defining a sequence of vectors $\{f_b^{(i)}, f_a^{(i)}, h^{(i)} \}_{i = q, q-1, \cdots, 0}$ satisfying the following conditions:
    \begin{enumerate}
        \item $f_b^{(q)} = f_b, f_a^{(q)} = f_a|_{\Gamma(\mb{R})}, h^{(q)} = 0$. 
        \item During step $i$, $|f_b^{(i)}|_{\mr{RW}} = i$. 
        \item For all $i$, we have \begin{equation}
            (I _{n_1}\otimes H_{2}) f_a^{(i)} + (H_1^T \otimes H_{2}) h^{(i)}  = (H_1^T \otimes I_{r_2}) f_b.
            \label{eq:constant_eq}
        \end{equation}
        \item The total weight of $f_a$ and $h$ is non-increasing from step-to-step, i.e. 
        \begin{equation}
            |f_a^{(i-1)}| + |h^{(i-1)}| \leq |f_a^{(i)}| + |h^{(i)}|.\label{eq:weight_decreasing}
        \end{equation}
    \end{enumerate}

    \qx{First, we check that Eq.~\ref{eq:constant_eq} is satisfied for the initial step $i = q$. By definitions of $f_a^{(q)}$ and $h^{q}$, we are checking $(I _{n_1}\otimes H_{2}) f_a|_{\Gamma(\mb{R})} = (H_1^T \otimes I_{r_2})f_b$, which holds since $\tilde{D}_X\left(\begin{array}{c}
            f_{a}\\
            f_{b}
        \end{array}\right) = (I _{n_1}\otimes H_{2}) f_a + (H_1^T \otimes I_{r_2})f_b = 0$, and $f_a|_{\mb{B}_1\backslash\Gamma(\mb{R})} = 0$ by the definition of $\mb{R}$.}
    We then construct the sequence iteratively by specifying $\{f_a^{(i-1)}, f_b^{(i-1)}, h^{(i-1)}\}$ based on $\{f_a^{(i)}, f_b^{(i)}, h^{(i)}\}$ at step $i$.
    
    During step $i$, assume that Eq.~\eqref{eq:constant_eq} is satisfied for $\{f_a^{(i)}, f_b^{(i)}, h^{(i)}\}$ and $|f_b^{(i)}|_{\mr{RW}} = i$. Let $\mb{R}^{(i)}$ be the row support of $f_b^{(i)}$. 
    As $|\mb{R}^{(i)}| \leq q \leq \gamma |\bs{C}_1|$, according to Corollary~\ref{corollary:maj_unique_neighbors}, we can find a row $r^{(i)} \in \mb{R}^{(i)}$ which has more unique neighbors in $\Gamma(\mb{R}^{(i)})$ than non-unique neighbors, i.e. $|\Gamma_u(r^{(i)}, \mb{R^{(i)}})| > |\Gamma(r^{(i)}) \backslash \Gamma_u(r^{(i)}, \mb{R^{(i)}})|$. 
    We further know that $h^{(i)}|_{r^{(i)}} = 0$ since otherwise we would have set $f_b^{(i)}|_{r^{(i)}} = 0$ at the previous step $i-1$ (this will become clear after going through the assignment at step $i$).
    Then, according to Eq.~\eqref{eq:constant_eq} --- $ (I _{n_1}\otimes H_{2}) f_a^{(i)} = (H_1^T \otimes I_{r_2}) ( (I_{r_1}\otimes H_{2}) h^{(i)} + f_b)$ --- we have that for each row $t \in \Gamma_u(r^{(i)},\mb{R}^{(i)})$,
    $$f_b|_{r^{(i)}} = H_{2} f_a^{(i)}|_t.$$
    In other words, we are guaranteed that $f_b|_{r^{(i)}}$ is a syndrome of $H_2$ that is generated by $\{f_a^{(i)}|_t\}_{t \in \Gamma_u(r^{(i)}, \mb{R}^{(i)})}$ .
    Let us select the minimum weight pattern, $g^{(i)} = f_a^{(i)}|_{t_{min}}$, where $t_{min} = \mathrm{argmin}_{t \in \Gamma_u(r^{(i)},\mb{R}^{(i)})} |f_a^{(i)}|_t |$.
    
    Then, we can update $h^{(i-1)}$ such that $h^{(i-1)}|_{r^{(i)}} = g^{(i)}$ and $h^{(i-1)}|_{r'} = h^{(i)}|_{r'}$ for all $r' \neq r^{(i)}$, 
    Meanwhile, we update $f_a^{(i-1)}$ in the following: we set its support on all of the unique neighbors of $r^{(i)}$ to zero, i.e. $f_a^{(i-1)}|_t = 0$ for all $t \in \Gamma_u(r^{(i)}, \mb{R}^{(i)})$; we add $g^{(i)}$ to $f_a^{(i-1)}$ on the remaining neighbors of $r^{(i)}$, i.e. 
    $f_a^{(i-1)}|_{t'} = f_a^{(i)}|_{t'} + g^{(i)}$ for all $t' \in \Gamma(r^{(i)}) \backslash \Gamma_u(r^{(i)}, \mb{R}^{(i)})$; finally, we keep $f_a^{(i-1)}$ the same for the remaining rows, i.e. $f_a^{(i-1)}|t^{\prime} = f_a^{(i)}|_{t^{\prime}}$ for all $t^{\prime} \in \Gamma(\mb{R}) \backslash \Gamma(r^{(i)})$. It is easy to verify that this guarantees Eq.~\eqref{eq:constant_eq} for step index $i-1$.
    In addition, this update rule guarantees the total weight of $f_a^{(i)}$ and $h^{(i)}$ is non-increasing from step $i$ to step $i-1$ (Eq.~\eqref{eq:weight_decreasing}).
    In particular, notice that
    \begin{align}
    |h^{(i-1)}| &= |h^{(i)}| + |g^{(i)}|, \\
    |f_a^{(i-1)}| &\leq |f_a^{(i)}| - |g^{(i)}|,
    \end{align}
    where the second inequality leverages the choice of $r^{(i)}$ such that $|\Gamma_u(r^{(i)},\mb{R}^{(i)})| \geq |\Gamma(r^{(i)})\backslash\Gamma_u(r,\mb{R}^{(i)})|+1$.

    Finally, we update $f_b^{(i-1)}$ in the following: we set $f_b^{(i-1)}|_{r^{(i)}} = 0$ and $f_b^{(i-1)}|_{r^{\prime}} = f_b^{(i)}|_{r^{\prime}}$ for all $r^{\prime} \neq r^{(i)}$.
    This guarantees that $|f_b^{(i-1)}| = i - 1$ and also that the unique neighbors of $r^{(i)}$, i.e. $\Gamma_u(r^{(i)}, \mb{R}^{(i)})$, will not be in $\Gamma(\mb{R}^{(i-1)})$. 
    
    We now prove that at the last step, when we go from $i=1$ to $i=0$, we have $f_a^{(0)} = 0$. According to the update rule for $f_a$, we have set $f_a^{(0)}|_t = 0$ if the row $t$ is ever a unique neighbor  of $r^{(i)}$, i.e. $t \in \Gamma_u(r^{(i)}, \mb{R}^{(i)})$ for some $r$. Let $\mb{S} := \bigcup_{i = 1}^q \Gamma_u(r^{(i)}, \mb{R}^{(i)})$, we then have $f_a^{(0)}|_{\mb{S}} = 0$. As such, we simply need to prove that $\mb{S} = \Gamma(\mb{R})$. Note that for any $r \in \Gamma(r^{(i)})\backslash \Gamma_u(r^{(i)}, \mb{R}^{(i)})$ that is a non-unique neighbor of $r^{(i)}$, it will become the unique neighbor of some $r^{(i^{\prime})}$ with $i^{\prime} < i$, i.e. $r \in \Gamma_u(r^{(i^{\prime})}, \mb{R}^{(i^{\prime})})$. Therefore, we have $\mb{S} \supseteq \bigcup_{i=1}^q \Gamma(r^{(i)}) = \Gamma(\mb{R})$.
    
    Then, $h^{(0)}$ satisfies  $(H_1^T \otimes I_{r_2}) [f_b + (I_{r_1}\otimes H_2)h^{(0)}] = 0$ (according to Eq.~\eqref{eq:constant_eq} at $i = 0$) and $|h^{(0)}| \leq |f_a|$ (according to Eq.~\eqref{eq:weight_decreasing}).   
    Since $|f_b + (I_{r_1}\otimes H_2)h^{(0)}|_{\mr{RW}} \leq |f_b| + |h^{(0)}| \leq |f|$, which is less than $d(H_1^T)$ (by assumption), we have $(I_{r_1}\otimes H_2)h^{(0)} = f_b$. 
    This indicates that $h^{(0)} \in [h]_{I_{r_1}\otimes H_2}$. Therefore, we have $[h]^r_{I_{r_1}\otimes H_2} \leq |h^{(0)}| \leq |f_a| \leq |f|$. 
    
\end{proof}

Note that Eq.~\eqref{eq:HGP_soundness} is not the same the standard soundness definition for HGP codes~\cite{campbell2019theory}. As we will show later, the detector-fault matrices Eqs.~\eqref{eq:X_detector_mat}\eqref{eq:Z_detector_mat} of the BSE gadget coincide with the check matrices of some HGP code, and we will use Eq.~\eqref{eq:HGP_soundness} to upper-bound the size of the effective residual error (corresponding to $h$) by that of the circuit fault (corresponding to $f$), which is the key ingredient behind the fault tolerance of the BSE gadget. 

\subsubsection{Proof for the residual $X$ syndrome error $r^Z_s$ \label{sec:r_Z_s}}
The $X$ detectors for correcting a $Z$-type error $e^Z = \left(\begin{array}{c}
     e^Z_d \\
     e^Z_s
\end{array}\right)$ are given by Eq.~\eqref{eq:X_detector_mat}. 
$D_X$ corresponds to the $X$ check matrix of a HGP code $\mc{C}^Z = \mr{HGP}(H_C^T, H_{q, X})$, whose $Z$ check matrix is
\begin{equation}
    \tilde{D}_Z = \left( H_C^T \otimes I_{n}, I_{n_C} \otimes H_{X}^T \right).
\end{equation}
By assumption, $H_C$ is full rank.
\qx{Then, accoring to Ref.~\cite{tillich2014quantum}, $\mc{C}^Z$ has dimension $k(H_C^T)k(H_{X}) + k(H_C)k(H_X^T) = k(H_C)k(H_X^T)$, where $k(H_i)$ denotes the dimension of the kernel of $H_i$. 
A logical $Z$ operator basis is given by $\left(0, \ker{H_C}\otimes \mr{im}^{\bullet}(H_X) \right)$~\cite{quintavalle2022partitioning, xu2024fast}. One can show that such a basis cannot be weight-reduced below $d_C$, and thus the $Z$ distance of $\mc{C}^Z$ is $d_C$.  
}

After a minimum-weight correction $e^{Z \prime} = \left(
\begin{array}{c}
     e^{Z \prime}_d \\
     e^{Z \prime}_s
\end{array}\right)$, the residual fault 
$f^Z = \left( \begin{array}{c}
     f^Z_d \\
     f^Z_s \\ 
\end{array}\right) := e^Z + e^{Z \prime}$ 
will be in the kernel of $D_X$ (i.e. $D_X f^Z = 0$), which could be either a $Z$ logical operator or a $Z$ check of $\mc{C}^Z$. 
If $|f^Z| < d_C$, then $f^Z$ has to be a $Z$ check, i.e.
$f^Z = \tilde{D}_Z^T h^Z$ for some $h^Z \in \mbb{F}_2^{n_C} \otimes \mbb{F}_2^{n}$. In addition, according to Lemma~\ref{lemma:residual_locally_stochastic}, $f^Z$ is locally stochastic with an error rate $2\sqrt{p}$.

Based on the assumption, $H_C$ is $(\gamma, \delta)$-expanding with $\delta < 1/4$. 
Then, according to Lemma.~\ref{lemma:HGP_soundness}, we have 
\begin{equation}
    |f^Z| \geq \min\{|h^Z|^r_{I_{n_C}\otimes H_X}, d_C, \gamma n_C\}= \min \{|h^Z|^r_{I_{n_C}\otimes H_X}, \Theta(d)\},
    \label{eq:X_error_soundness}
\end{equation}
\qx{where we implicitly use that $d_C = \Theta(d)$ and $\gamma n_C = \Theta(d)$ for a good $[n_C = \Theta(d_C), k_C = \Theta(d_C), d_C = \Theta(d)]$ classical code $H_C$. }

Let $h^Z_m$ be a minimum-weight representative among $[h^Z]_{I_{n_C}\otimes H_X}$. 
This indicates that 
the residual syndrome error $f^Z_s$ is the boundary of some effective data error $h^Z_m$ on the blocks $\bs{B}_0\cup \bs{B}_1$, whose weight is upper bounded by the weight of the fault $f^Z$ in the circuit provided that $f^Z$ is not too large. 

Now, we prove that the distribution of $h^Z_m$ is locally stochastic. 
Let $f^Z = \left(
\begin{array}{c}
     f^{Z}_d \\
     f^{Z}_s
\end{array}\right)$ be one fault configuration that gives rise to a target $h^Z_m$. 
Clearly, any other fault $f^{Z \prime}$ that also gives rise to the same $h^Z_m$ belong to the class $\left(
\begin{array}{c}
[f^{Z}_d]_{I_{\sigma_C}\otimes H_X} \\
     f^{Z}_s
\end{array}\right)$, which we simply denote as $[f^Z]_{I_{\sigma_C}\otimes H_X}$, and it satisfies $|f^{Z \prime }| \geq \min\{|h^Z_m|, \Theta(d)\}$.
This indicates that 
\begin{equation}
|f^Z|^r_{I_{\sigma_C}\otimes H_X} \geq \min \{|h^Z_m|, \Theta(d)\}.
\label{eq:Z_equivalent_class_weight}
\end{equation}
Therefore, we have
\begin{equation}
    \mr{Pr}(f^Z_s = (I_{n_C}\otimes H_{X})h^Z_m) \leq \mr{Pr}([f^Z]_{I_{\sigma_C}\otimes H_{X}}) \leq \mc{O}(\sqrt{p})^{|f^Z_d|_{I_{\sigma_C}\otimes H_{X}}^r} \leq \mc{O}(p)^{\min\{|h^Z_m|, \Theta(d) \}/2},
\end{equation}
where the second equality utilizes Lemma.~\ref{lemma:prob_equivalent_classes}, and the third equality utilizes Eq.~\eqref{eq:Z_equivalent_class_weight}. This proves the property of the residual syndrome error in Eq.~\eqref{eq:th_residual_syndrome} in the limit $d \rightarrow \infty$.

\subsubsection{Proof for the residual $X$ data error $r^X$ \label{sec:r_X}}
The $Z$ detectors for correcting an error $e^X = \left(\begin{array}{c}
     e^X_d \\
     e^X_s
\end{array}\right)$
are given by Eq.~\eqref{eq:Z_detector_mat}.
$D_Z$ corresponds to the $X$ check matrix of a HGP code $\mc{C}^X = \mr{HGP}(H_C^1, H_{Z})$, whose associated $Z$ check matrix is
\begin{equation}
    \tilde{D}_X = \left(H_C^1 \otimes I_{n}, I_{\sigma_C} \otimes H_{Z}^{T} \right).
\end{equation}
By assumption, $H_C^1$ is full rank, and thus $\mc{C}^X$ encodes no logical qubits and has a distance $\infty$. 

After a perfect minimum-weight correction $e^{X \prime} = (e^{X \prime}_d, e^{X \prime}_s)$, the residual fault 
$f^X = \left( \begin{array}{c}
     f^X_d \\
     f^X_s \\ 
\end{array}\right) := e^X + e^{X \prime}$ 
will be in the kernel of $D_Z$ (i.e. $D_Z f^X = 0$) and satisfies
$f^X = \tilde{D}_X^T h^X$ for some $h^X \in \mbb{F}_2^{\sigma_C} \otimes \mbb{F}_2^{n}$. In addition, according to Lemma~\ref{lemma:residual_locally_stochastic}, $f^X$ is locally stochastic with an error rate $2\sqrt{p}$.

By assumption, $(H_C^{1})^T$ is $(\gamma, \delta)$-left expanding with $\delta < 1/4$. According to Lemma.~\ref{lemma:HGP_soundness}, we have
\begin{equation}
    |f^X| \geq \min\{|h^X|^r_{I_{\sigma_C}\otimes H_{Z}}, d((H_C^1)^T)) = \infty, \gamma r_C\} = \min\{|h^X|^r_{I_{\sigma_C}\otimes H_{Z}}, \Theta(d)\}.
\end{equation}

Let $h^X_m$ be the minimum-weight representative among the equivalent class $[h^X]_{I_{\sigma_C}\otimes H_{Z}}$, we then have $|f^X| \geq \min\{ |h^X_m|, \Theta(d)\}$. 
Here, we further introduce $h^{X \prime}_m := \min_{s^{\prime} \in \im{I_{\sigma_C \otimes H_X}}}h^X + s^{\prime}$, which is the minimum weight representative of the coset of $h^X$ with respect to the $X$ checks. Note that $h^X_m$ and $h^{X \prime}_m$ could differ by a logical $X$ operator. 
Nevertheless, we now prove that $h^X_m = h^{X\prime}_m$ if $|f^X| < d/(2\Delta)$, where $\Delta$ denotes the maximum row- and column-weight of $H_C$.
We prove this by contradiction.
Assume that $h^X = h^X_m + h^{\prime}$, where $h^{\prime} \in \ker{I_{\sigma_C}\otimes H_{Z}}$ contains some nontrivial logical $X$ operators $\mc{Q}$ distributed on set of blocks in $\bs{A}$, we have 
\begin{equation}
    |f^X_d| = |((H_C^1)^T\otimes I_n)(h^X_m + h^{\prime})| \geq |((H_C^1)^T\otimes I_n)h^{\prime}| - |((H_C^1)^T\otimes I_n)h^{X}_m| \geq |((H_C^1)^T\otimes I_n)h^{\prime}| - \Delta |h^X_m| > |((H_C^1)^T\otimes I_n)h^{\prime}| - d/2,
\end{equation}
where the second last inequality uses that $|((H_C^1)^T\otimes I_n)h^{X}_m| \leq \Delta |h^{X}_m|$ since the row- and column-weight of $(H_C^1)^T$ is bounded by $\Delta$ and the last equality uses $|h^X_m| \leq |f^X| < d/(2\Delta)$.
Now, since $\ker{(H_C^1)^T\otimes I_n} = 0$ as $H_C^1$ is full rank, $((H_C^1)^T\otimes I_n)h^{\prime}$ contains at least one nontrivial logical $X$ operators of $\mc{Q}$ distributed in some block of $\bs{A}$. Therefore, we end up with $|f^X_d| > |((H_C^1)^T\otimes I_n)h^{\prime}| - d/2 > d - d/2 = d/2$, which contradicts the assumption that $|f^X_d| \leq |f^X| < d/(2\Delta)$ for $\Delta \geq 1$.  
Thus, we conclude that $h^X = h^X_m + s$ for some $s \in \im{I_{\sigma_C}\otimes H_X}$, which indicates that $h^{X \prime}_m = h^X_m$. Consequently, we have $|f^X| \geq \min\{|h^{X \prime}_m|, \Theta(d)\}$

Now, we prove that $h^{X \prime}_m$ is subject to a locally stochastic distribution.
Let $f^X = \left(
\begin{array}{c}
     f^{X}_d \\
     f^{X}_s
\end{array}\right)$ be one fault configuration that gives rise to a target $h^{X \prime}_m$. 
Clearly, any other fault $f^{X \prime}$ that also gives rise to the same $h^{X \prime}_m$ belong to the class $\left(
\begin{array}{c}
[f^{X}_d]_{I_{\sigma_C}\otimes H_Z} \\
     f^{X}_s
\end{array}\right)$, which we simply denote as $[f^X]_{I_{\sigma_C}\otimes H_Z}$, and it satisfies $|f^{\prime X}| \geq \min\{|h^{X \prime}_m|, \Theta(d)\}$.
This indicates that 
\begin{equation}
|f^X|^r_{I_{\sigma_C}\otimes H_Z} \geq \min \{|h^{X \prime}_m|, \Theta(d)\}.
\label{eq:X_equivalent_class_weight}
\end{equation}
As such, we have
\begin{equation}
    \mr{Pr}(h^{X \prime}_m) \leq \mr{Pr}([f^X]_{I_{\sigma_C}\otimes H_{Z}}) \leq \mc{O}(\sqrt{p})^{|f^X|_{I_{\sigma_C}\otimes H_Z}^r} \leq \mc{O}(p)^{\min\{|h^{\prime X}_m|, \Theta(d)\}/2},
\end{equation}
where the second equality utilizes Lemma.~\ref{lemma:prob_equivalent_classes} and the third equality utilizes Eq.~\eqref{eq:X_equivalent_class_weight}. 
This indicates that $h^{X\prime}_m$ is locally stochastic in the limit $d \rightarrow \infty$.

Finally, we prove that the residual output error $r^X$ on $\bs{B}_0$ can be written as the propagation of $h^{X \prime}_m$ through a constant depth circuit, thereby also being locally stochastic.

The output residual error $r^X \in \mbb{F}_2^{k_C}\otimes \mbb{F}_2^{n}$ on the block $\mb{B}_0$ is given by 
\begin{equation}
    r^X = ((H_C^0)^T \otimes I_n) h^X = ((H_C^0)^T \otimes I_n)(h^{X \prime}_m + s^{\prime}).
    \label{eq:residual_error_1}
\end{equation} 
for some $s^{\prime} \in \im{I_{\sigma_C}\otimes H_X}$.
Here, the first equality can be seen from the following: the residual error $r^X$ is propagated from the residual fault $f^X_d$ and $f^X_s$. The syndrome fault $f^X_s$ is equivalent to two layers of data errors $h^X$ before and after the $X$ syndrome extraction on $\bs{A}$. The error $h^X$ before the syndrome extraction can be absorbed to the initial $X$-basis preparation of $\bs{A}$ and thus will not contribute to $r^X$, whereas the error $h^X$ after the syndrome extraction will propagate though the transversal CNOTs to (1) $\bs{B}_1$ in the form $((H_C^1)^T\otimes I_n)h^X$, which cancels $f^X_d$ (2) $\bs{B}_0$ in the form of $((H_C^0)^T \otimes I_n) h^X$, which gives the residual error $r^X$.

Since the BSE gadget measures a round of $X$ checks of the quantum code, i.e. ${I}_{k_C}\otimes H_{X}$, at the beginning, the $\mb{B}_0$ block will be in the eigenstates of ${I}_{\sigma_C}\otimes H_{X}$. 
As such, we are free to remove $s^{\prime}$ in Eq.~\eqref{eq:residual_error_1} and set $r^X = ((H_C^0)^T \otimes I_n)h^{X \prime}_m$. 
This indicates that the residual error $r^X$ is propagated from $h^{X\prime}$ via a constant depth circuit. 
Finally, according to Lemma 5.2 of Ref.~\cite{nguyen_qldpc_2024}, $r^X$ is also locally stochastic if $h^{X \prime}$ is locally stochastic.

\subsubsection{Existence of $H_C$}
Our proofs of the fault tolerance of BSE in Theorem~\ref{theorem:formal_BSE} relied on the existence of an LDPC classical code $H_C$ with certain expansion properties. In particular, we need $H_C$ to be an expander code, and also need a sub-matrix of its transpose, $(H_C^1)^{T}$, to be an expander code.
In this section, we discuss how to ensure these conditions are satisfied, and argue that an asymptotically good family of classical random codes can be easily constructed and made to satisfy these conditions (see Prop.~\ref{prop:existence_H_c}).

\begin{definition}[Random regular bipartite graph~\cite{sipser1996expander} (configuration model)]
    To construct a bipartite graph with a bit degree $c$ and a check degree $r$, we first choose a random matching between $N = c r m$ ``left" nodes and $N$ ``right" nodes. We then collapse consecutive sets of $c$ left nodes to form the $r m$ degree-$c$ bits and collapse consecutive sets of $r$ right nodes to form the $c m$ degree-$r$ checks.
    This gives the random bipartite graph $G(\mb{V_C}\cup \mb{V_V}, \mb{E})$. The bipartite adjacency matrix of $G$ gives the check matrix $H_C \in \mbb{F}_2^{(c m)\times (r m)}$ of the constructed classical code. We denote such a random ensemble as $\bs{\mr{LDPC}}(m, r, c)$. 
    \label{def:random_bipartite_graph}
\end{definition}

We construct the target $H_C = (H_C^0, H_C^1)$ by taking $H_C^0$ and $H_C^1$ independently from the ensemble in Def.~\ref{def:random_bipartite_graph}:
\begin{equation}
    H_C^0 \in \bs{\mr{LDPC}}(m, \Delta_c^{0}, \Delta_v); \quad H_C^1 \in \bs{\mr{LDPC}}(m, \Delta_v, \Delta_v).
    \label{eq:sub_mat_construction}
\end{equation}
Then, $H_C \in \mbb{F}_2^{m\Delta_v \times m(\Delta_c^0 + \Delta_v)}$ with a check degree $\Delta_c^0 + \Delta_v$ and a bit degree $\Delta_v$.
Denoting $r_c = m\Delta_v$, $n_c^0 = m\Delta_c^0$ and $n_c^1 = m\Delta_v$, we have $H_C^0 \in \mbb{F}_2^{r_c \times n_c^0}$ and $H_C^1 \in \mbb{F}_2^{r_c \times n_c^1}$. 

\begin{proposition}[Existence of classical code $H_c$ for Theorem~\ref{theorem:formal_BSE}]
    A random classical code $H_c$ constructed using Eq.~\eqref{eq:sub_mat_construction} satisfies the requirements in Theorem~\ref{theorem:formal_BSE} --- $H_C$ and $(H_C^1)^T$ are $(\gamma, \delta)$-left expanding for some constant $\gamma$ and $\delta < 1/4$ and $H_C^1$ is full rank --- with high probability when $\Delta_v \gg 1$. 
    \label{prop:existence_H_c}
\end{proposition}
\begin{proof}
    We first prove that $H_C^1 \in \mbb{F}_2^{r_c \times n_c^1}$, with $r_c = n_c^1$, is full rank with high probability. According to Ref.~\cite{coja2020rank}, 
    \begin{equation}
        \mr{lim}_{n_c^1 \rightarrow \infty} \mr{Rank}(H_C^1)/n_c^1 = 1 - \max_{\beta \in [0, 1]} \Phi(\beta),
    \end{equation}
    with high probability, where 
    \begin{equation}
        \Phi(\beta) := D_c\left(1-D_v^{\prime}(\beta) / \Delta_v\right)-\left(1-D_v(\beta)-(1-\beta) D_v^{\prime}(\beta)\right),
    \end{equation}
    where $D_c(x)$ and $D_v(x)$ are polynomials in $x$ with coefficients being the check- and bit-degree distribution of $H_C^1$, respectively. 
    In our case, we have $D_c(x) = D_v(x) = x^{\Delta_v}$, and $\max_{\beta \in [0, 1]}\Phi(\beta) = 0$~\cite{coja2018rank}, which concludes that $H_C^1$ has full rank with high probability.

    Next, we prove that both $(H_C^1)^T$ and $H_C$ are left expanding with high probability. Consider the bipartite Tanner graph $G(\bs{C} \cup \bs{B}, \bs{E})$ of $H_C$, whose bit nodes $\bs{B}$ are further decomposed into $\bs{B}^0 \cup \bs{B}^1$, where $\bs{B}^0$ with $|\bs{B}^0| = n_c^0$ and $\bs{B}^1$ with $|\bs{B}^1| = n_c^1$ are associated with the columns of $H_C^0$ and $H_C^1$, respectively.
    Given any $\bs{S} \subseteq \bs{B}$ (resp. $\bs{S}_c \subseteq \bs{C}$), we denote its neighbor in $\bs{C}$ (resp. $\bs{B}$) as $\Gamma(\bs{S})$ (resp. $\Gamma_c(\bs{S}_c)$).
    In addition, give any $\bs{S} \subseteq \bs{B}$, we denote its restriction on $\bs{B}^1$ as $\bs{S}|_{\bs{B}^1}$.
    We first show the following properties for $G$:
    \begin{enumerate}
    \item For any $\mb{S}_c \subset \bs{C}$ with $|\mb{S}_c| = \alpha_c |\bs{C}|$, where $0< \alpha_c < 1$, we have 
    \begin{equation}
        |\Gamma_c(\mb{S}_c)|_{\bs{B}^1}| \geq |\bs{C}|\Delta_v\left(\frac{1}{\Delta_v
    }\left(1-(1-\alpha_c)^{\Delta_v}\right)-\sqrt{2 \alpha_c H(\alpha_c) / (\Delta_v \log_2 e})\right),
    \label{eq:expansion_eq1}
    \end{equation} 
    where $H(\bullet)$ denotes the binary entropy function.
    \item For any $\bs{S} = \bs{S}^0 \cup \bs{S}^1$ with $|\bs{S}| = \alpha |\bs{B}|$, where $\bs{S}^0 \subseteq \bs{B}^0$ with $|\bs{S}^0| = \alpha_0 |\bs{B}^0|$ and $\bs{S}^1 \subseteq \bs{B}^1$ with $|\bs{S}^1| = \alpha_1 |\bs{B}^1|$, we have
     \begin{equation}
        |\Gamma(\mb{S})| \geq |\bs{B}|\Delta_v \left[\frac{|\bs{C}|}{|\bs{B}|\Delta_v}\left( 1 - (1 - \alpha_0)^{\Delta_c^0} (1 - \alpha_1)^{\Delta_v}\right) -  \sqrt{2 \alpha H(\alpha) / (\Delta_v \log_2 e})\right].
        \label{eq:expansion_eq2}
    \end{equation}
\end{enumerate}
The first property is related to the expansion of $(H_C^1)^T$ and is proved in Ref.~\cite{sipser1996expander} as $H_C^1$ is constructed independently. Here, we extend the proof in Ref.~\cite{sipser1996expander} for the second property, which is related to the expansion of the joint matrix $H_C$. 

We first consider the probability that a given check node $c$ in $\bs{C}$ is not a neighbor of $\mb{S}$. Each neighbor of $c$ in $\bs{B}^0$ (resp. $\bs{B}^1$) is in the set $\bs{S}^0$ (resp. $\bs{S}^1$) with probability $\alpha_0$ (resp. $\alpha_1$). Thus the probability that $c \notin \Gamma(\bs{S})$ is 
    \begin{equation}
        \left(\prod_{i = 0}^{\Delta_c^0 - 1} \frac{n_c^0 - \alpha_0 n_c^0 - i}{n_c^0 - i}\right) \left(\prod_{i = 0}^{\Delta_v - 1} \frac{n_c^1 - \alpha_1 n_c^1 - i}{n_c^1 - i}\right)  \approx (1 - \alpha_0)^{\Delta_c^0} (1 - \alpha_1)^{\Delta_v},
    \end{equation}
    as $n_c^0, n_c^1 \rightarrow \infty$. 
    This implies that the expected value of $|\bs{C} \backslash \Gamma(\bs{S})|$ is 
    $|\bs{C}| (1 - \alpha_0)^{\Delta_c^0} (1 - \alpha_1)^{\Delta_v}$, and consequently, the expected value of $|\Gamma(\bs{S})|_{\bs{B}^1}|$ is 
     \begin{equation}
        |\bs{C}|\left( 1 - (1 - \alpha_0)^{\Delta_c^0} (1 - \alpha_1)^{\Delta_v}\right).
    \end{equation}
    In the following, we prove that with high probability, $|\Gamma(\bs{S})|_{\mb{V_1}}|$ only deviates from this expected value by a small amount using a martingale argument~\cite{sipser1996expander}. 

    Each node in $\bs{S}$ will have $\Delta_v$ outgoing edges, so there are in total $N := \Delta_v |\bs{S}|$ outgoing edges from $\bs{S}$. We consider the random process in which the destinations of these edges are revealed one at a time. We let $X_i$ be the random variable equal to the expected size of the set of neighbors of $\bs{S}$ given that the first $i$ edges leaving $\bs{S}$ have been revealed. Namely, let $x_i \in \{0, 1\}$ be the random variable indicating if revealing the $i$-th edge adds a new neighbor in $\bs{C}$ to the known neighbors of $\bs{S}$ in $\bs{C}$, then we have $X_i = \sum_{j = 1}^{i} x_j + E[\sum_{j = i + 1}^{N} x_j]$. 
    In particular, we have $X_0 = E[\sum_{j = 1}^N x_j]$, which is the expected value of $|\Gamma(\bs{S})|$, and $X_N = \sum_{j = 1}^{N} x_j$, which is the actual value of $|\Gamma(\bs{S})|$. 
    $\{X_i\}_{i = 0}^N$ forms a martingale since 
    \begin{equation}
        E[X_i | \{x_j\}_{j = 1}^{i - 1}] = E[\sum_{j = 1}^{i} x_j + E[\sum_{j = i + 1}^{N} x_j] | \{x_j\}_{j = 1}^{i - 1}] =  \sum_{j = 1}^{i - 1} x_j + E[\sum_{j = i}^{N} x_j] = X_{i - 1}.
    \end{equation}
    Moreover, $|X_{i + 1} - X_i| \leq 1$ for all $0 \leq i < N$. Thus by Azuma’s Inequality~\cite{dubhashi2009concentration}, we have
    \begin{equation}
        \mr{Pr}(|X_N - X_0| > \lambda \sqrt{N}) < e^{-\lambda^2/2}, 
    \end{equation}
    Since there are only $\left( \begin{array}{c}
         |\bs{B}| \\
         \alpha |\bs{B}|
    \end{array}\right)$ choices for $\mb{S}$, it suffices to choose $\lambda$ so that 
    \begin{equation}
        \binom{|\bs{B}|}{\alpha|\bs{B}|} e^{-\lambda^2 / 2} \ll 1.
    \end{equation}
    Using Stirling's formula, this holds for large $n_c$ if $\lambda$ satisfies 
    \begin{equation}
        \frac{|\bs{B}| H(\alpha)}{\log_2^e}<\frac{\lambda^2}{2} \Rightarrow \lambda > \sqrt{2 |\bs{B}| H(\alpha) / \log_2^e},
    \end{equation}
    which gives the second property (Eq.~\eqref{eq:expansion_eq2}).

Now, we prove that $(H_C^1)^T$ and $H_C$ can be both $(\gamma, \delta)$-left expanding using Eq.~\eqref{eq:expansion_eq1} and Eq.~\eqref{eq:expansion_eq2}, respectively. Following Ref.~\cite{sipser1996expander}, we assume that $\alpha_c = |\bs{S}_c|/|\bs{C}|$ and $\alpha = |\bs{S}|/|\bs{B}|$ are constants. In addition, we assume $\Delta_c^0 < \Delta_v$, which corresponds to the case where the rate of $H_C$ is less than $1/2$. When $\alpha_c \Delta_v$ and $\alpha \Delta$ is much smaller than $1$, Eq.~\eqref{eq:expansion_eq1} and Eq.~\eqref{eq:expansion_eq2} are approximated by 
\begin{equation}
    \begin{aligned}
        |\Gamma_c(\bs{S}_c)|_{\bs{B}^1}|/(\Delta_v|\bs{S}_c|) & \geq 1 - \sqrt{\frac{2H(\alpha_c)}{\log_2^e \alpha_c \Delta_v}}, \\
        |\Gamma(\bs{S})|/(\Delta_v |\bs{S}|) & \geq 1 - \sqrt{\frac{2H(\alpha)}{\log_2^e \alpha \Delta_v}},\\
    \end{aligned}
    \label{eq:expansion_eq}
\end{equation}
where we have used that $(1 - \alpha_c)^{\Delta_v} \approx 1 - \Delta_v \alpha_c$, $(1 - \alpha_0)^{\Delta_c^0} \approx 1 - \Delta_c^0 \alpha_0$, and $(1 - \alpha_1)^{\Delta_v} \approx 1 - \Delta_v\alpha_1$.
This indicates that both $H_C$ and $(H_C^1)^T$ are $(\gamma, \delta)$-expanding when $\Delta_v = \Omega(1/\delta^2)$ and $\gamma = \mc{O}(1/\Delta_v)$. We note that such proof of expansion from Ref.~\cite{sipser1996expander}, which utilizes the lower bounds in Eq.~\eqref{eq:expansion_eq}, implicitly relies on that we only consider constant fractions of graph nodes as $H(\alpha)/\alpha$ diverges as $\alpha \rightarrow 0$. For smaller subsets, e.g. $\alpha_c = \mc{O}(1/r_c)$ and $\alpha = \mc{O}(1/n_c)$, the lower bound here is likely loose and one would need to use tighter element-counting lower bounds(e.g. the proof for Theorem 8.7 in ~\cite{richardson2008modern}) for the neighbors.

Since $H_C^1$ is full-rank with high probability, this family has a constant rate $\Delta_c^0/(\Delta_c^0 + \Delta_v)$. In addition, the expansion of $H_C$ implies that these codes have a distance that scale linearly with the block length~\cite{sipser1996expander}. 
As such, we can increase the distance of $H_c$ until it matches with the quantum-code distance by increasing the block length.
\end{proof}

\subsection{Connection to cluster-state schemes
\label{sec:cluster_state}}
\begin{figure}
    \centering
    \includegraphics[width=1\linewidth]{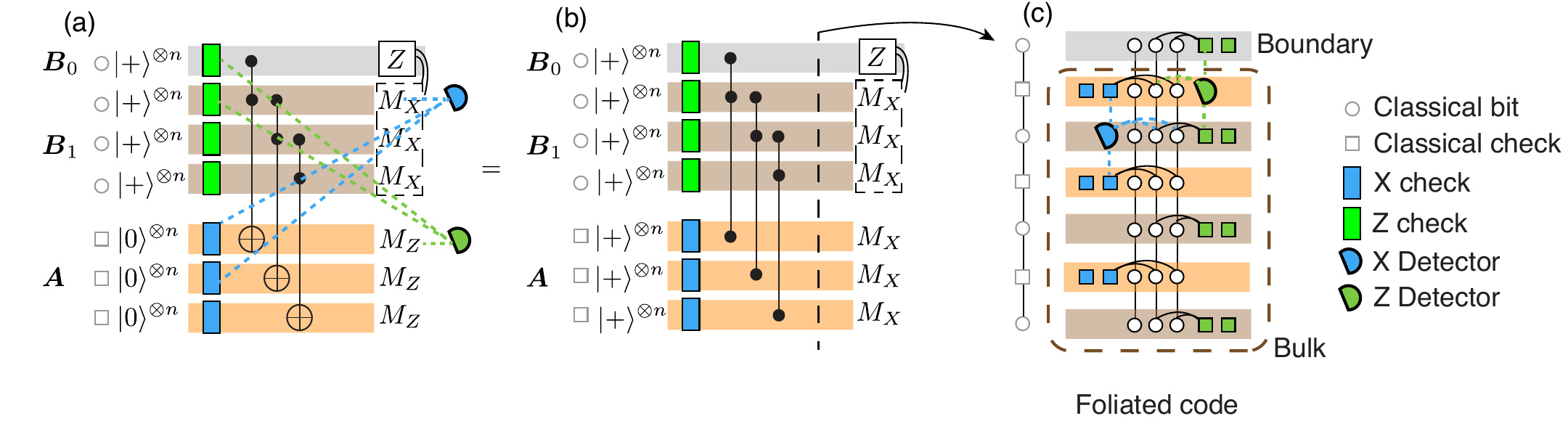}
    \caption{\textbf{Connection between the BSE gadget and the cluster-state protocol in Ref.~\cite{bergamaschi2024fault}}. A BSE gadget for preparing the logical $\overline{\ket{+}}$ states of the blocks $\bs{B}_0$ using a repetition code (a) is equivalent to the circuit in (b), where the $\bs{A}$ blocks are initialized and measured in the $X$ basis and the $X$ checks on the $\bs{A}$ are measured in the $Z$ basis, i.e. measuring $H^{\otimes n } S_X H^{\otimes n}$. The circuit in (b) exactly prepares the foliated-code cluster state (c) before the final transversal $X$ measurements. By then performing the transversal $X$ measurements on blocks $\bs{B}_1$ and $\bs{A}$ in (a) and (b), which corresponds to measuring out the bulk of the 3D cluster state in (c), the protocol outputs the blocks $\bs{B}_0$ encoded in the logical $\overline{\ket{+}}^{\otimes k}$ of the target $[[n, k, d]]$ quantum code (with a feedback $Z$ correction based on the measurement outcomes on $\bs{B}_1$). The primal and dual layer of the foliated code in (c) corresponds to the bits and the checks of the classical repetition code, and the nearest-block connections of the foliated code correspond to the connections of the Tanner graph of the repetition code. The $X$ and $Z$ detectors of the BSE gadget in (a) and (b) correspond to the primal and dual stabilizers of the foliated code, respectively. Therefore, the cluster-state protocol in Ref.~\cite{bergamaschi2024fault} corresponds to a special-case BSE gadget using a repetition code. A high-rate BSE gadget is equivalent to preparing a generalized foliated code involving a $[n_C, k_C, d_C]$ classical code $\mc{C}$ with $k_C > 1$, measuring out its ``bulk" corresponding to the $n_C - k_C$ bits and the $n_C - k_C$ checks of $\mc{C}$, and outputting its ``boundary" corresponding to the $k_C$ bits of $\mc{C}$. }
    \label{fig:BSE_cluster_state}
\end{figure}

As illustrated in Fig.~\ref{fig:BSE_cluster_state}, the BSE gadget for measuring $Z$ checks can be viewed as the generalization of the foliated cluster-state protocol in Ref.~\cite{bergamaschi2024fault} by replacing the underlying repetition code with a high-rate classical code. See the caption of Fig.~\ref{fig:BSE_cluster_state} for details. 
\qx{Note that the repetition-code construction in Ref.~\ref{fig:BSE_cluster_state} indicates that the classical code being expanding is only a sufficient condition for guaranteeing the fault tolerance of the BSE gadget, and one might be table to find weaker conditions.}

According to Ref.~\cite{hillmann2024single}, the fault tolerance of a foliated cluster state, associated with a quantum code $\mc{Q}$, can be captured by a fault complex that is formed by taking a tensor product of the chain complexes of $\mc{Q}$ and a repetition code. 
Based on the analysis in Fig.~\ref{fig:BSE_cluster_state}, we show in the following that the fault-detector structure of our BSE gadget can be captured by a (generalized) fault complex that is the tensor product of $\mc{Q}$ and a high-rate classical code:

Given a CSS quantum code $\mc{Q}: \begin{tikzcd} 
C_2 & C_1 & C_0
\arrow["{H_Z^T}", from=1-1, to=1-2]
\arrow["{H_X}", from=1-2, to=1-3]
\end{tikzcd}$ and a classical code $\mc{R}: \begin{tikzcd} 
R_1 & R_0
\arrow["{H_C^T}", from=1-1, to=1-2]
\end{tikzcd}$, the (generalized) fault complex is defined by taking the tensor product of 
$\mc{Q}$ and $\mc{R}$~\cite{hillmann2024single}:
\begin{equation}
    \begin{tikzcd}
    	F_3 & & {R^1 \otimes C_2} & \\
    	F_2 & R_1 \otimes C_1 & & R_0 \otimes C_2 \\
     F_1 & R_0\otimes C_1 & & R_1\otimes C_0\\
     F_0 & & R_0\otimes C_0 & \\
    	\arrow["{D_Z^T}", from=1-1, to=2-1]
            \arrow["{G}", from=2-1, to=3-1]
            \arrow["{D_X}", from=3-1, to=4-1]
            \arrow["{I\otimes H_Z^T}", from=1-3, to=2-2]
            \arrow["{I\otimes H_Z^T}", from=1-3, to=2-2]
            \arrow["{H_C^T \otimes I}", from=1-3, to=2-4]
            \arrow["{H_C^T \otimes I}", from=2-2, to=3-2]
            \arrow["{I \otimes H_Z^T}", from=2-4, to=3-2]
            \arrow["{I \otimes H_X}", from=2-2, to=3-4]
            \arrow["{I \otimes H_X}", from=3-2, to=4-3]
            \arrow["{H_C^T \otimes I}", from=3-4, to=4-3]
    \end{tikzcd}
    \label{eq:HGP_complex}
\end{equation}
$F_0$ and $F_3$ are associated with the primal and dual stabilizers of the foliated codes, which correspond to the $X$ detectors and the $Z$ detectors of the BSE gadget, respectively. $F_1$ is associated with the primal faults of the foliated codes, which correspond to the $Z$-type data errors on $\bs{B}_0$ and $\bs{B}_1$ ($e_d^Z \in R_0 \otimes C_1$) and the $X$ syndrome errors on $\bs{A}$ ($e_s^Z \in R_1 \otimes C_0$). 
$F_2$ is associated with the dual faults of the foliated codes, which correspond to the $X$-type data errors on $\bs{A}$ ($e_d^X \in R_1\otimes C_1$) and the $Z$ syndrome errors on $\bs{B}_0$ and $\bs{B}_1$ ($e_s^X \in R_0 \otimes C_2$). The matrix forms of the boundary maps are given by: 
\begin{equation}
D_Z^T =\left(\begin{array}{c}
I\otimes H_Z^T \\
H_C^T \otimes I
\end{array}\right), \quad 
G =\left(\begin{array}{c c}
H_C^T\otimes I & I \otimes H_X \\
0 & I \otimes H_Z^t
\end{array}\right), \quad
D_X =\left(\begin{array}{c c}
I\otimes H_X & H_C^T \otimes I \\
\end{array}\right), 
\end{equation}
where $D_X$ and $D_Z$ are identical to Eq.~\eqref{eq:X_detector_mat} and Eq.~\eqref{eq:Z_detector_mat}, respectively, explaining the product structure of the BSE gadget. 

\subsection{Single-shot code surgery using BSE}
\begin{figure}
    \centering
    \includegraphics[width=1\linewidth]{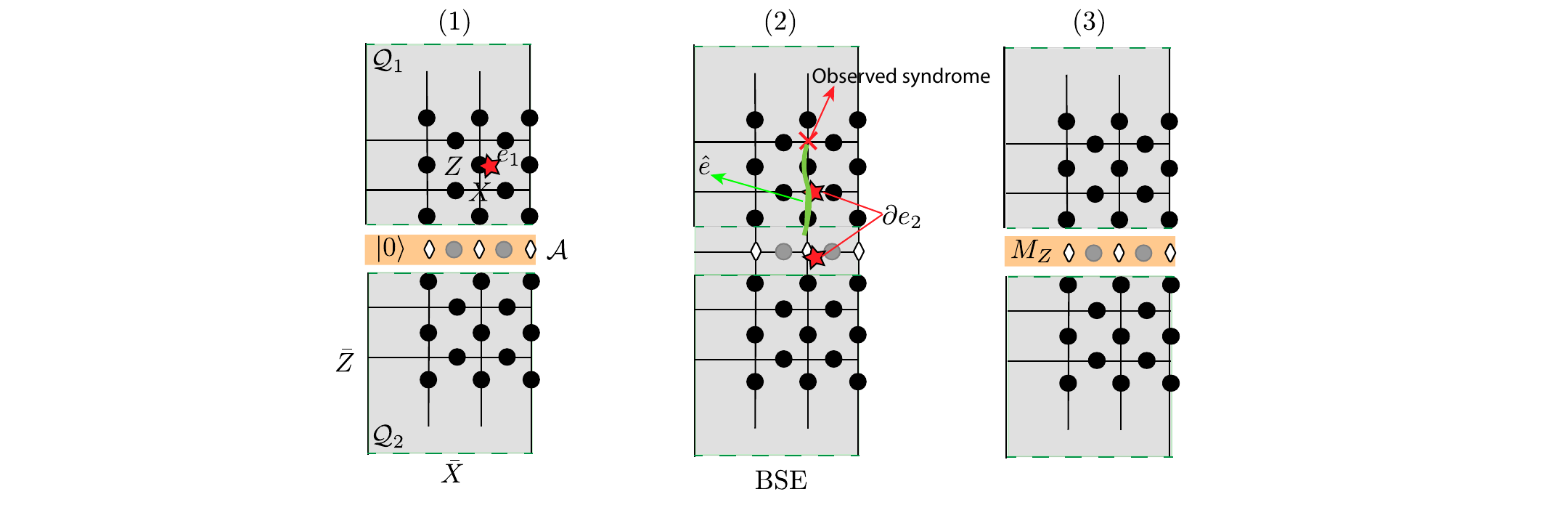}
    \caption{\textbf{Illustration of single-shot lattice surgery using BSE}. To perform $XX$ measurements on a batch of  paired surface codes, we (1) prepare the ancilla qubits (the grey qubits on the orange strip $\mc{A}$) in $\ket{0}$, (2) measure all the $X$ stabilizers on the merged codes using the BSE gadget, (3) meausre the ancilla qubits in the $Z$ basis (with $X$ feedback on the output codes). }
    \label{fig:lattice_surgery_BSE}
\end{figure}
Here, we show how the BSE gadget can be used for performing batched code surgeries in single shot. 
We first illustrate this for performing standard lattice surgery between two surface code patches~\cite{horsman2012surface} and then generalize it to the recent surgery techniques for generic qLDPC codes~\cite{cohen2022low, cross2024improved}.

As shown in Fig.~\ref{fig:lattice_surgery_BSE}, we first consider two surface code patches $\mathcal{Q}_1$ and $\mathcal{Q}_2$ with distance $d$ and perform a standard $\bar{X}_1\bar{X}_2$ logical measurement by introducing a thin strip of ancilla system $\mc{A}$ at the interface of $\mc{Q}_1$ and $\mc{Q}_2$~\cite{horsman2012surface}.
The $\bar{X}_1\bar{X}_2$ measurement outcome is obtained by taking the product of the interface $X$ checks $\mc{S}_m$ on $\mc{A}$, which are part of the stabilizers of the merged code (see step (2) in Fig.~\ref{fig:lattice_surgery_BSE}).
Conventionally, we need to repeat $\mc{O}(d)$ rounds of stabilizer measurements for the merged code to reliably infer the values of $\mc{S}_m$ so that we obtain the correct $\bar{X}_1\bar{X}_2$ measurement.
Here, we show that a single round suffices if we measure the merged code stabilizers using the BSE gadget.

Here we consider a phenomenological noise model and focus on $Z$-type errors, i.e. $Z$ data errors and syndrome errors on $X$ checks since the analysis for the other type of error is trivial. 
We assume that the two individual surface code patches have been fault-tolerantly prepared, i.e. the stabilizers have all been fixed to $+1$, with some residual noise $e_1$ (see Fig.~\ref{fig:lattice_surgery_BSE}(1)) that is local stochastic with parameter $p$. Although we assume here that all stabilizers have been fixed to $+1$, other patterns can be easily incorporated via standard Pauli frame tracking methods. 
After performing the syndrome correction using the BSE gadget, we obtain some syndrome $S = S|_{\mc{Q}_1\cup \mc{Q}_2} \cup S|_{\mc{A}}$, whose component $S|_{\mc{Q}_1\cup \mc{Q}_2}$ on the data codes should be deterministic while component $S|_{\mc{A}}$ on the ancilla system is non-deterministic, in the absence of errors.
We then perform a final correction $\hat{e}$ based on $S|_{\mc{Q}_1\cup \mc{Q}_2}$ assuming only data errors using, e.g. the matching decoder. 
We aim to show that, after such corrections, we can obtain the true measurement outcome for $\bar{X}_1\bar{X}_2$ with high probability $1-\mc{O}(p^{d/4})$.

According to Theorem~\ref{theorem:formal_BSE}, the syndromes of the merged code are measured fault-tolerantly, up to residual syndrome errors that form the boundary of some locally stochastic $Z$ data errors. 
Therefore, we can write the measured syndrome as $S=S_0\oplus \partial(e_1) \oplus \partial(e_2)$, where $S_0$ are the true syndrome values in the absence of errors, which satisfies $S_0|_{\mathcal{Q}_1,\mathcal{Q}_2}=0$,
and $e_2$ is the equivalent local stochastic data noise of the BSE syndrome measurements (see Fig.~\ref{fig:lattice_surgery_BSE}(2)).
Here, $\partial$ denotes the boundary map of the merged code associate with the $X$ checks.
Based on $S$, we infer a data-error correction $\hat{e}$ supported on $\mc{Q}_1 \cup \mc{Q}_2$ satisfying $(\partial(\hat{e}) + S)|_{\mc{Q}_1 \cup \mc{Q}_2} = 0$.
Therefore, we have that $\partial(\hat{e}\oplus e_1\oplus e_2)|_{\mathcal{Q}_1,\mathcal{Q}_2}=0$.
Since $e_1$ and $e_2$ are local stochastic with parameter $p$, $e_1\oplus e_2$ is local stochastic with parameter $\sqrt{p}$. Since the recovery is given by a minimum weight decoder, $|\hat{e}|\leq |e_1\oplus e_2|$. Therefore, with probability $1-\mc{O}(p^{d/4})$, $|\hat{e}\oplus e_1\oplus e_2|<d$.
Since $\partial(\hat{e}\oplus e_1\oplus e_2)|_{\mathcal{Q}_1,\mathcal{Q}_2}=0$, we conclude that with high probability, $(\hat{e}\oplus e_1\oplus e_2)|_{\mathcal{Q}_1,\mathcal{Q}_2}$ must act as a stabilizer on $\mathcal{Q}_1,\mathcal{Q}_2$.

The true logical measurement is obtained by taking the product of syndromes in $S_0|_{\mc{A}}$. 
The actual (corrected) logical measurement result returned from our procedure is the product of $(S\oplus \partial \hat{e})|_{\mc{A}}=S_0|_{\mc{A}} \oplus (\partial(e_1\oplus e_2 \oplus \hat{e}))|_{\mc{A}}$. 
Therefore, our deduced logical measurement result differs from the ideal result by the product of $\partial(e_1\oplus e_2\oplus \hat{e})|_{\mc{A}}$. 
Since $e_1\oplus e_2 \oplus \hat{e}$ acts as a stabilizer on $\mathcal{Q}_1,\mathcal{Q}_2$ with high probability, its action on those qubits must commute with all stabilizers of the merged code, including $\bar{X}_1 \bar{X}_2$, and thus not affect the logical measurement result. 
The only remaining contribution must come from its restriction to the qubits on $\mc{A}$. 
However, since the product of the checks on $\mc{A}$ is simply $\bar{X}_1 \bar{X}_2$ and does not support on $\mc{A}$, it will also not affect the logical measurement. Therefore, with probability $1-\mc{O}(p^{d/4})$, the logical measurement result will be correct.

We can also generalize the preceding analysis to more general surgery constructions. Consider the homological measurement framework for general code surgeries~\cite{ide2024fault,williamson2024low,cross2024improved,swaroop2024universal} and a surgery protocol that involves two qlDPC codes $\mc{Q}_1$ and $\mc{Q}_1^{\prime}$ along with an adapter system $\mc{A}$ in Def.~\ref{def:cone_code}.
The preceding discussion implies that the ideal logical measurement result and inferred logical measurement result differ by certain products of the syndromes in $\partial(e_1\oplus e_2\oplus \hat{e})|_{\mc{A}}$.

In the language of Def.~\ref{def:cone_code}, we can write the effective data error $e_1\oplus e_2\oplus \hat{e}$ as
\begin{align*}
\begin{pmatrix}
    a \\ b \\ c
\end{pmatrix},
\end{align*}
where the three components act on the data qubits in $\mathcal{Q}_1$, $\mathcal{A}$, $\mathcal{Q}_1'$ respectively.
The boudary map $\partial$ is given by the $X$ check matrix of the merged code:
\begin{equation}
    \Tilde{H}_X = \left(
    \begin{array}{ccc}
       H_x  & 0 & 0 \\
       f_1^T & \partial_1^T & f_1^{\prime T} \\
       0 & 0 & H_x^{\prime} \\
    \end{array}
    \right),
    \label{eq:cone_x}
\end{equation}
where the three rows are associated with the $X$ syndromes on $\mc{Q}_1$, $\mc{A}$, and $\mc{Q}_2$, respectively.

The logical measurements $\{\bar{X}_i \bar{X}_i^{\prime}\}_{i \in [k]}$ (for codes encoding $k$ logical qubits) are obtained by the parities of the basis vectors $\{v_i\}$ of $\ker{\partial_1}$, which satisfies $f_i v_i = \bar{X}_i$ and $f_i^{\prime} v_i = \bar{X}_i^{\prime}$ (see Def.~\ref{def:cone_code} for details).
From the preceding discussion, the effective data error does not trigger any stabilizers in the original data codes $\mathcal{Q}$ and $\mathcal{Q}'$, i.e. $H_xa=0$, $H_x'c=0$.
This implies that $a$ and $c$ act, with high probability due to the weight bound from local stochasticness, as $Z$ stabilizers on the original data codes, and therefore also commute with the $X$ logical operators of the respective codes, $\langle \bar{X}_i, a\rangle=0$.
We thus have $\langle f_1v_i, a\rangle=0$. Taking the adjoint of this inner product, we have $\langle v_i, f_1^T a\rangle=0$.
Similarly, we have $\langle v_i, f_1'^T c\rangle=0$.
Furthermore, since $v_i\in \ker{\partial_1}$, $\partial_1 v_i=0$ and $\langle \partial_1 v_i, b\rangle=0$. Taking the adjoint, we have $\langle v_i, \partial_1^T b\rangle=0$.
Plugging these into Eq.~\ref{eq:cone_x}, we have that
\begin{equation}
    \langle \left[\tilde{H}_X \begin{pmatrix}
    a \\ b \\ c
\end{pmatrix}\right]|_{\mc{A}}, v_i\rangle = \langle f_1^T a + \partial_1^T b + f_1^{\prime T} c, v_i\rangle = 0,
\end{equation}
which indicates that $\partial(e_1\oplus e_2\oplus \hat{e})|_{\mc{A}}$ does not flip the relevant product of stabilizers that product the logical measurement outcomes.
Therefore, when using the BSE procedure to perform fault-tolerant lattice surgery with the homological measurement construction, the logical error rate is exponentially suppressed in the code distance.

\subsection{Numerical simulations
\label{sec:BSE_numerics}}
\begin{figure}
    \centering
    \includegraphics[width=1\linewidth]{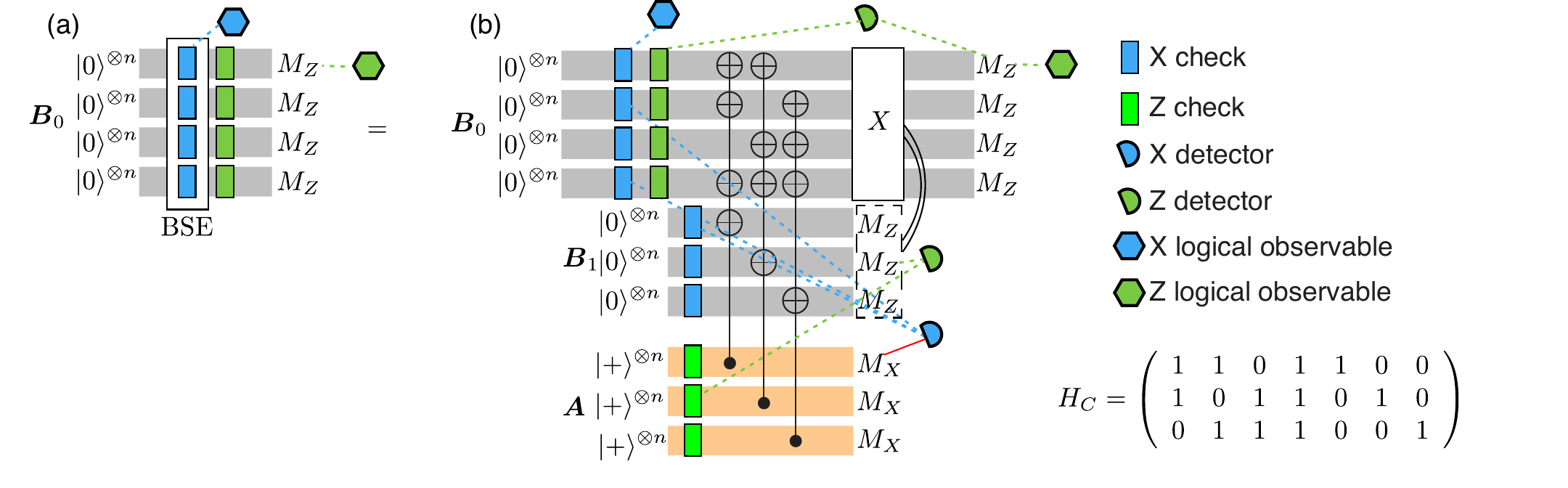}
    \caption{\textbf{Circuit for numerically simulating the BSE gadget}. }
    \label{fig:BSE_numerics}
\end{figure}

Here, we present the details of numerically simulating the logical error rates of the BSE gadgets. 
As illustrated in Fig.~\ref{fig:BSE_numerics}, given a $[[n, k, d]]$ quantum code $\mc{Q}$ and a $[n_C, k_C, d_C]$ classical code $\mc{C}$, we simulate a circuit on a batch of $k_C$ $\mc{Q}$ codes that implements a combination of the memory experiment~\cite{acharya2022suppressing} and the stability experiment~\cite{gidney2022stability}. Concretely, we initialize the physical qubits of $\mc{Q}^{\otimes k_C}$ in the product state $\ket{0}^{\otimes n}$, measure one round of $X$ checks using a BSE gadget associated with $\mc{C}$ and one round of $Z$ checks directly, and finally transversally measure them in the $Z$ basis. 

As a memory experiment, the logical qubits should all be measured to give logical $+1$, so we set $k_C\times k$ $Z$ logical observables to be the values of the $k_C \times k$ logical $Z$ operators of $\mc{Q}^{\otimes k_C}$, which are constructed by combinations of the final transversal physical $Z$ measurements. The flip of these $Z$ logical observables characterize the space-like logical errors of the protocol. 

Let $H_X \in \mbb{F}_2^{r_X\times n}$ be the $X$ check matrix of $\mc{Q}$. We consider $\mc{Q}$ with redundant $X$ checks, i.e. $\ker{H_X^T} \neq 0$. Let $\bs{M}_X = \{m_{x, i} \in \mbb{F}_2^{r_X}\}$ be a set of basis vectors spanning $\ker{H_X^T}$, these are the set of ``meta X checks" of $\mc{Q}$, each being a product of a subset of $X$ checks indexed by $m_{x, i}$, that should, in the absence of syndrome errors, have deterministic value $+1$ even if all the constituting $X$ checks are non-deterministic.  As such, we have a stability experiment by setting $k_C \times |\bs{M}_X|$ $X$ logical observables to the values of $\bs{M}_X^{\otimes k_C}$, which are constructed by combinations of the $X$ syndromes (extracted using BSE gadget). The flip of these $X$ logical observables characterize the time-like logical errors of the BSE gadget. As a concrete example, taking $\mc{Q}$ to be the toric code, we have $\bs{M}_X = \{(1, 1, \cdots, 1)\}$ since the product of all the $X$ checks is $I$. 

The circuit for measuring the $X$ checks using the BSE gadget and how the QEC is performed within the BSE gadget are described in the main text as well as the previous sections. As a concrete example, the circuit using the $[7, 4, 3]$ Hamming code is shown in Fig.~\ref{fig:BSE_numerics}. 

We note that there is one type of (undetectable) residual error that is not captured by the above numerical experiment: any residual syndrome error on the initial $X$ checks on $\bs{B}_0$ that is the boundary of some $Z$ data error $e_Z$ will commute with all the logical observables, and thus will not contribute to the reported logical error rates. After fixing the $X$ checks to $+1$ using only the information from the BSE gadget, the prepared state is close to the logical zero states of the underlying quantum code up to $e_Z$. Therefore, these uncaptured errors do affect how far we are from the codespace. Nevertheless, as entailed from the structure of the errors analyzed in Sec.~\ref{sec:proof_BSE_formal}, there is also some residual $X$ error on $\bs{X}_0$ ($r^X$ in Theorem~\ref{theorem:formal_BSE}), which has essentially the same error mechanisms (it arises from the residual syndrome error on the $Z$ checks on $\bs{A}$) as those for $e_Z$ and is captured by the numerics here (it anti-commutes with the space-like $Z$ logical observables). As such, the report logical error rates do characterize how far we are from the codespace using the single-shot BSE gadget.

\section{Batched code switching}
Here, we provide the detailed construction of the BCS gadget and prove its fault tolerance.

Given two CSS qLDPC code $\mc{Q}_1$ and $\mc{Q}_2$ 
with parameters $[[n_1, k_1, d_1]]$ and $[[n_2, k_2, d_2]]$, respectively. Let $r_{X, i}$ and $r_{Z, i}$ be the rank of the $X$ and $Z$ parity check, respectively, for $\mc{Q}_i$ (for $i = 1, 2$).
According to Def. 6 in the main text, both $\mc{Q}_1$ and $\mc{Q}_2$ admit a canonical logical operator basis:
\begin{equation}
L_Z^i = 
\begin{blockarray}{ccccc}
 & \bs{L}_i & \bs{Z}_i & \bs{X}_i \\
\begin{block}{c(ccc)c}
 & I_{k_i} & A_i & 0 & \\
\end{block}
\end{blockarray}, \quad 
L_X^i = 
\begin{blockarray}{ccccc}
 & \bs{L}_i & \bs{Z}_i & \bs{X}_i \\
\begin{block}{c(ccc)c}
 & I_{k_i} & 0 & B_i & \\
\end{block}
\end{blockarray},
\label{eq:canonical_logical_basis}
\end{equation}
for $i = 1, 2$, where $A_i \in \mbb{F}_2^{k_i \times r_{X,i}}$ and $B_i \in \mbb{F}_2^{k_i \times r_{Z,i}}$. The qubits $[n_i]$ are partitioned into three zones with sizes $|\bs{L}_i| = k_i$, $|\bs{Z}_i| = r_{X,i}$, and $|\bs{X}_i| = r_{Z,i}$, respectively. 

We perform the code switching $\mc{Q}_1^{\otimes k_2} \rightarrow \mc{Q}_2^{\otimes k_1}$ using the following protocol:
\begin{enumerate}
    \item Prepare $r_{X, 2}$ (resp. $r_{Z, 2}$) blocks of $\mc{Q}_1$ associated with the $\bs{Z}_2$ (resp. $\bs{X}_2$) zone of $\mc{Q}_2$ in logical $\overline{\ket{0}}^{\otimes k_1}$ (resp. $\overline{\ket{+}}^{\otimes k_1}$) using BSE gadgets. 
    \item Measure the stabilizers of $\mc{Q}_2^{\otimes n_1}$ using BSE gadgets.
    \item Measure transversally the $r_{X, 1}$ (resp. $r_{Z, 1}$) blocks of $\mc{Q}_2$ associated with the $\bs{Z}_1$ (resp. $\bs{X}_1$) zone of $\mc{Q}_1$ in the $Z$ (rep. $X$) basis.
    \item Apply Pauli feedback on the output $\mc{Q}_2^{\otimes k_1}$ block based on the transversal measurement outcomes in step 3 in the following way: 
    Indexing the $\mc{Q}_2$ blocks on the $\bs{L}_1$, $\bs{Z}_1$, and $\bs{X}_1$ zones of $\mc{Q}_1$ by $[k_1]$, $[r_{Z, 1}]$ and $[r_{X, 1}]$, respectively. Apply a classically controlled transversal CNOT between the $i$-th $\bs{L}_1$ block (target) and the $j$-th $\bs{Z}_1$ block (control) if $A_1(i, j) = 1$; Apply a classically controlled transversal CZ between the $i$-th $\bs{L}_1$ block (target) and the $j$-th $\bs{X}_1$ block if $B_1(i, j) = 1$. 
\end{enumerate} 

Now, we prove the fault tolerance of the above protocol.
We adopt the standard symplectic representation of the a $n$-qubit Pauli operator $P \in \mc{P}^n$ (up to phases) with a vector $(a \mid b) \in \mbb{F}_2^{2n}$, where $a_i = 1$ if $Z$ or $Y$ acts on qubit $i$ and $b_j = 1$ if $X$ or $Y$ acts on qubit $j$.  Let $\mb{v} = \{v_i \in \mbb{F}_2^{2n}\}$ be a set of vectors, we denote $\langle \mb{v}\rangle$ as the group generated by the Pauli operators represented by $\mb{v}$, or equivalently, the subspace of $\mbb{F}_2^{2n}$ that is spanned by $\mb{v}$. 

We arrange the $n_1\times n_2$ physical qubits of $\mc{Q}_1^{\otimes n_2}$ (or $\mc{Q}_2^{\otimes n_1}$) on a $[n_1]\times[n_2]$ rectangle, where $\mc{Q}_1^{\otimes n_2}$ are defined on the $n_2$ columns and $\mc{Q}_2^{\otimes n_1}$ are defined on the $n_1$ rows. 
We denote an operator in the form $(a_1 \otimes a_2 \mid 0)$, where $a_1 \in \mbb{F}_2^{n_1}$ and $a_2 \in \mbb{F}_2^{n_2}$, as a Pauli $Z$ operator supported on the rectangle $\mr{supp}(a_1)\times \mr{supp}(a_2)$. Let $\mb{a}_1$ and $\mb{a}_2$ be a set of vectors in $\mbb{F}_2^{n_1}$ and $\mbb{F}_2^{n_2}$, respectively, with a slight abuse of notation, we denote $\mb{a}_1 \otimes \mb{a}_2$ as $\{a_1\otimes a_2\}_{a_1 \in \mb{a}_1, a_2 \in \mb{a}_2}$. 
Given a set of qubits $\bs{R} \subseteq [n_1]\times [n_1]$, we denote $\overline{\bs{R}} := [n_1]\times[n_2]\backslash \bs{R}$ as complementary set of qubits of $\bs{R}$, and $l|_{\bs{R}}$ the restriction of an operator $l$ on $\bs{R}$. 
We also denote $|l|_{\mr{RW}}$ as the row weight of $l$ --- the number of rows that $l$ has nonzero support on.

Let $\bs{s}_z^i$ and $\bs{s}_x^i$ be the set of $Z$ and $X$ stabilizer generators of $\mc{Q}_i$. 
Let $e_m^{n_i}$ denote a $n_i$-dimensional unit vector with the $m$-th entry being $1$, and $\bs{e}^{n_i} := \{e_m^{n_i}\}_{m \in [n_i]}$ be the set of all unit vectors in $\mbb{F}_2^{n_i}$.
For ease of notation, we also use $0$ to denote a zero vector or matrix, whose dimension should be clear from the context.
The stabilizer group of $\mc{Q}_1^{\otimes n_2}$ and $\mc{Q}_2^{\otimes n_1}$ are given by
\begin{equation}
\begin{aligned}
    \mc{S}_1 & = \langle (\bs{s}_z^1\otimes \bs{e}^{n_2} \mid 0), (0 \mid \mb{s}_x^1 \otimes \mb{e}^{n_2})\rangle, \\
    \mc{S}_2 & = \langle (\bs{e}^{n_1}\otimes \bs{s}_z^2 \mid 0), (0 \mid \bs{e}^{n_1} \otimes \mb{s}_x^2)\rangle,
\end{aligned}
\end{equation}
respectively.

Let $\bs{l}_z^i$ and $\bs{l}_x^i$ be the canonical logical $Z$ and $X$ operators of $\mc{Q}_i$, corresponding to the rows of $L_Z^i$ and $L_X^i$ in Eq.~\eqref{eq:canonical_logical_basis}, respectively. Eq.~\eqref{eq:canonical_logical_basis} indicates that we can partition the physical qubits of $\mc{Q}_i$, $[n_i]$, into three zones, $[n_i] = \bs{L}_i\cup \bs{Z}_i\cup \bs{X}_i$, where zone $\bs{Z}_i$ only supports $\bs{l}_z^i$, zone $\bs{X}_i$ only supports $\bs{l}_x^i$, and zone $\bs{L}_i$ supports the intersection of $\bs{l}_z^i$ and $\bs{l}_x^i$. Let $\bs{e}_L^{n_i}$, $\bs{e}_Z^{n_i}$ and $\bs{e}_X^{n_i}$ denote the set of unit vectors associated with the zone $\bs{L_i}$, $\bs{Z}_i$ and $\bs{X}_i$, respectively.

After step 1 of the BCS protocol, the $\mc{Q}_1$ blocks associated with the columns indexed by $\bs{Z}_2$ ($\bs{X}_2$) are initialized in logical $\overline{\ket{0}}$ (logical $\overline{\ket{+}}$). The system is then stabilized by
\begin{equation}
    \mc{S}_o = \langle (\bs{s}_z^1\otimes \bs{e}^{n_2} \mid 0), (\bs{l}_z^1\otimes\bs{e}_Z^{n_2} \mid 0), (0 \mid \bs{s}_x^1\otimes\bs{e}^{n_2}), (0 \mid \bs{l}_x^1\otimes\bs{e}_X^{n_2})\rangle,
\end{equation}
corresponding to further adding the canonical logical $Z$ ($X$) operators of the $\mb{Z}_2$-column ($\mb{X}_2$-column) $\mc{Q}_1$ blocks to $\mc{S}_1$. Using the BSE gadget, these stabilizers $\mc{S}_o$ are deterministically established up to locally stochastical data errors.

After step 3 of the BCS protocol, the encoding is switched to $\mc{Q}_2^{\otimes n_1}$ with the $\mc{Q}_2$ blocks associated with the rows indexed by $\mb{Z}_1$ ($\mb{X}_1$) measured in the $Z$ ($X$) basis. Accordingly, the codes are now stabilized by 
\begin{equation}
    \mc{S}_n = \langle (\bs{e}^{n_1}\otimes \bs{s}_z^2 \mid 0), (\bs{e}_Z^{n_1}\otimes\bs{l}_z^2 \mid 0), (0 \mid  \bs{e}^{n_1}\otimes\bs{s}_x^2), (0 \mid \bs{e}_X^{n_1} \otimes \bs{l}_x^2)\rangle,
\end{equation}
corresponding to further adding the canonical logical $Z$ ($X$) operators of the $\mb{Z}_1$-row ($\mb{X}_2$-row) $\mc{Q}_2$ blocks to $\mc{S}_2$. Similarly, we also reliably establish these stabilizers $\mc{S}_n$ up to locally stochastic data errors due to the fault tolerance of the BSE gadget and the transversal $Z$-($X$-)basis measurements. 

It remains to prove that we have a large effective code distance \emph{during} the code deformation process $\mc{S}_o \rightarrow  \mc{S}_n$. According to Ref.~\cite{vuillot2019code}, this is equivalent to proving that the parent subsystem code of the deformation procedure, defined by the gauge group $\mc{G} = \langle \mc{S}_o, \mc{S}_n\rangle$, has a distance $\geq \min\{d_1, d_2\}$. 

The subsystem code has stabilizers
\begin{equation}
    \mc{S} = Z(\mc{G}) = \langle (\bs{s}_z^1\otimes \bs{s}_z^2 \mid 0), (0 \mid \bs{s}_x^1 \otimes \bs{s}_x^2)\rangle,
\end{equation}
which are used to detect and correct errors during the deformation process. $\mc{S}$ is LDPC since the stabilizers of both $\mc{Q}_1$ and $\mc{Q}_2$ are LDPC. The undetectable logical errors are given by the dressed logical operators of the subsystem code
\begin{equation}
    \bs{L}_d = C(\mc{S})\backslash \mc{G}.
\end{equation}
Equivalently, they are given by the bare logical operators
\begin{equation}
    \bs{L}_b = C(\mc{G})\backslash \mc{S} = \langle(\bs{l}_z^1 \otimes \bs{l}_z^2 \mid 0), (0 \mid \bs{l}_x^1 \otimes \bs{l}_x^2 \mid 0)\rangle\backslash (0 \mid 0),
\end{equation}
multiplied by the gauge operators in $\mc{G}$.

Since the code is CSS, it suffices to prove its $Z$ distance (the proof for the $X$ distance follows).
Since a bare logical $Z$ operator $l^b_z \in \bs{L}_b$ can be viewed either as a set of column vectors each being a logical operator in $\bs{l}^1_Z$ or a set of row vectors each being a logical operator in $\bs{l}^2_Z$, we have $|l^b_z| \geq \max\{d_1, d_2\}$.
Thus, it remains to prove that any bare $Z$ logical operator cannot be weight reduced below $\min\{d_1, d_2\}$ by the $Z$ gauge operators in $\mc{G}$.

Let $v_0 = (0 \mid 0) \in \mbb{F}_2^{2n}$ denote the zero vector. 
Let $l_z^b \in \langle (\bs{l}_z^1 \otimes \bs{l}_z^2 \mid 0)\rangle \backslash v_0$ be a nontrivial bare logical  $Z$ operator, $s_z \in \langle  (\bs{s}_z^1\otimes \bs{e}^{n_2} \mid 0)\rangle$ a $Z$ stabilizer of $\mc{Q}_1^{\otimes n_2}$ or $v_0$, $l_z \in \langle (\bs{l}_z^1\otimes\bs{e}_Z^{n_2} \mid 0) \rangle$ a logical $Z$ operator of $\mc{Q}_1^{\otimes r_{X, 2}}$ or $v_0$, $s_z^{\prime} \in (\bs{e}^{n_1}\otimes \bs{s}_z^2 \mid 0)$ a $Z$ stabilizer of $\mc{Q}_2^{\otimes n_1}$ or $v_0$, and $l_z^{\prime} \in $ a $Z$ logical operator $(\bs{e}_Z^{n_1}\otimes\bs{l}_z^2 \mid 0)$ of $\mc{Q}_2^{\otimes r_{X, 1}}$ or $v_0$. 
We aim to prove that 
\begin{equation}
    |l_z^b + l_z + s_z + l_z^{\prime} + s_z^{\prime}| \geq \min\{d_1, d_2\}.
\end{equation}. 

We prove this by contradiction. Assume that $|l_z^b + l_z + s_z + l_z^{\prime} + s_z^{\prime}| < d_1$ and $|l_z^b + l_z + s_z + l_z^{\prime} + s_z^{\prime}| < d_2$. 

First, we consider the case where $l_z \neq v_0$, and show how this leads to $|l_z^b + l_z + s_z + l_z^{\prime} + s_z^{\prime}| \geq d_1$, which causes a contradiction. In this case, there exists a column $q \in \bs{Z}_2$ such that $l_z$ has a component $(l_z^1 \otimes e_q)$, where $l_z^1$ is a nontrivial logical operator in $\bs{l}^1_z$.

To simplify the proof, we first define the ``reduced weight" of an operator $l$ with respect to $\langle (\bs{e}^{n_1}\otimes \bs{s}_z^2\mid 0), (\bs{e}^{n_1}\otimes \bs{l}_z^2 \mid 0)\rangle$:
\begin{equation}
    |l|^r := \min_{v_z^{\prime} \in \langle (\bs{e}^{n_1}\otimes \bs{s}_z^2\mid 0), (\bs{e}^{n_1}\otimes \bs{l}_z^2 \mid 0)\rangle} |l + v_z^{\prime}|,
\end{equation}
which is the minimum weight of $l$ modulo all possible $Z$ stabilizers and $Z$ logical operators of $\mc{Q}_2^{\otimes n_1}$. We then have 
\begin{equation}
    |l_z^b + l_z + s_z + l_z^{\prime} + s_z^{\prime}| \geq |l_z^b + l_z + s_z|^r.
\end{equation}
The reduced weight has a property $|l + l^{\prime}|^r = |l|^r$ if $|l^{\prime}|^r = 0$. Therefore, we further have
\begin{equation}
    |l_z^b + l_z + s_z + l_z^{\prime} + s_z^{\prime}| \geq |l_z + s_z|^r,
    \label{eq:inter_eq1}
\end{equation}
since $|l_z^b|^r = 0$.

Now, we prove the following fact: Let $\bs{R} := [n_1]\times \bs{Z}_2$, then for any operator $l$ that is supported only on $\bs{R}$, we have $|l|^r \geq |l|_{\mr{RW}}$. 
To see this, we utilize the canonical form of the stabilizers of $\mc{Q}_2$. 
Analogous to Eq.~\eqref{eq:canonical_logical_basis}, we can find a basis of the stabilizers of $\mc{Q}_i$~\cite{gottesman1997stabilizer}
\begin{equation}
S_Z^i = 
\begin{blockarray}{ccccc}
 & \bs{L}_i & \bs{Z}_i & \bs{X}_i \\
\begin{block}{c(ccc)c}
 & B_i^T & C_i & I & \\
\end{block}
\end{blockarray}, \quad 
S_X^i = 
\begin{blockarray}{ccccc}
 & \bs{L}_i & \bs{Z}_i & \bs{X}_i \\
\begin{block}{c(ccc)c}
 & A_i^T & I & D_i & \\
\end{block}
\end{blockarray},
\label{eq:canonical_stabilizer_basis}
\end{equation}
whose rows correspond to $\bs{s}_z^i$ and $\bs{s}_x^i$, respectively. 
According to Eq.~\eqref{eq:canonical_stabilizer_basis}, $S_X^2$ takes a form of $I$ on $\bs{Z}_2$, indicating that $\langle \bs{e}_Z^{n_2} \rangle$ is not in the kernel of $S^2_X$. 
Then, for any row vector $v = (e_i^{n_1}\otimes v_2 \mid 0)$ (for some $i$), where $v_2$ has and only has nontrivial support on $\bs{Z}_2$, $v$ is not equivalent to some vector in $\langle (\bs{e}^{n_1}\otimes \bs{s}_z^2\mid 0), (\bs{e}^{n_1}\otimes \bs{l}_z^2 \mid 0)\rangle$  (which would be in the kernel of $S^2_X$) and therefore, $|v|^r > 0$.
This indicates that, for any operator $l$ supported only on $\bs{R}$, which can be viewed as a set of row vectors each supported only on $\bs{Z}_2$ and not eliminable by $\langle (\bs{e}^{n_1}\otimes \bs{s}_z^2\mid 0), (\bs{e}^{n_1}\otimes \bs{l}_z^2 \mid 0)\rangle$, its weight cannot be reduced below its row weight. 

Note that in Eq.~\eqref{eq:inter_eq1}, $l_z$ is already supported on $\mb{R}$. We now aim to move $s_z$ also to $\bs{R}$ by adding an operator $v_z \in \langle (\bs{e}^{n_1}\otimes \bs{s}_z^2\mid 0), (\bs{e}^{n_1}\otimes \bs{l}_z^2 \mid 0)\rangle$, and show that $|l_z + s_z|^r = |l_z + (s_z + v_z)|^r \geq |l_z + (s_z + v_z)|_{\mr{RW}} \geq d_1$. To do this, we first clean out the support of $s_z$ on $[n_1]\times \bs{X}_2$ by applying some stabilizer $s_{z, 0} \in \langle (\bs{e}^{n_1}\otimes \bs{s}_z^2\mid 0)\rangle$ (this utilizes the canonical form of $S^2_Z$ in Eq.~\eqref{eq:canonical_stabilizer_basis}, which acts as $I$ on $\bs{X}_2$). Now, $s_z + s_{z, 0}$ is supported only on $[n_1] \times (\bs{L}_2 \cup \bs{Z}_2)$. 
We can further clean out its support on $[n_1]\times \bs{L}_2$ by adding another operator $l_{z, 0} \in (\bs{e}^{n_1}\otimes \bs{l}_z^2 \mid 0)\rangle$ (this utilizes the canonical form of $L^2_Z$ in Eq.~\eqref{eq:canonical_logical_basis}, which has an $I$ on $\bs{L}_2$). Denote $s_z + s_{z,0} + l_{z,0}$ as $s_{z, R}$, which only supports on $\bs{R}$. 
We can achieve this in a way that $s_{z, R}$ is still a set of column vectors each being a stabilizer of $\mc{Q}_1$ since each stabilizer generator $s_z^1 \otimes e^{n_2}_j$ of $\mc{Q}_1^{\otimes n_2}$ on a column $j$ can be cleaned to an operator in the form $s_z^1 \otimes v_R$ for some $v_R$ supported only on $\bs{Z}_2$. 
Now, we have
\begin{equation}
    |l_z + s_z|^r = |l_z + s_{z, R}|^r \geq |l_z + s_{z, R}|_{\mr{RW}} \geq d_1,
\end{equation}
where the last equality utilizes that $l_z + s_{z, R}$ contains at least a column of nontrivial operator of $\mc{Q}_1$.
This contradicts the assumption and thus $l_z$ must be $v_0$. 

By symmetry, we must have $l_z^{\prime} = v_0$, too. 
Therefore, it remains to analyze the case when $l_z = l_z^{\prime} = v_0$. 
In this case, we aim to show that 
\begin{equation}
    |l^b_z + s_z + s_z^{\prime}| \geq d_1,
    \label{eq:inter_eq2}
\end{equation}
yet contradicting the assumption.
To this end, we define another reduced weight of an operator $l$ with respect to $\langle (\bs{e}^{n_1}\otimes \bs{s}_z^2\mid 0)$
\begin{equation}
    |l|^{r^{\prime}} := \min_{v_z^{\prime} \in \langle (\bs{e}^{n_1}\otimes \bs{s}_z^2\mid 0)\rangle} |l + v_z^{\prime}|,
\end{equation}
which is the minimum weight of $l$ modulo the $Z$ stabilizers of $\mc{Q}_2^{\otimes n_1}$. 
We now aim to prove that $|l^b_z + s_z|^{r^{\prime}} \geq d_1$.
Similar to the previous arguments, we have the following fact: for any operator $l$ supported only on $\bs{R}^{\prime} := [n_1]\times (\bs{L}_2 \cup \bs{Z}_2)$, we have $|l|^{r^{\prime}} \geq |l|_{\mr{RW}}$. Again, we can clean $s_z$ to $\bs{R}^{\prime}$ by adding an operator $s_{z, 0} \in \langle (\bs{e}^{n_1}\otimes \bs{s}_z^2\mid 0)\rangle$ while  having that $s_{z, R^{\prime}} := s_z + s_{z,0}$ is still a $Z$ stabilizer of $\mc{Q}_1^{\otimes n_2}$.
Then,
\begin{equation}
    |l^b_z + s_z + s_z^{\prime}| \geq |l^b_z + s_z|^{r^{\prime}} = |l^b_z + s_{z, R^{\prime}}|^{r^{\prime}} \geq d_1,
\end{equation}
which contradicts the assumption.

Based on the above analysis, we conclude that $|l_z^b + l_z + s_z + l_z^{\prime} + s_z^{\prime}| \geq \min\{d_1, d_2\}$, and thus the distance of the subsystem code is at least $\min\{d_1, d_2\}$.

Finally, we bound the logical error rate of the BCS gadget using the standard percolation analysis in Ref.~\cite{gottesman2013fault}. Since the entire protocol remains LDPC and has a large effective distance $\Omega(\min\{{d_1, d_2}\})$, we can bound the logical error rate by 
\begin{equation}
    \mc{O}\left(n_1 n_2 (p/p_{\mr{th}})^{\Omega(\min\{{d_1, d_2}\})}\right),
\end{equation}
according to Eq.(3) of Ref.~\cite{gottesman2013fault}. 

\subsection{Reducing Overhead Constant Factors with Hypergraph Product Codes}
In this section, we describe how the qubit overhead for performing batched code switching can be further reduced. While the general procedure described above achieves constant overhead, for$k_1k_2$ logical qubits, it still requires $n\sim n_1n_2/r_c$ qubits, where $n_1$ and $n_2$ are the block lengths of the quantum code and $r_c$ is the encoding rate of the classical code used in BSE. In practice, the classical code can achieve fairly high rate using modest batch size, but the quantum codes often achieve a rate of only a few percent, leading to significant overhead. We now show how utilizing additional structure of hypergraph (HGP) codes, the same procedure can be performed using $n\sim(n_1+n_2)/r_c$ qubits, representing a sizable reduction for large $n_1$ and $n_2$. Thus, instead of a multiplicative, the overhead is now additive, alleviating the qubit requirements.
	
We first consider the qubit requirements for the standard BCS gadget described above, when switching from $k_2$ copies of a $[[n_1,k_1,d_1]]$ quantum code $\mathcal{Q}_1$ to $k_1$ copies of a $[[n_2,k_2,d_2]]$ quantum code $\mathcal{Q}_2$. As shown in the main text, the protocol involves growing the code into a joint $n_1n_2$-qubit intermediate code.

In the protocol, we use a BCS gadget to perform a round of reliable syndrome measurement. The largest instantaneous qubit usage occurs when performing syndrome measurements on the joint intermediate code. In particular, we will need to perform syndrome measurements on $n_1$ copies of $\mathcal{Q}_2$ in parallel. To do so, we can use a $[n_1/r_c, n_1, d]$ classical code and a BSE gadget, resulting in a total of $(2/r_c-1)n_1n_2$ data qubits for $k_1k_2$ logical qubits. While this is a constant, the prefactor is somewhat large, as the number of qubits is multiplied between the two codes.

We now describe how we can utilize the structure of certain quantum codes, such as HGP codes, to reduce this overhead. The key idea is to utilize homomorphic measurement gadgets to perform row-wise and column-wise switching separately, leading to an overhead that scales as the sum instead of product of the individual codes, reducing the costs.

More specifically, let $\mathcal{Q}_1$ be the HGP of classical codes $[n_{a1}, k_{a1}, d]\times [n_{b1}, k_{b1}, d]$, and let $\mathcal{Q}_2$ be the HGP of classical codes $[n_{a2}, k_{a2}, d]\times [n_{b2}, k_{b2}, d]$. We assume that the dual code of the classical codes contain no logical qubits. We perform code switching in two steps: We start with $k_2=k_{a2}k_{b2}$ copies of $[n_{a1}, k_{a1}, d]\times [n_{b1}, k_{b1}, d]$, switch into $k_{a1}k_{b2}$ copies of $[n_{a2}, k_{a2}, d]\times [n_{b1}, k_{b1}, d]$, and finally switch into $k_1=k_{a1}k_{b1}$ copies of $[n_{a2}, k_{a2}, d]\times [n_{b2}, k_{b2}, d]$. This corresponds to first performing switching along the rows, and then the columns.

We now explain how to perform the switching along the rows, and estimate its overhead. The column switching can be done similarly. Let $A$ be the code with parameters $[n_{a1}k_{a2}+n_{a2}k_{a1}, 2k_{a1}k_{a2}, d]$ formed by appending $k_{a2}$ copies of $[n_{a1}, k_{a1}, d]$ and $k_{a1}$ copies of $[n_{a2}, k_{a2}, d]$. We perform a homomorphic logical measurement on the code $A\times [n_{b1}, k_{b1}, d]$, with added ancilla qubits so that in the horizontal direction, a logical $ZZ$ measurement is performed between the $i$th qubit in the $j$th copy of $[n_{a1}, k_{a1}, d]$ and the $j$th qubit in the $i$th copy of $[n_{a2}, k_{a2}, d]$. Following the transversal $X$-basis measurement of the $[n_{a1}, k_{a1}, d]$ blocks, this allows the fault-tolerant teleportation of information to the intermediate code, as long as the homomorphic logical measurement is performed fault-tolerantly. We can perform the homomorphic logical measurements using BSE gadgets, and the fact that we have $k_{b2}$ copies in the vertical direction naturally provides a suitable batch size: we can just use a $[n_{b2},k_{b2},d]$ classical code, although since there are fewer requirements on this code, more efficient choices will likely exist.

The total number of data qubits required in the above steps can be estimated as follows: we require $(n_{a1}k_{a2}+n_{a2}k_{a1}+k_{a1}k_{a2})n_{b1}$ qubits for each code. Denoting the rate of the classical code used in the BSE gadget as $r_c'$, we will require $(2/r_c'-1)k_{b2}$ copies in the vertical direction. If we assume $n_{a1}=n_{b1}=\sqrt{n_1}$ and $n_{a2}=n_{b2}=\sqrt{n_2}$ for simplicity, then the number of data qubits required can be simplified to $n\sim (2/r_c'-1)\sqrt{n_1k_2}\qty(\sqrt{n_1k_2}+\sqrt{n_2k_1}+\sqrt{k_1k_2})$, which is approximately $(2/r_c'-1)\cdot 2nk$ when the code parameters are equal. Thus, compared to the general BCS construction, we can reduce the overhead by a factor of $n/k$, corresponding to the encoding rate of an individual quantum code.

\section{Batched addressable Cliffords}

Here, we prove Theorem 3 of the main text by explicitly constructing the batched addressable Clifford (BAC) gadgets for any $[[n_1, k_1, d_1]]$ CSS qLDPC code $\mc{Q}_1$. 
Concretely, we present a scheme that implements any $k_1$-qubit Clifford circuit $C_k$ on a batch of $k_2 = \Omega(d_1^2)$ $\mc{Q}_1$ blocks using at most $\mc{O}(k_2)$ ancillary $\mc{Q}_1$ blocks (therefore, with a constant space overhead) and in a depth of at most $\mc{O}\left(D_{P}(C_k)\right)$, where $D_{P}(C_k)$ denotes the depth of implementing $C_k$ with physical circuits. 

As discussed in the main text, we can decompose $C_k$ into layers of circuits each consisting of only $H$, $S$, or CNOT gates with a constant overhead. Thus, we present the implementation of $C_k$ of these three gate types separately. Let $\mc{Q}_2$ be a symmetric HGP code (i.e. the two base classical codes of $\mc{Q}_2$ are identical, good classical LDPC codes) with parameters $[[n_2 = \Theta(k_2), k_2, d_2 = \Theta(\sqrt{k_2}) \geq d_1]]$. 
The general strategy is to code switch from $\mc{Q}_1^{\otimes k_2}$ to $\mc{Q}_2^{\otimes k_1}$ using a BCS gadget, perform transversal Clifford gates on $\mc{Q}_2^{\otimes k_1}$, and code switch back to $\mc{Q}_1^{\otimes k_2}$ using another BCS gadget.

First, we present the implementation of a $k_1$-qubit CNOT circuit $C_{\mr{CNOT}}$ on $k_2$ blocks of $\mc{Q}_1$, i.e. $\overline{C_{\mr{CNOT}}}^{\otimes k_2}$. $C_{\mr{CNOT}}$ can be decomposed into $D_P$ layers of physical CNOTs, $C_{\mr{CNOT}} = \prod_{t = 1}^{D_P}(\bigotimes_m \mr{CNOT(p^{(t)}_m, q^{(t)}_m)})$, where $\{(p^{(t)}_m, q^{(t)}_m)\}$ index pairs of qubits that each physical CNOT gate is acting on, which are distinct for different $m$ at the same time step $t$.
Let $\bar{Q}_{i, j}$ be the $i$-th logical qubit of the $j$-th $\mc{Q}_1$ block. 
We first switch from $\mc{Q}_1^{\otimes k_2}$ to $\mc{Q}_2^{\otimes k_1}$ using the BCS gadget. This routes $\bar{Q}_{i, j}$ to the $j$-th logical qubit of the $i$-th $\mc{Q}_2$ block. We then implement $D_P$ layers of transversal CNOTs on the $\mc{Q}_2$ blocks, where the $t$-th layer consists of the following: for each $m$, apply a transversal CNOT between the $p^{(t)}_m$-th $\mc{Q}_2$ block and the $q^{(t)}_m$-th $\mc{Q}_2$ block. 
These transversal CNOT gates then implement the desired logical circuit, $\overline{C_{\mr{CNOT}}}^{\otimes k_2}$, via $\prod_{t =1}^{D_P} \bigotimes_{j \in [k_2]}\bigotimes_m \overline{\mr{CNOT}}(\bar{Q}_{p_m^{(t)}, j}, \bar{Q}_{q_m^{(t)}, j})$. Finally, we switch back to $\mc{Q}_1^{\otimes k_2}$ again using the BCS gadget. Clearly, the entire protocol involves only $\mc{O}(n_2) = \mc{O}(k_2)$ blocks of $\mc{Q}_1$.
In addition, since each of the BCS gadget has a depth $\mc{O}(1)$ and the transversal CNOTs (interleaved with syndrome extractions on $\mc{Q}_2$) have a depth $\mc{O}(D_P)$, the protocol has a depth $\mc{O}(D_P)$.  

Second, we implement a $k_1$-qubit $H$ circuit $C_H$ on $k_2$ blocks of $\mc{Q}_1$, i.e. $\overline{C_H}^{\otimes k_2}$. Any $C_H = \bigotimes_{m \in \mb{\Lambda} \subseteq [k_1]} H(m)$, which applies $H$ to a subset $\mb{\Lambda}$ of the $k_1$ qubits, can be implemented with a physical depth $1$. Therefore, we aim to implement $\overline{C_H}^{\otimes k_2} = \bigotimes_{j \in [k_2]}\bigotimes_{m \in \mb{\Lambda}}\bar{H}(\bar{Q}_{m,j})$ with a constant depth. We start, again, by switching to $\mc{Q}_2^{\otimes k_1}$ using the BCS gadget. Then, we apply the fold-transversal $H-\mr{SWAP}$ gates~\cite{quintavalle2022partitioning} on a subset of the $\mc{Q}_2$ blocks indexed by $\mb{\Lambda}$. This implements the following gates:
\begin{equation}
    \bigotimes_{m \in \mb{\Lambda}} \left(\bigotimes_{j \in [k_2]}\bar{H}(\bar{Q}_{m,j}) \bigotimes_{\alpha} \overline{\mr{SWAP}}(\bar{Q}_{m, p_{\alpha}}, \bar{Q}_{m, q_{\alpha}})\right),
    \label{eq:H_SWAP_circuit}
\end{equation}
where $\{p_{\alpha}, q_{\alpha}\}$ index pairs of logical qubits of $\mc{Q}_2$ that are mirrored along the diagonal, i.e. $\{(i, j), (j,i)\}_{i \neq j}$. 
These implements the desired gates $\overline{C_H}^{\otimes k_2}$ up to some extra SWAP operations within each $\mc{Q}_2$, which are the same across all the $|\mb{\Lambda}|$ $\mc{Q}_2$ blocks. Let $\mr{SWAP}^{\prime}$ denote the $k_2$-qubit SWAP circuit on each $\mc{Q}_2$ block in Eq.~\eqref{eq:H_SWAP_circuit}, the extra swap operations are simply $\overline{\mr{SWAP^{\prime}}}^{\otimes |\mb{\Lambda}|}$ being applied to the $\mb{\Lambda}$ subset of the $\mc{Q}_2$ blocks, where the encoded gates are with respect to the $\mc{Q}_2$ encoding.
We can then cancel these extra swap operations by introducing $k_1 - |\mb{\Lambda}|$ dummy $\mc{Q}_2$ blocks, applying the batched SWAP gates $\overline{\mr{SWAP}}^{\otimes k_1}$ in a similar way as to implement the batched CNOT circuits described above -- switching from $\mc{Q}_2^{\otimes k_1}$ to $\mc{Q}_1^{\otimes k_2}$, swapping the $\mc{Q}_1$ blocks according to $\mr{SWAP}^{\prime}$, and switching back to $\mc{Q}_2^{\otimes k_1}$. After cancelling these extra swap operations, we have implemented the desired $\overline{C_H}^{\otimes k_2}$. Finally, we switch back to $\mc{Q}_1^{\otimes k_2}$ using another BCS gadget. Clearly, the entire protocol involves only $\mc{O}(n_2) = \mc{O}(k_2)$ $\mc{Q}_1$ blocks and has a constant depth (since the intermediate SWAP circuit has a constant depth). 

\begin{figure}
    \centering
    \includegraphics[width=1\linewidth]{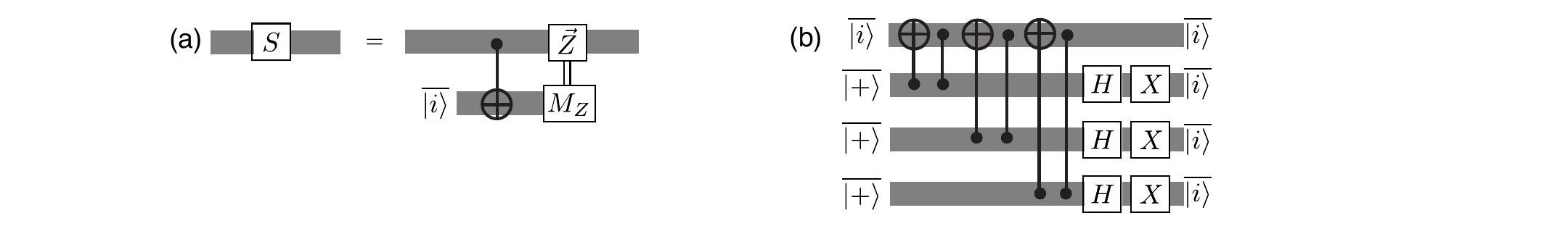}
    \caption{\textbf{Circuits for teleporting global $S$ gates using a block of $\ket{Y}$ states (a) and catalyzing more blocks of $\ket{Y}$ states using one block of $\ket{Y}$ states~\cite{litinski2022active} (b)}. }
    \label{fig:Y_states}
\end{figure}

Finally, we implement a $k_1$-qubit $S$ circuit $C_S$ on $k_2$ blocks of $\mc{Q}_1$, i.e. $\overline{C_S}^{\otimes k_2}$. Same as a $H$ circuit, any $C_S = \bigotimes_{m \in \mb{\Lambda} \subseteq [k_1]} S(m)$, which applies $S$ to a subset $\mb{\Lambda}$ of the $k_1$ qubits, can be implemented with a physical depth $1$. Therefore, we aim to implement $\overline{C_S}^{\otimes k_2} = \bigotimes_{j \in [k_2]}\bigotimes_{m \in \mb{\Lambda}}\bar{S}(\bar{Q}_{m,j})$ with a constant depth. Again, we can implement $\overline{C_S}^{\otimes k_2}$ by switching to $\mc{Q}_2^{\otimes k_1}$, applying global $\bar{S}$ gates on a subset of the $\mc{Q}_2$ blocks indexed by $\mb{\Lambda}$, and then switching back to $\mc{Q}_1^{\otimes k_2}$. The main challenge is then to implement $\overline{S}^{\otimes k_2}$ for some $\mc{Q}_2$ blocks with a low space-time overhead. We implement such global $S$ gates using the gate teleportation circuit in Fig.~\ref{fig:Y_states} that consumes a $\mc{Q}_2$ block of $\overline{\ket{i}}^{\otimes k_2}$ states. To have sufficient copies of $\overline{\ket{i}}$ states, we first prepare one block of $\overline{\ket{i}}$ states with some space overhead $W(k_2)$ and time overhead $T(k_2)$, and then catalyze more blocks of $\overline{\ket{i}}$ states using the circuit in Fig.~\ref{fig:Y_states}(b). 
Note that the global $H$ gates on a $\mc{Q}_2$ block in Fig.~\ref{fig:Y_states}(b) can be implemented efficiently using the batched $H$ gadget described above and the global $CZ$ gates between two $\mc{Q}_2$ blocks can be implemented by conjugating a transversal CNOT with two batched global $H$ gates. As the catalyst circuit and the $S$-teleportation circuit can all be implemented with a constant space overhead and in a constant depth, the average space-time overhead over $D$ uses of the batched $S$ gadget is constant if the initial preparation of a block of $\overline{\ket{i}}$ states has a space overhead $W(k_2) = \mc{O}(k_2)$ and $T(k_2) = \mr{O}(D)$. Using, e.g. the protocol in Ref.~\cite{xu2024fast}, we can achieve $W(k_2) = \mc{O}(1)$ and $T(k_2) = \tilde{\mc{O}}(\sqrt{k_2})$. Therefore, we achieve a constant space-time overhead on average for any logical circuits that apply $D = \tilde{\Omega}(\sqrt{k_2})$ batched $S$ gadgets.

\section{Parallel non-Clifford gates via distilled $T$ measurements}
\begin{figure}
    \centering
    \includegraphics[width=1\linewidth]{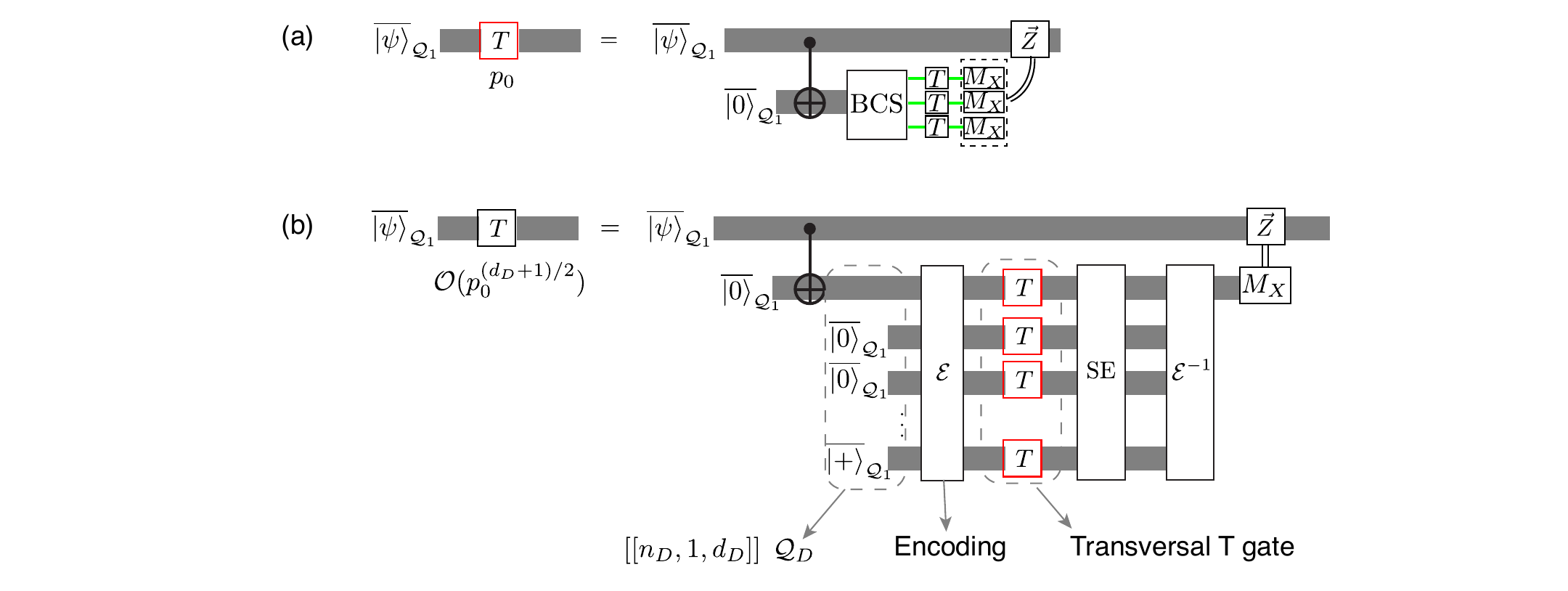}
    \caption{\textbf{Distillation of $T$ measurement}. (a) Implementation of a global noisy $T$ gate on a $[[n_1, k_1, d_1]]$ CSS code $\mc{Q}_1$ with a bounded marginal error rate $p_0$ using faulty $T$ measurements on an ancilla $\mc{Q}_1$ code. The faulty $T$ measurements can be implemented by code switching to $k_1$ $[[n_3, 1, d_3]]$ 3D color codes (indicated by the green lines) using the BCS gadget. Following the $T$ measurements, only random Pauli fixups are needed on the data $\mc{Q}_1$ code. (b) Implementation of a lower-error-rate global $T$ gate on the $\mc{Q}_1$ code by distilling a higher-fidelity $T$ measurement on the ancilla code utilizing a $[[n_D, 1, d_D]]$ distillation code $\mc{Q}_D$ from many copies of the noisy $T$ measurements in (a). }
    \label{fig:MSD}
\end{figure}
Here, we present a scheme that distills higher-fidelity $T$ measurements from lower-fidelity ones. 

As shown in Fig.~\ref{fig:MSD}(a), given a $[[n_1, k_1, d_1]]$ CSS qLDPC code $\mc{Q}_1$, we can perform noisy global $T$ gates on $\mc{Q}_1$ by coupling it to an ancilla $\mc{Q}_1$ code, performing a faulty $T$ measurement~\cite{litinski2019magic} on the ancilla code, and applying feedback $Z$ corrections on the data code. 
As presented in the main text, the faulty $T$ measurements on the ancilla code can be implemented by switching to $k_1$ copies of $[[n_3, 1, d_3]]$ 3D color codes using the BCS gadget and then performing the $T$ measurements on the color codes. By choosing $d_3 = \mathcal{O}(\log n_1)$, these $T$ measurements have a marginal error rate $p_0$ that is independent of $n_1$. Consequently, the noisy global $T$ gates on $\mc{Q}_1$ have a bounded marginal error rate $p_0$. Crucially, by implementing the $T$ gates using $T$ measurements instead of $|T\rangle$ states, we avoid the Clifford fixups on the data $\mc{Q}_1$ code (instead, we only have Pauli fixups), which could induce a large compilation overhead when implementing them with the BAC gadget. 

In Fig.~\ref{fig:MSD}(b), we show how to obtain lower-error-rate $T$ measurements on $\mc{Q}_1$ by consuming many copies of the noisy $T$ measurements in Fig.~\ref{fig:MSD}(a). The idea is to replace the faulty $T$ measurement on the ancilla code in Fig.~\ref{fig:MSD}(a) with an error-corrected $T$ measurement using a ``distillation"  code $\mc{Q}_D$. Without loss of generality, we assume that $\mc{Q}_D$ only encodes $1$ logical qubit and has parameters $[[n_D, 1, d_D]]$. In addition, it supports a transversal $T$ gate. Then, we perform the distillation following the protocol in Ref.~\cite{bravyi2012magic}: (1) prepare $n_D - 1$ extra $\mc{Q}_1$ codes in either logical $0$ or $+$, (2) perform an encoding circuit $\mathcal{E}$ that encodes the ancilla $\mc{Q}_1$ code into the concatenated code with the outer code being $\mc{Q}_D$ (e.g. by using the measurement-based encoding protocol in Def. 6 of the main text), (3) apply the transversal $T$ gate on $\mc{Q}_D$, where each $T$ gate on a $\mc{Q}_1$ block is implemented using the noisy $T$ measurement scheme in Fig.~\ref{fig:MSD}(a), (4) perform a round of syndrome extraction for the $\mc{Q}_D$ code to correct the errors associated with the preceding noisy $T$ gates on $\mc{Q}_1$, (5) apply a decoding circuit $\mc{E}^{-1}$ (using e.g. the measurement-based decoding protocol in Def. 6 of the main text) that decodes back into the ancilla $\mc{Q}_1$ code, (6) perform a transversal $X$ measurement on the ancilla $\mc{Q}_1$ code. In the regime where $\mc{Q}_1$ is large, the only errors of the above protocol come from the noisy $T$ measurements on the $\mc{Q}_1$ codes, which have a bounded error rate $p_0$. As such, the final $T$ gates in Fig.~\ref{fig:MSD}(b) have a marginal error rate $\mc{O}\left(p_0^{(d_D + 1)/2}\right)$. For instance, using a $[[15, 1, 3]]$ quantum Reed-Muller code, we can suppress the error to $\mc{O}(p_0^2)$. By induction, we can implement global $T$ gates on $\mc{Q}_1$ with lower and lower error rates by concatenating the above distillation scheme and implementing a multi-level distillation. Crucially, we only need random Pauli fixups on $\mc{Q}_1$ codes as opposed to Clifford fixups throughout the protocol, and thus do not introduce extra compilation overhead. 

Finally, using high-rate distillation codes in Ref.~\cite{nguyen_qldpc_2024}, we can distill $T$ gates $\mc{Q}_1$ codes in batches with a space-time overhead $k_1^{o(1)}$, where $o(1) \sim 1/\log \log k_1$. 

\section{Magic state cultivation with BB codes \label{app:magic_state_cultivation}}
Here, we detail the magic state cultivation/teleportation protocols for the self-dual BB codes in Table I of the main text.

In Sec.~\ref{sec:merged_CSBB_code}, we present the details of the merged color-surface-BB code, where the cultivated magic states are eventually encoded in. In particular, we prove that it has a distance lower bounded by the distance of the BB code. In Sec.~\ref{sec:code_merging}, we detail the protocol that merges small color codes with a BB code into the final color-surface-BB code and analyze its logical error rates. 
In Sec.~\ref{sec:homo_CNOT}, we show how to couple the merged color-surface-BB code with a standard BB code using a homomorphic CNOT, which is a variant of the standard transversal CNOT.
In Sec.~\ref{sec:logical_translation}, we show how to implement the logical translation gadget for a BB code, which is utilized in the main text for simulating lattice Hamiltonians.
Finally, in Sec.~\ref{sec:Clifford_fixups}, we describe how to perform the Clifford fixups in the $T$-teleportation circuit in Fig. 6(c) of the main text. 

\subsection{Merged color-surface-BB code
\label{sec:merged_CSBB_code}}
Here, we adopt the standard chain complex representations of classical codes and quantum codes: 
Given a classical linear code $\mc{C}$ with a check matrix $H_C \in \mbb{F}_2^{r_C \times n_C}$, we associate it with a two-term complex $\mc{C}: \begin{tikzcd}
            {C_1} & {C_{0}}
    	\arrow["{H_C}", from=1-1, to=1-2]
\end{tikzcd}$,
where the basis elements of $C_1 = \mbb{F}_2^{n_C}$ and $C_0 = \mbb{F}_2^{r_C}$ are associated with the checks and the bits of $\mc{C}$, respectively;
Given a CSS quantum code $\mc{Q}$ with check matrices $H_X \in \mbb{F}_2^{r_X\times n}$ and $H_z \in \mbb{F}_2^{r_Z \times n}$, we associate it with a three-term chain complex:
$\mc{Q}: \begin{tikzcd}
            {Q_2} & {Q_1} & {Q_{0}}
    	\arrow["{H_X^T}", from=1-1, to=1-2]
            \arrow["{H_Z}", from=1-2, to=1-3]
\end{tikzcd}$,
where the basis elements of $Q_2 \in \mbb{F}_2^{r_X}$, $Q_1 \in \mbb{F}_2^n$, and $Q_0 \in \mbb{F}_2^{r_Z}$ are associated with the $X$ stabilizer generators, the qubits, and the $Z$ stabilizer generators of $\mc{Q}$, respectively. 
We refer the readers to Ref.~\cite{breuckmann2021quantum} for more background on the homological-algebra description of codes. 
In addition, we will also adopt the chain complex representations of generalized lattice surgery protocols introduced in Ref.~\cite{ide2024fault}. See also Ref.~\cite{he2025extractors} for a review of the recent developments of the generalized lattice surgery protocols and Ref.~\cite{swaroop2024universal} for the detailed construction of adapters. 

\begin{figure}[h!]
    \centering
    \includegraphics[width=1\linewidth]{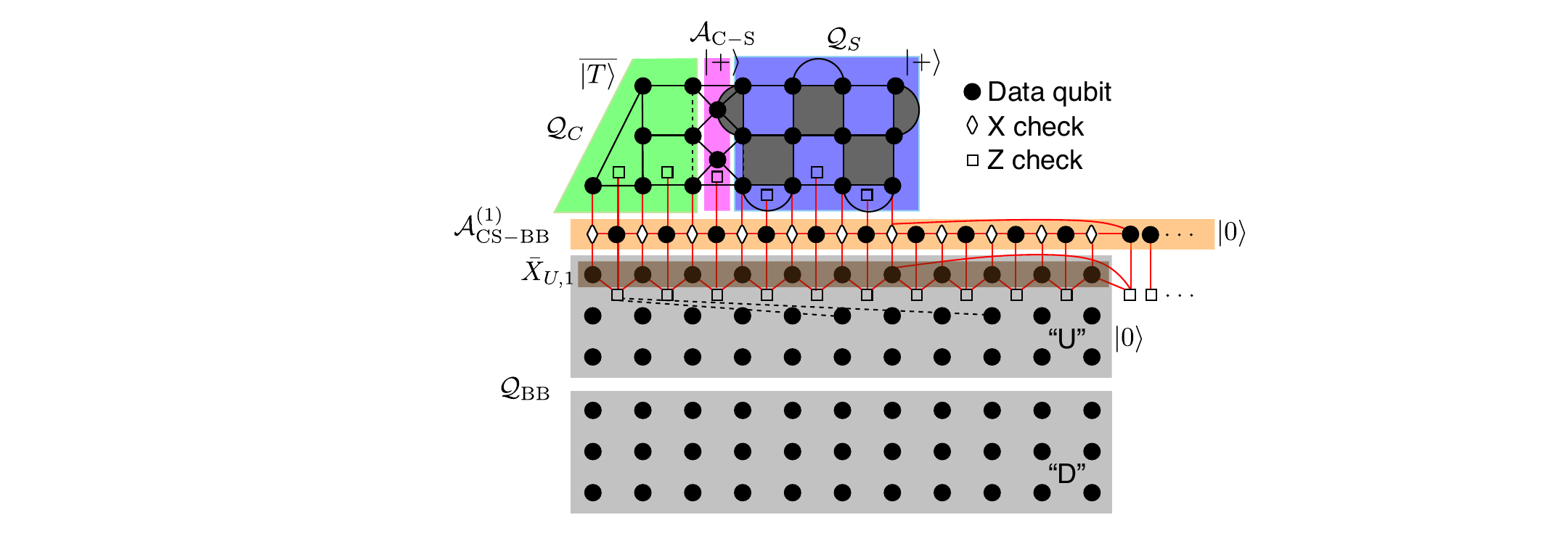}
    \caption{\textbf{Illustration of the merged color-surface-BB code}}
    \label{fig:BB_merging}
\end{figure}

The construction of a merged color-surface-BB code is built upon the following gadget that merges two CSS quantum codes using an adapter system~\cite{swaroop2024universal}:

\begin{definition}[Merging two codes through an adapter]
Given two quantum codes $\mc{Q}$ and $\mc{Q}^{\prime}$ with parameters $[[n, k, d]]$ and $[[n^{\prime}, k^{\prime} = k, d^{\prime}]]$, respectively, we define their merged code through an adapter system 
$\mc{A}: \begin{tikzcd}
            {A_1} & {A_0} & {A_{-1}}
    	\arrow["{\partial_1}", from=1-1, to=1-2]
            \arrow["{\partial_0}", from=1-2, to=1-3]
\end{tikzcd}$, where $A_1$, $A_0$, and $A_{-1}$ are associated with the $X$ checks, the qubits, and the $Z$ checks of the adapter, respectively, and $\ker{\partial_1^T} = \im{\partial_0^T}$, 
as the cone code~\cite{ide2024fault} $\begin{tikzcd} 
\Tilde{C}_2 & \Tilde{C_1} & \Tilde{C_0}
\arrow["{\Tilde{H}_X^T}", from=1-1, to=1-2]
\arrow["{\Tilde{H}_Z}", from=1-2, to=1-3]
\end{tikzcd}$ defined from the following complex: 
    \begin{equation}
    \begin{tikzcd}
    	{Q_2} & {Q_1} & {Q_0} \\
            {A_1} & {A_0} & {A_{-1}} \\
            {Q^{\prime}_2} & {Q^{\prime}_1} & {Q^{\prime}_0} \\
    	\arrow["{H_X^T}", from=1-1, to=1-2]
            \arrow["{H_Z}", from=1-2, to=1-3]
            \arrow["{H_X^{\prime T}}", from=3-1, to=3-2]
            \arrow["{H^{\prime}_Z}", from=3-2, to=3-3]
            \arrow["{\partial_1}", from=2-1, to=2-2]
            \arrow["{\partial_0}", from=2-2, to=2-3]
            \arrow["{f_1}", from=2-1, to=1-2]
            \arrow["{f_0}", from=2-2, to=1-3]
            \arrow["{f^{\prime}_1}", from=2-1, to=3-2]
            \arrow["{f^{\prime}_0}", from=2-2, to=3-3]
    \end{tikzcd},
    \label{eq:cone_complex}
    \end{equation}
where we define $\tilde{C}_i := Q_i\oplus A_{i - 1}\oplus Q^{\prime}_{i}$ for $ i = 0, 1, 2$ and
\begin{equation}
    \Tilde{H}_X = \left(
    \begin{array}{ccc}
       H_x  & 0 & 0 \\
       f_1^T & \partial_1^T & f_1^{\prime T} \\
       0 & 0 & H_x^{\prime} \\
    \end{array}
    \right),
    \quad
    \Tilde{H}_Z = \left(
    \begin{array}{ccc}
       H_z  & f_0 & 0 \\
       0 & \partial_0 & 0 \\
       0 & f_0^{\prime} & H_z^{\prime} \\
    \end{array}
    \right).
    \label{eq:check_mats_cone_code}
\end{equation}
Let $\{\bar{X}_i \in Q_1\}_{i = 1}^k$ and $\{\bar{X}_i^{\prime} \in Q_1^{\prime}\}_{i = 1}^k$ be a basis for the first homological group of $\mc{Q}$ and $\mc{Q}^{\prime}$, respectively. Both $f_1$ and $f_1^{\prime}$ have a maximum row- and column-weight of $1$ and there exists a basis $\{v_i\}_{i = 1}^k$ for $\ker{\partial_1}$ that satisfies 
\begin{equation}
    f_1 v_i = \bar{X}_i, \quad f_1^{\prime} v_i = \bar{X}_i^{\prime},
    \label{eq:partial_1_condition}
\end{equation}
for $i \in [k]$.
\label{def:cone_code}
We denote the merged code as $\tilde{\mc{Q}}(\mc{Q}, \mc{Q}^{\prime}, \mc{A})$.
\end{definition}

The above protocol can be viewed as merging the two codes $\mc{Q}$ and $\mc{Q}^{\prime}$ through $k$ parallel lattice-surgery logical measurements, $\{\bar{X}_i \bar{X}^{\prime}_{i}\}_{i \in [k]}$, which are obtained by measuring the new added $X$ checks in $A_1$.  

Now, we describe how to construct the merged color-surface-BB code using the gadget in Def.~\ref{def:cone_code}.

Let $H_{\mr{rep}}(n) \in \mbb{F}_2^{(n-1)\times n}$ denotes the standard repetition-code check matrix with $n$ bits and $n-1$ checks, and $H_{\mr{cyc}}(P, n) \in \mbb{F}_2^{n\times n}$, where $P \in \mbb{F}_2(z)/(z^n - 1)$ a cyclic matrix with a generator $P$ represented by a polynomial in $z$. Note that $H_{\mr{rep}}(n)$ corresponds to $H_{\mr{cyc}}(1 + z, n)$ without the periodical boundary condition. 

Given a $[[n_c, 1, d_c]]$ color code $\mc{Q}_{C}$ and a $[[n_s, 1, d^{\prime}_{s, X} = d_c, d^{\prime}_{s, Z}]]$ rotated surface code $\mc{Q}_S$, we construct a repetition-code adapter $\mc{A}_{\mr{C-S}}$ with $A_{-1} = 0$ and $\partial_1 = H_{\mr{rep}}(d_c)$. The $d_c$ $X$ checks in $A_1$ are connected transversally to the weight-$d_c$ boundary logical $X$ operators of $\mc{Q}_C$ and $\mc{Q}_S$ ($f_1$ and $f_1^{\prime}$ take the form of the identity matrix when restricted to the support of the logical operators), respectively. 
Let $(\bar{X}_C, \bar{Z}_C)$ and $(\bar{X}_S, \bar{Z}_S)$ be the logical operators of $\mc{Q}_C$ and $\mc{Q}_S$, respectively.
The merged code $\tilde{\mc{Q}}(\mc{Q}_C, \mc{Q}_S, \mc{A}_{\mr{C-S}})$ is merged through a $\bar{X}_C \bar{X}_S$ lattice-surgery measurement. In our case, we take the merged color-surface code $\mc{Q}_{\mr{CS}}$ as the dual of $\tilde{\mc{Q}}(\mc{Q}_C, \mc{Q}_S, \mc{A}_{\mr{C-S}})$, i.e. by swapping the $X$ and $Z$ checks in $\tilde{\mc{Q}}(\mc{Q}_C, \mc{Q}_S, \mc{A}_{\mr{C-S}})$, which is equivalent to merging $\mc{Q}_C$ and $\mc{Q}_S$ through a $\bar{Z}_C \bar{Z}_S$ measurement. As proved in Lemma~\ref{lemma:distance_color_surface}, $\mc{Q}_{\mr{CS}}$ has a $Z$ distance matching that of the color code $d_c$ and a boosted $X$ distance $d_c + d_{s, X}$, where $d_{s, X} = d^{\prime}_{s, Z}$ is the $Z$ distance of $\mc{Q}_S$. 
We illustrate the adapter $\mc{A}_{\mr{C-S}}$ for $d_c = 3$ with the purple strip in Fig.~\ref{fig:BB_merging}. Note that this adapter construction is identical to that in Ref.~\cite{poulsen2017fault}. 

Next, we merge $k_b$ copies of the color-surface code $\mc{Q}_{\mr{CS}}$ with a $[[n_b, k_b, d_b]]$ BB code $\mc{Q}_{\mr{BB}}$.
To obtain a merged code with distance $\geq d_b$ (which will be proved later), 
we choose the $X$ distance of the thin surface code such that $d_c + d_{s, X} \geq d_b$. We then merge $\mc{Q}_{\mr{CS}}^{\otimes k_b}$ with $\mc{Q}_{\mr{BB}}$ through an adapter $\mc{A}_{\mr{CS-BB}}$ described in the following: $\mc{A}_{\mr{CS-BB}}$ consists of $k_b$ disjoint and identical copies $\{\mc{A}_{\mr{CS-BB}}^{(i)}\}_{i = 1}^{k_b}$, where each $\mc{A}_{\mr{CS-BB}}^{(i)}$ couples to the $X$ logical operator of the $i$-th logical qubit in $\mc{Q}_{\mr{BB}}$ and the $X$ logical operator of a copy of $\mc{Q}_{\mr{CS}}$. In other words, 
\begin{equation}
    \mc{Q}_{\mr{CSBB}} = \tilde{Q}(\mc{Q}_{\mr{BB}}, \mc{Q}_{\mr{CS}}^{\otimes k_b}, \bigotimes_{i=1}^{k_b} \mc{A}^{(i)}_{\mr{CS-BB}}).
\end{equation}
We illustrate $\mc{A}_{\mr{CS-BB}}^{(1)}$, which couples to the first logical operator $\bar{X}_{U, 1}$ in the $U$ block of $\mc{Q}_{\mr{BB}}$, with the orange strip in Fig.~\ref{fig:BB_merging}. $\mc{A}_{\mr{CS-BB}}^{(1)}$ is simply taken as the gauging ancilla system for measuring $\bar{X}_{U, 1}$, which is constructed in Ref.~\cite{williamson2024low}. Here, we briefly describe the construction and refer technical readers to Ref.~\cite{williamson2024low} for more details. 

We construct $\mc{A}_{\mr{CS-BB}}^{(1)}: A_1 \xrightarrow{\partial_1} A_0 \xrightarrow{\partial_0} A_{-1}$ using a graph $G(V, E)$, where the vertices $V$ and the edges $E$ are associated with basis elements of $A_1$ ($X$ checks) and $A_0$ (qubits), respectively, and a cycle basis of $G$ are associated with the basis elements of $A_{-1}$ ($Z$ checks). 
We introduce a vertex for each qubit in the support of $\bar{X}_{U, 1}$, thus $|V| = |\bar{X}_{U, 1}| = l$ for a $\mc{Q}_{\mr{BB}} = \mr{BB}(c, d; R_{l, m})$ code. This gives a bijection $\tau: \mr{supp}(\bar{X}_{U, 1}) \rightarrow V$. Then, for each $Z$ check $s_Z$ that has a nontrivial support $s_Z|_{\mr{supp}(\bar{X}_{U, 1})}$ on $\mr{supp}(\bar{X}_{U, 1})$, add $|s_Z|_{\mr{supp}(\bar{X}_{U, 1})}|/2$ edges that give a perfect matching of $\tau(s_Z|_{\mr{supp}(\bar{X}_{U, 1})})$ to $E$. 
In our case, for the special BB codes in Table I of the main text with the same polynomail $c = 1 + y^3 + x(y^2 + y^4)$, there are $2l$ $Z$ checks with non-trivial support on $\mr{supp}(\bar{X}_{U, 1})$, all of weight $2$. 
Thus, $E$ with $|E| = 2l$ simply consists of all edges associated with $\{\tau(s_Z|_{\mr{supp}(\bar{X}_{U, 1})})\}$. Moreover, $\partial_1$, which is given by the edge-vertex adjacency matrix of $G$, takes the following form:
\qx{
\begin{equation}
    \partial_1 = \left( \begin{array}{c}
         H_{\mr{cyc}}(1 + z^3, l)\\
         H_{\mr{cyc}}(z^2 + z^4, l)\\ 
    \end{array}\right),
    \label{eq:adapter_X_checks_1}
\end{equation}
with two sets of checks associated with the rows of $H_{\mr{cyc}}(1 + z^3, l)$ and $H_{\mr{cyc}}(z^2 + z^4, l)$, respectively. 
Equivalently, we can write Eq.~\eqref{eq:adapter_X_checks_1} as 
\begin{equation}
     \partial_1 = \left( \begin{array}{c}
         H_{\mr{cyc}}(1 + z, l)\\
         H_{\mr{cyc}}(1 + z^{\beta}, l)\\ 
    \end{array}\right),
    \label{eq:adapter_X_checks}
\end{equation}
where $\beta$ depends on $l$, by permuting the bits and the checks. For instance, we have $\beta = 3$ for the $[[66, 6, 8]]$-code with $l = 11$. In fact, we can equivalently write $c = 1 + y + x(y^5 + y^8)$ for this code, which is what we use in Fig.~\ref{fig:BB_merging}. 
}

Due to the structure of Eq.~\eqref{eq:adapter_X_checks}, we can find a low-weight cycle basis for $G$ that give low-weight $Z$ checks ($A_{-1}$) of $\mc{A}_{\mr{CS-BB}}^{(1)}$. The cycle basis of $G$ is given by a basis of $\ker{\partial_1^T} \subset \mbb{F}_2^{2l}$, which has dimension $l + 1$. Taking $\beta = 3$ as an example, a natural basis of $\ker{\partial_1^T}$ is
\begin{equation}
    \ker{\partial_1^T} = \mr{span}\{\{S_i[(1,1, 1, 0, \cdots, 0), (1, 0, \cdots, 0)]^T\}_{i = 0}^{l - 1} \cup \{[(1,1,\cdots, 1), (0, 0, \cdots, 0)]^T\}\},
    \label{eq:cycle_basis}
\end{equation}
where $[u, v]$ denotes a vector in $\mbb{F}_2^{2l}$ with two blocks $u \in \mbb{F}_2^{l}$ and $v \in \mbb{F}_2^l$, and $S_i$ denotes a cyclic permutation matrix by $i$ in $\mbb{F}_2^{2l \times 2l}$. The first $l$ basis vectors in Eq.~\eqref{eq:cycle_basis} have weight $4$ whereas the last basis vector has a large weight $l$. Nevertheless, we can reduce the weight of the last basis vector by adding the first $l$ low-weight vectors. For instance, for $l = 11$, we can reduce the last vector to weight $5$. 

Now, we specify how the $X$ checks in $\mc{A}_{\mr{CS-BB}}^{(1)}$ are connected to $\mc{Q}_{\mr{BB}}$ and $\mc{Q}_{\mr{CS}}$ according to $f_1$ and $f_1^{\prime}$ in Eq.~\eqref{eq:cone_complex}, respectively. The way that the ancilla graph $G(V, E)$ is constructed simply gives $f_1 = \tau^T$, i.e. $f_1$ takes a transversal form of $\left(\begin{array}{c}
     I_{l} \\
     0 \\ 
\end{array}\right)$, where the first $l$ rows are associated with $\mr{supp}(\bar{X}_{U, 1})$. Note that when constructing $G$, we add a unique edge for each $s_Z$ of $\mc{Q}_{\mr{BB}}$ that has a nontrivial support on $\mr{supp}(\bar{X}_{U, 1})$. As such, we set $f_0$ similarly in a transversal form that connects each qubit (associated with an edge) in $A_0$ with its corresponding $Z$ check in $Q_0$. So far, the connection between $\mc{A}_{\mr{CS-BB}}^{(1)}$ and $\mc{Q}_{\mr{BB}}$ simply follows the standard construction in Ref.~\cite{williamson2024low} since $\mc{A}_{\mr{CS-BB}}^{(1)}$ is nothing but the standard ancilla system constructed for measuring $\bar{X}_{U, 1}$ of $\mc{Q}_{\mr{BB}}$. 
The connection between $\mc{A}_{\mr{CS-BB}}^{(1)}$ and $\mc{Q}_{\mr{CS}}$, on the other hand, leverages the structure that $\mc{A}_{\mr{CS-BB}}^{(1)}$ has some components that simply correspond to a repetition code, which exactly matches the structure of $\bar{X}_{\mr{CS}}$ and its incident $Z$ checks in $\mc{Q}_{\mr{CS}}$. As shown in Fig.~\ref{fig:BB_merging}, by choosing $\bar{X}_{\mr{CS}}$ as the bottom boundary of $\mc{Q}_{\mr{CS}}$, we can construct a graph $G^{\prime}(V^{\prime}, E^{\prime})$, where $V^{\prime}$ with $|V^{\prime}| = d_c + d_{s, X}$ are associated with $\mr{supp}(\bar{X}_{\mr{CS}})$ and $E^{\prime}$ with $|E^{\prime}| = d_c + d_{s, X} - 1$ are associated with the $Z$ checks of $\mc{Q}_{\mr{CS}}$ that each has weight-$2$ support on $V^{\prime}$. Since $G^{\prime}$ has takes the form of a repetition code, similar to a subset of the graph $G$ with only checks in the first block $H_{\mr{cyc}}(1 + z, l)$ of Eq.~\eqref{eq:adapter_X_checks}, we can choose $f_1^{\prime}$ and $f_0^{\prime}$ both in a transversal form that connects pairs of vertices in $V$ and $V^{\prime}$ and pairs of edges in $E$ and $E^{\prime}$, respectively. We illustrate the above connection of $\mc{A}_{\mr{CS-BB}}^{(1)}$ with $\mc{Q}_{\mr{BB}}$ and $\mc{Q}_{\mr{CS}}$ in Fig.~\ref{fig:BB_merging}. 

In the following, we prove that the merged code $\mc{Q}_{\mr{CSBB}}$ constructed above has a distance $\geq d_b$ if $d_{c} + d_{s, X} \geq d_b$.

\begin{definition}[Relative Cheeger constant]
    Given a check matrix $H \in \mbb{F}_2^{m \times n}$, a subset of column indices $\mb{S} \subseteq [n]$, and an integer $t$, the relative Cheeger constant $\beta_t(H, \mb{S})$ is defined as 
    \begin{equation}
        \beta_t(H, \mb{S}) := \min_{v \in \mbb{F}_2^n, |v| \leq t} \frac{|H v|}{|v|_{\mb{S}}|}.
    \end{equation}
    We define $\beta(H, \mb{S})$ as $\beta_{t = \infty}(H, \mb{S})$.
\end{definition}

\begin{lemma}[Distance guarantee of the merged code in Def.~\ref{def:cone_code}]
    Let $d_X$ and $d_Z$ (resp. $d_X^{\prime}$ and $d_Z^{\prime}$) be the $X$ and $Z$ distance of $\mc{Q}$ (resp. $\mc{Q}^{\prime}$). 
    The $Z$ distance of the merged code $\tilde{\mc{Q}}(\mc{Q}, \mc{Q}^{\prime}, \mc{A})$ in Def.~\ref{def:cone_code} is naturally guaranteed: $\Tilde{d}_Z \geq d_Z + d_Z^{\prime}$. 
    Let $H_g := \left( 
    \begin{array}{cc}
       \partial_1  & 0  \\
       f_1^{\prime} & H_x^{\prime T} \\
    \end{array}
    \right)$ and $\mb{C}_A$ denotes its first block of columns, the $X$ distance of the merged code satisfies $\Tilde{d}_X \geq d_X$ if 
    \begin{equation}
        \beta(H_g, \mb{C}_A) \geq 1.
        \label{eq:cheeger_constant}
    \end{equation}
    \label{lemma:distance_merged_code}
\end{lemma}
\begin{proof}
    We first prove the $Z$ distance. Let $v = \left( \begin{array}{c}
        v_B \\
        v_A \\
        v_C \\
    \end{array}\right)$ be a $Z$ operator, where $v_B \in Q_1$, $v_A \in A_0$, and $v_C \in Q_1^{\prime}$, that commutes with the merged $X$ checks, i.e. $\Tilde{H}_X v = 0$. We have $H_X v_B = H_X^{\prime}v_C = 0$. Thus, $v_B$ (resp. $v_C$) has to be either a $Z$ check or a $Z$ logical operator of $\mc{Q}$ (resp. $\mc{Q}^{\prime}$). We now prove that if $v$ is a nontrivial logical operator of the merged code, $v_B$ and $v_C$ have to take the form of $\bar{Z} \bar{Z}^{\prime}$ for some logical $Z$ operator $\bar{Z}$ of $\mc{Q}$ and some logical $Z$ operator $\bar{Z}^{\prime}$ of $\mc{Q}^{\prime}$. 
    We prove this by contradiction. First, assume that both $v_B$ and $v_C$ are stabilizers, i.e. $v_B = H_Z^T w_B$ and $v_C = H_Z^{\prime T} w_C$ for some $w_B \in Q_0$ and $w_C \in Q_0^{\prime}$. 
    From the condition that $\Tilde{H}_X v = 0$, we have 
    \begin{equation}
        f_1^T v_B + f_1^{\prime T} v_C + \partial_1^T v_A = 0.
        \label{eq:middle_eq}
    \end{equation}
    From the commutativity of the diagram in Eq.~\eqref{eq:cone_complex}, we have $f_1^T H_Z^T = \partial_1^T f_0^T$ and $f_1^{\prime T} H_Z^{\prime T} = \partial_1^T f_0^{\prime T}$. Then, Eq.~\eqref{eq:middle_eq} becomes
    \begin{equation}
        \partial_1^T(f_0^T w_B + f_0^{\prime T} w_C + v_A) = 0.
    \end{equation}
    Thus, $f_0^T w_B + f_0^{\prime T} w_C + v_A \in \ker{\partial_1^T}$. Since $\ker{\partial_1^T} = \im{\partial_0^T}$, we have $f_0^T w_B + f_0^{\prime T} w_C + v_A \in \ker{\partial_1^T} = \partial_0^T w_A$ for some $w_A \in A_{-1}$. This leads to 
    $v = \Tilde{H}_Z^T 
    \left( \begin{array}{c}
        w_B \\
        w_A \\
        w_C \\
    \end{array}\right)$, which indicates that $v$ is a $Z$ stabilizer of the merged code, thus contradicting the assumption. 
    Second, assume that either $v_B$ or $v_C$ is a $Z$ check. Without loss of generality, assume that $v_B$ is a $Z$ check of $\mc{Q}$ while $v_C$ is a $Z$ logical operator of $\mc{Q}^{\prime}$. Let $w_A \in \mr{ker}(\partial_1)$ and $s_X = \Tilde{H}_X^T  \left( \begin{array}{c}
        0 \\
        w_A \\
        0 \\
    \end{array}\right)$ be a $X$ check of the merged code. From the condition in Eq.~\eqref{eq:partial_1_condition}, we know that $s_X = \left( \begin{array}{c}
        \bar{X} \\
        0 \\
        \bar{X}^{\prime} \\
    \end{array}\right)$ for some $X$ logical operator $\bar{X}$ of $\mc{Q}$ and some $X$ logical operator $\bar{X}^{\prime}$ of $\mc{Q}^{\prime}$. Since $v_B = H_Z^T w_B$ for some $w_B \in Q_0$, we have $s_X^T v = \bar{X}^T v_B + \bar{X}^{\prime T} v_C = \bar{X}^T H_Z^T w_B + \bar{X}^{\prime T} v_C$ = $\bar{X}^{\prime T} v_C$. Since $v_C$ is a $Z$ logical operator of $\mc{Q}^{\prime}$, there exists some choice of $s_X$ and $\bar{X}^{\prime}$ such that $s_X^T v = \bar{X}^{\prime T} v_C \neq 0$, which violates the assumption that $v$ is a logical $Z$ operator of the merged code. 

    Based on the above analysis, both $v_B$ and $v_C$ have to be a $Z$ logical operator of $Q$ and $Q^{\prime}$, respectively, which guarantees that $|v| \geq |v_B| + |v_C| \geq d_Z + d_Z^{\prime}$.

    Finally, we prove the $X$ distance of the merged code utilizing the condition that $H_g$ has a large (relative) Cheeger constant (Eq.~\eqref{eq:cheeger_constant}). It is easy to check that the logical $X$ operators of the merged code can be represented purely by those of the code $\mc{Q}$, i.e. $\{\bar{X}_i\}_{i = 1}^k$. We now prove that any operator in $\mr{span}\{\bar{X}_i\}_{i = 1}^k$ cannot be weight reduced below $d_X$ by applying the $X$ checks of the merged code. Let $L = \left(\begin{array}{c}
        L_B \\
        0 \\
        0 \\
    \end{array}\right)$ be such a merged-code $X$ logical operator, where $L_B \in \mr{span}\{\bar{X}_i\}_{i = 1}^k$. Let $s_X = \Tilde{H}_X^T \left(\begin{array}{c}
        w_B \\
        w_A \\
        w_C \\
    \end{array}\right)$ be an arbitrary $X$ check of the merged code. We have $L + s_X = \left(\begin{array}{c}
        L_B + H_X^T w_B + f_1 w_A \\
        H_g w_{AC} \\
    \end{array}\right)$, where $w_{AC} := \left(\begin{array}{c}
        w_A \\
        w_C \\
    \end{array}\right)$. Since $f_1$ has a maximum row- and column-weight of $1$, $|f_1 w_A| \leq |w_A|$. Using the relative expansion of $H_g$ with respect to $\mb{C}_A$, we have $|H_g w_{AC}| \geq |w_{A}|$. Then, $|L + s_X| \geq |L_B + H_X^T w_B| - |f_1 w_A| + |H_g^T w_{AC}| \geq |L_B + H_X^T w_B| \geq d_X$.
\end{proof}

\begin{figure}[h!]
    \centering
    \includegraphics[width=1\linewidth]{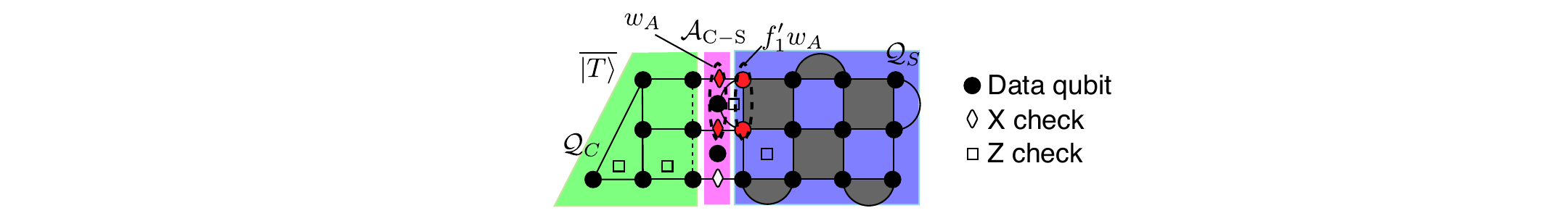}
    \caption{\textbf{Illustration of the proof for Lemma~\ref{lemma:distance_color_surface}.}}
    \label{fig:proof_color_surface}
\end{figure}

\begin{lemma}[Distance of the merged color-surface code]
    Given a $[[n_c, 1, d_c]]$-color code and a surface code with a $Z$ distance $d_c$ and a $X$ distance $d_{s, X}$, the merged color-surface code $\mc{Q}_{\mr{CS}}$ has a $Z$ distance $d_c$ and a $X$ distance $d_c + d_{s, X}$. 
    \label{lemma:distance_color_surface}
\end{lemma}
\begin{proof}
    The dual of the color-surface code $\mc{Q}_{\mr{CS}}$, $\mc{Q}^{\prime}_{\mr{CS}}$ with a $X$ distance $d_X^{\prime}$ and a $Z$ distance $d_Z^{\prime}$, is a merged code in Def.~\ref{def:cone_code} through a repetition code adapter, where $\mc{Q}$ is given by the color code and $\mc{Q}^{\prime}$ is given by the dual of the surface code. We simply need to prove that $d_Z^{\prime} = d_c + d_{s, X}$ and $d_X^{\prime} \geq d_c$. 
    According to Lemma.~\ref{lemma:distance_merged_code}, we are naturally guaranteed that $d_Z^{\prime} \geq d_c + d_{s, X}$, and it remains to show that $H_g$, which is associated with the $X$ checks on the adapter and the surface code, satisfies the large-Cheeger-constant condition in Eq.~\eqref{eq:cheeger_constant}. 
    We now prove this based on the ``topological" property of the adapter and the surface code.
    Let $w = \left(\begin{array}{c}
         w_A\\
         w_S 
    \end{array}\right)$ be a vector on the space of the $X$ checks, where $w_A$ and $w_S$ are associated with the $X$ checks on the adapter and the surface code, respectively. We have $H_g w = \left(\begin{array}{c}
         \partial_1 w_A\\
         f_1^{\prime} w_A + H_x^{\prime T} w_S
    \end{array}\right) $. Now, we prove that $|H_g w| \geq |f_1^{\prime} w_A + H_x^{\prime T} w_S| \geq |w_A|$. As illustrated in Fig.~\ref{fig:proof_color_surface}, $w_A$ is supported on a column of interface $X$ checks and on $|w_A|$ rows. $f_1^{\prime} w_A$ is supported on a column of the surface code data qubits spanning the same set of rows as $w_A$ does. Based on the structure of the surface code, applying $H_x^{\prime T} w_S$ does not reduce the row weight of $f_1 w_A$, i.e. $f_1^{\prime} w_A + H_x^{\prime T} w_S$ always has a nontrivial support on a row $R$ whenever $f_1^{\prime} w_A$ has support on $R$. As such, we have $|f_1^{\prime} w_A + H_x^{\prime T} w_S| \geq |f_1^{\prime} w_A| = |w_A|$. Therefore, we have $d^{\prime}_X \geq d_c$ as the boundary Cheeger constant of $H_g$ is larger than $1$, according to Lemma.~\ref{lemma:distance_merged_code}.  
\end{proof}

\begin{proposition}[Distance of the merged color-surface-BB code]
    Given $k_b$ copies of $[[n_c, 1, d_c]]$ color codes and $[[n_s, 1, d_{s, X}, d_{s,Z} = d_c]]$ thin surface codes and a $[[n_b, k_b, d_c]]$ BB code, where $d_c + d_{s, X} \geq d_b$, the merged code $\mc{Q}_{\mr{CSBB}}$ has a distance $\geq d_b$. 
    \label{th:fault_tolerance_merging}
\end{proposition}
According to Lemma~\ref{lemma:distance_color_surface}, each color-surface code $\mc{Q}_{\mr{CS}}$ has a $Z$ distance $d_{cs, Z} = d_c$ and a $X$ distance $d_{cs, X} = d_c + d_{s, X} \geq d_b$. The color-surface-BB code is a merged code $ \mc{Q}_{\mr{CSBB}} = \tilde{Q}(\mc{Q}_{\mr{BB}}, \mc{Q}_{\mr{CS}}^{\otimes k_b}, \bigotimes_{i=1}^{k_b} \mc{A}^{(i)}_{\mr{CS-BB}})$ defined in Def.~\ref{def:cone_code}, where $\mc{Q}$ is given by $\mc{Q}_{\mr{BB}}$ and $\mc{Q}^{\prime}$ is given by $\mc{Q}_{\mr{CS}}^{\otimes k_b}$. According to Lemma~\ref{lemma:distance_merged_code}, $\mc{Q}_{\mr{CSBB}}$ has a $Z$ distance $\geq d_b + d_{\mr{cs, Z}} = d_b + d_c > d_b$ as well as a $X$ distance $\geq d_b$ if the corresponding $H_g$, which is associated with the $X$ checks of $\otimes_{i = 1}^{k_b}\mc{A}_{\mr{CS-BB}}^{(i)}$ and $\mc{Q}_{\mr{CS}}^{\otimes k_b}$, has a large relative Cheeger constant (see eq.~\eqref{eq:cheeger_constant}). Since the $k_b$ copies of the adapter $\mc{A}_{\mr{CS-BB}}^{(i)}$ and the color-surface code $\mc{Q}_{\mr{CS}}$ have disjoint support, it suffices to prove the corresponding $H_g^{(1)}$ when merging the BB code with only one color code through an adapter $\mc{A}_{\mr{CS-BB}}^{(1)}$ has a large boundary Cheeger constant. Since $\mc{Q}_{\mr{CS}}$ has a topological structure, we conjecture that one can find an analytical proof that is analogous to that for Lemma.~\ref{lemma:distance_color_surface}. Here, for simplicity, we simply numerically verified that $\beta(H_g^{(1) T}, \mb{C}_A) \geq 1$ holds for all the BB codes listed in Table I of the main text.

Note that in practice, we might choose a smaller ancillary surface code such that the $X$ distance of the merged color-surface code, $d_c + d_{s, X}$, is smaller than $d_b$. In this case, the distance of the merged color-surface-BB code, will be determined by $d_c + d_{s, X}$. For instance, to minimize the overall logical error rates, we choose $d_c + d_{s, X} = 7$ for the $[[66, 6, 8]]$-BB code, which will result in a merged code with distance $7$. 

\subsection{Code-merging protocol 
\label{sec:code_merging}}
In Alg~\ref{alg:code_merging}, we present a code deformation protocol that starts with $k_b$ copies of color codes, $\mc{Q}_{C}^{\otimes k_b}$, and ends in one merged color-surface-BB code, $\mc{Q}_{\mr{CSBB}}$. 
At the high level, the protocol is analogous to the growth of surface codes~\cite{li2015magic, lao2022magic} and it grows the logical operators of the small color code in two steps: (i) grow the $X$ logical operators to weight $d_c + d_{s, X} \geq d_b$ by merging with thin surface codes (ii) grow the $Z$ logical operators to weight $\geq d_c + d_b$ by merging with the BB code.  

\begin{algorithm}[h!]
\caption{A protocol for merging $\mc{Q}_C^{\otimes k_b}$ with $\mc{Q}_{\mr{BB}}$: $\mc{Q}_C^{\otimes k_b} \rightarrow \mc{Q}_{\mr{CSBB}}$}\label{alg:code_merging}
\Input{$k_b$ copies of $[[n_c, 1, d_c]]$-color code $Q_{C}$. }
\Output{A merged code $\mc{Q}_{\mr{CSBB}}$ encoding the same information as that in $\mc{Q}_C^{\otimes k_b}$. }

Construct the merged code $\mc{Q}_{\mr{CSBB}}$ as $\tilde{Q}(Q_{\mr{BB}}, Q_{\mr{CS}}^{\otimes k_b}, \mc{A}_{\mr{CS-BB}})$ according to Def.~\ref{def:cone_code}, where $Q_{\mr{CS}}$ is constructed as $\tilde{Q}(\mc{Q}_C, \mc{Q}_S, \mc{A}_{\mr{C-S}})$. 
See Sec.~\ref{sec:merged_CSBB_code} for the details of the construction.\\

Prepare the data qubits of the surface codes $\mc{Q}_S^{\otimes k_b}$ and the color-surface adapters $\mc{A}_{\mr{C-S}}^{\otimes k_b}$ in $\ket{+}$ states and measure the stabilizers of $\mc{Q}_{\mr{CS}}$ for one round. \\

Prepare the data qubits of the BB code $\mc{Q}_{\mr{BB}}$ and the color-surface-BB adapter $\mc{A}_{\mr{CS-BB}}$ in $\ket{0}$ states and measure the stabilizers of $\mc{Q}_{\mr{CSBB}}$ for $t_m$ rounds. \\

Restart the protocol when certain detectors (see Fig.~\ref{fig:code_merging_detectors} and the text for details) are triggered.\\

Calculate soft information (complementary gap) of the decoder correction for the merged code and restart when it is unreliable.
\end{algorithm}

\begin{figure}
    \centering
    \includegraphics[width=1\linewidth]{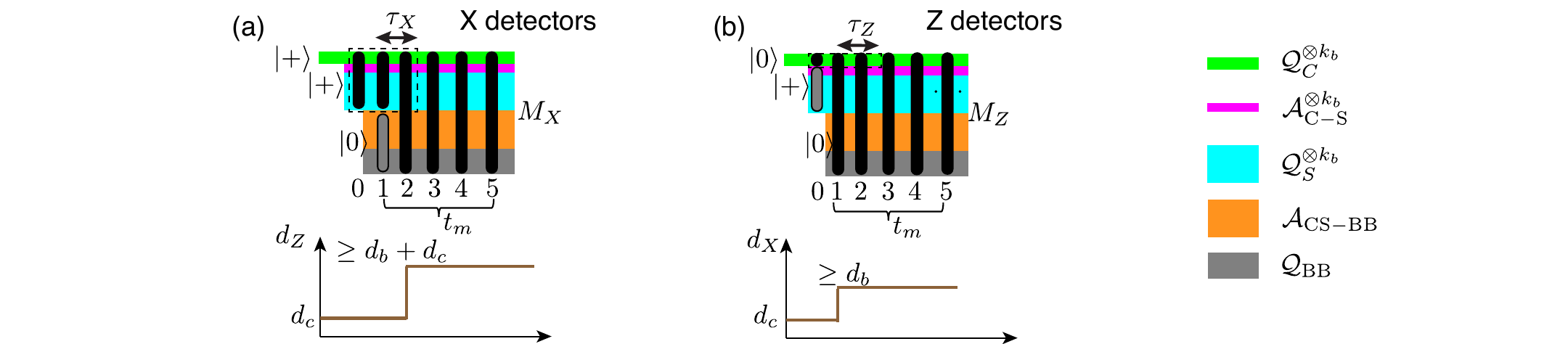}
    \caption{\textbf{Illustration of the $X$ (a) and $Z$ (b) detectors of the code merging protocol in Alg.~\ref{alg:code_merging}.} The colored area of each diagram represents 
    different codes/adapter systems at different code cycles (time goes from left to right). 
    The protocol starts with the $0$-th code cycle on the merged color-surface code and lasts for another $t_m$ code cycles on the merged color-surface-BB code.
    A black bar supported on a given area at a given code cycle represents the detectors constructed by taking either the value of the stabilizers on the same area at the same time step (if they are deterministic) or the parity of the above stabilizers with the stabilizers on the same area but at an earlier time step. 
    For example, the first black bar in (a) represents the $X$ detectors associated with the first round of $X$ stabilizer measurements of the merged color-surface code $\mc{Q}_{\mr{CS}}$ in step 2 of Alg.~\ref{alg:code_merging}, which are deterministic as the qubits of $\mc{A}_{\mr{C-S}}^{\otimes k_b}$ and $\mc{Q}_S^{\otimes k_b}$ are prepared in $\ket{+}$ states. 
    A grey bar, on the other hand, represents stabilizer measurements that, nevertheless, are non-deterministic and do not form detectors (by comparing against proceeding cycles).
    The dashed boxes represent the detectors that we post-select on. We also show the growth of the $X$ and $Z$ space-distance of the protocol below the detector diagrams with aligned time axes. 
    }
    \label{fig:code_merging_detectors}
\end{figure}

We illustrate the stabilizer measurements and the detectors of the code merging protocol in Alg.~\ref{alg:code_merging} in Fig.~\ref{fig:code_merging_detectors}. 
Counting the round of stabilizer measurements on the merged color-surface code $\mc{Q}_{\mr{CS}}$ in the step 2 of Alg.~\ref{alg:code_merging} as the zero-th code cycle, the protocol establishes the full set of deterministic $X$ stabilizers (resp. $Z$ stabilizers) on the merged code $\mc{Q}_{\mr{CSBB}}$ and gains sufficiently large $Z$ distance (resp. $X$ distance) at the second (first) code cycle. 
We need to post-select on the detectors before this stage, where the space distance of the protocol is still small, to prevent logical errors from low-weight space-like faults. In addition, we also post-select on part of the detectors after reaching the full space distance to prevent logical errors from the combination of space-like and time-like errors. 
As shown in Fig.~\ref{fig:code_merging_detectors}(a) and (b), respectively, we post-select on the $X$ and $Z$ detectors only on the merged color-surface codes $\mc{Q}_{\mr{CS}}^{\otimes k_b}$ and the color codes $\mc{Q}_C^{\otimes k_b}$, starting from the zero-th code cycle to up to $\tau_X$ and $\tau_Z$ code cycles on the merged code $\mc{Q}_{\mr{CSBB}}$, respectively. We provide the choice of $\tau_X$ and $\tau_Z$ that balances the space-like logical error rates and the success rate of the protocol in Table~\ref{tab:space-like_error_rates}.
\qx{Finally, we perform light post-selection on the merged code based on the soft information (aka complimentary gap~\cite{gidney2024magic}) quantifying the reliability of the decoder corrections to further reduce the error rates of the final $\ket{T}$ states. See Methods in Ref.~\cite{sales2025experimental} for obtaining such soft information using an integer-programming decoder and upcoming work~\cite{bienias2025upcoming} for doing so with the BP+OSD decoder (used in this work). 
}

\begin{figure}
    \centering
    \includegraphics[width=1\linewidth]{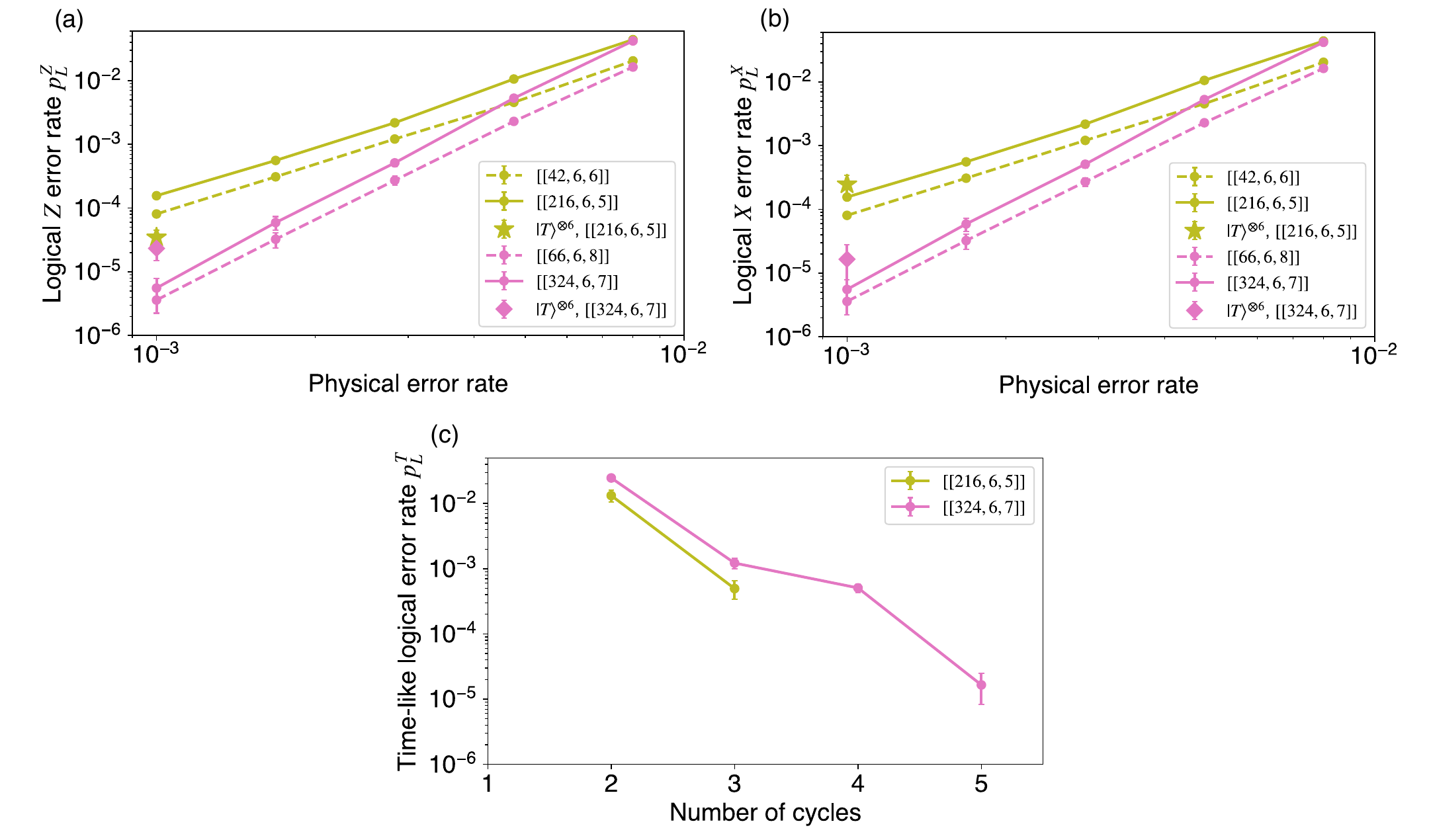}
    \caption{\textbf{The space-like $X$ (a) and $Z$ (b) logical error rates as well as the time-like logical error rate (c) of the $\ket{T}$ states}, compared against memory error rates (dashed lines) of the corresponding BB codes and merged CSBB codes (solid lines). 
    The memory simulations take $\tau_M = 3$ code cycles before the final transversal measurements and are decoded using the relay-BP decoder~\cite{muller2025improved}. Both the memory and the stability experiments (c) are decoded using the relay-BP decoder~\cite{muller2025improved} and the $\ket{T}$-states simulation are decoded using the BP-LSD decoder~\cite{hillmann2025localized}.
    Denote the probability that any logical error occurs as $P$ for these experiments, the reported logical error rate $p_L = 1 - (1 - P)^{1/m}$ is normalized by $m = k_b\times \tau_M$ for the memory experiments (corresponding to logical error rates per logical qubit per code cycle) and $m = k_b$ for the stability experiments and the $\ket{T}$-states simulations (corresponding to logical error rates per logical qubit). The error bars are calculated using the standard derivation $\delta p_L = \frac{(1 - P)^{1/m - 1}}{m} \sqrt{\frac{P(1-P)}{N_0}}$, where $N_0$ denotes the number of binary samples collected for estimating $P$.
    }
    \label{fig:space_like_error_rates}
\end{figure}

We probe the space-like logical error rates and the time-like logical error rates of the code merging protocol of Alg.~\ref{alg:code_merging} separately using the memory experiment and the stability experiment~\cite{gidney2022stability}, respectively. 
For the memory experiment, we initialize the data qubits of the $k_b$ color codes $\mc{Q}_C^{\otimes k_b}$ in $\ket{+}$ (resp. $\ket{0}$) states, simulate the merging protocol in Alg.~\ref{alg:code_merging}, and measure the data qubits of the final merged code $\mc{Q}_{\mr{CSBB}}$ in the $X$ (resp. $Z$) basis, and estimate the logical $Z$ (resp. $X$) error rate by checking if values of the logical $X$ (resp. $Z$) operators of $\mc{Q}_{\mr{CSBB}}$ are all $+1$. 
We refer to the logical failure probability (LFP) as the probability that any of the logical operators are measured in $-1$ and approximately estimate the logical failure rate per logical qubit as $\mr{LFR} = \mr{LFP}/k_b$. We plot the logical $Z$ and $X$ error rate versus the physical error rate in Fig.~\ref{fig:space_like_error_rates}(a)-(b).

For the stability experiment, we utilize the redundancy of the $X$ checks $H_X \in \mbb{F}_2^{r\times n}$ of $\mc{Q}_{\mr{CSBB}}$, i.e. $\ker{H_X^T} \neq 0$. 
In the code-merging protocol in Alg.~\ref{alg:code_merging}, the $X$ syndromes $\{s_X(t) \in \mbb{F}_2^{r}\}_{t = 1}^{t_m}$ on the merged code $\mc{Q}_{\mr{CSBB}}$ for different code cycles $t$ are non-deterministic (but should be the same across different cycles) in the absence of errors. However, due to the redundancy of $H_X$, there are certain combinations of the syndromes at a given round, i.e. $\mb{S} = \{a^T s_X(t) \mid a \in \ker{H_X^T}\}$, that should be deterministically $+1$. 
Denote the time-like logical failure probability $\mr{LFP}^T$ as the probability that any element in $\mb{S}$ takes a $-1$ value after the error correction, we characterize the time-like logical error rate per logical qubit as $p^T_L = \mr{LFP}^T/k_b$. 
These time-like logical failures characterize the failures of obtaining the correct $X$ stabilizer frame of $\mc{Q}_{\mr{CSBB}}$ in Alg.~\ref{alg:code_merging} when growing the logical $Z$ operators by appending the block of qubits of $\mc{Q}_{\mr{BB}}$ in the $\ket{0}$ states during the step 3. 
Note that we only need to reliably obtain the $X$ checks by repeating multiple code cycles when considering the teleportation circuit in Fig. 6(c) of the main text since the merging protocol is followed by transversal $Z$ measurements.
We plot the time-like logical failure rate $p_L^T$ of the merged $[[216, 6, 5]]$ code and the $[[324, 6, 7]]$ code for different choices of $t_m$ in Fig.~\ref{fig:space_like_error_rates}(c).

Finally, we estimate the fidelity and the space-time costs for the final $\ket{T}$ states hosted in the $[[324, 6, 7]]$ merged code $\mc{Q}_{\mr{CSBB}}$.
\qx{Since the initial cultivation steps on the $d=3$ color codes~\cite{gidney2024magic} has an negligible error rates and space-time costs compared to our code-merging protocol here, we simply assume that we start from $k_b = 6$ copies of perfect $\ket{T}$ states encoded in the color codes. }
Let $N_{\mr{CSBB}} = 648$ denote the total number of qubits for $\mc{Q}_{\mr{CSBB}}$, which includes both the data qubits and the ancilla qubits. 
We estimate the fidelity of the final $\ket{T}$ states by summing up the space-like logical error rates (in Fig.~\ref{fig:space_like_error_rates}(a)-(b)) and the time-like logical error rates (Fig.~\ref{fig:space_like_error_rates}(c)), $p_L(\ket{T}) \approx p_L^X + p_L^Z + p_L^T \approx 2\times 10^{-5}$ (see Table~\ref{tab:space-like_error_rates}). 
Clearly, the space-time costs are dominated by the $t_m$ code cycles (including retries) spent on the merged $\mc{Q}_{\mr{CSBB}}$ code. 
The post-selection is primarily applied on the first $\tau_X = 2$ code cycles. We assume that we restart the protocol whenever we detect any nontrivial syndromes after each of the $\tau_X$ code cycles. 
Let $\eta = \eta_Z \eta_X = 0.65$ be the success rate per cycle. 
It then takes on average $t_0 = (1 + \eta)/\eta^2 \approx 3.9$ cycles to complete the first two (post-selected) code cycles. 
The total number of code cycles for the entire merging protocol is then roughly given by $T \approx (t_0 + t_m - 2)/(\eta_X^s \eta_Z^s) \approx 8$. 
The space-time cost (per kept $\ket{T}$ state) is then given by $W(\ket{T}) \approx N_{\mr{CSBB}}*T/6 \approx 860\ \mr{qubit*cycles}$. 

\begin{table}[h!]
    \centering
    \begin{tabular}{c|c|c|c|c|c|c|c|c|c|c|c|c}
    \hline
    \hline
          $\mc{Q}_{\mr{BB}}$ & $\mc{Q}_S$ & $\mc{Q}_{\mr{CSBB}}$ & $t_m$ & $\tau_X$ & $\eta_X$ & $\eta^{s}_X$ & $p_L^Z$ & $\tau_Z$ & $\eta_Z$ & $\eta^s_Z$ & $p_L^X$ & $p_L^T$ \\
          \hline
          $[[42, 6, 6]]$ & $[[6, 1, d_X = 2, d_Z = 3]]$ & $[[216, 6, 5]]$ & $3$ & $2$ & $0.66$ & $0.99$ & $3\times 10^{-5}$ & $0$ &  $1$ & $0.9$ &  $2\times 10^{-4}$ & $5\times 10^{-4}$ \\
          \hline
          $[[66, 6, 8]]$ & $[[12, 1, d_X = 4, d_Z = 3]]$& $[[324, 6, 7]]$ & $5$ & $2$ & $0.65$ & $0.94$ & $1.6\times 10^{-5}$ & $0$ &  $1$ & $0.89$ & $2.9\times 10^{-5}$ & $8\times 10^{-6}$\\
    \hline
    \hline
    \end{tabular}
    \caption{\textbf{Simulation results of the code-merging protocols of Alg.~\ref{alg:code_merging} at a $10^{-3}$ physical error rate.} We list the parameters of the BB code $\mc{Q}_{\mr{BB}}$, the thin surface code $\mc{Q}_{S}$, the merged color-surface-BB code $\mc{Q}_{\mr{CSBB}}$, the number of code cycles spent on $\mc{Q}_{\mr{CSBB}}$ $t_m$, the number of $X$ (resp. $Z$) code cycles $\tau_X$ (resp. $\tau_Z$) on $\mc{Q}_{\mr{CSBB}}$ that we post-select on, the success rate $\eta_X$ (resp. $\eta_Z$) of each post-selected $X$ (resp. $Z$) code cycle, the final success rate $\eta_X^s$ (resp. $\eta_Z^s$) for post-selecting the $X$ (resp. $Z$) detectors based on the soft information, the logical $Z$ (resp. $X$) error rate $p_L^Z$ (resp. $p_L^X$) of the merging protocol, and the time-like logical error rate $p_L^T$. }
    \label{tab:space-like_error_rates}
\end{table}

\subsection{Homomorphic CNOT
\label{sec:homo_CNOT}}
Here, we prove that the block of the cultivated magic states $\overline{\ket{T}}^{\otimes k_b}$ in the merged code $\mc{Q}_{\mr{CSBB}}$ can be used to teleport parallel $\bar{T}^{\otimes k_b}$ gates on a data BB code $\mc{Q}_{\mr{BB}}$ (see Fig.5(c) of the main text) by implementing a homomorphic CNOT between $\mc{Q}_{\mr{CSBB}}$ and $\mc{Q}_{\mr{BB}}$. 
Specifically, we prove in Proposition~\ref{prop:homomorphic_cnot} that pairs of physical CNOTs that each couples a data qubit of $\mc{Q}_{\mr{CSBB}}$ and the corresponding data qubit of the BB component of $\mc{Q}_{\mr{CSBB}}$ (see the gray patch in Fig.~\ref{fig:BB_merging}) implement parallel logical $\mr{CNOT}^{\otimes k_b}$ between pairs of logical qubits of $\mc{Q}_{\mr{CSBB}}$ and $\mc{Q}_{\mr{BB}}$. Moreover, this homomorphic CNOT is trivially fault-tolerant since it only involves transversal physical CNOTs that do not spread errors within a code block. 

\begin{proposition}[Homomorphic CNOT from the merged code to the BB code]
The $n_b$ pairs of physical CNOTs that are each controlled by a data qubit of $\mc{Q}_{\mr{BB}}$ and targeting the corresponding data qubit of the BB component of $\mc{Q}_{\mr{CSBB}}$ implement parallel logical CNOTs, $\overline{\mr{CNOT}}^{\otimes k_b}$, controlled by $\mc{Q}_{\mr{BB}}$ and targeting $\mc{Q}_{\mr{CSBB}}$.
\label{prop:homomorphic_cnot}
\end{proposition}
\begin{proof}
Let $\mc{Q}_{\mr{BB}}: \begin{tikzcd} C_2 & C_1 & C_0
\arrow["{H_X^T}", from=1-1, to=1-2]
\arrow["{H_Z}", from=1-2, to=1-3]
\end{tikzcd}$ be the chain complex of the BB code. 
Let  $\mc{Q}_{\mr{CSBB}}: \begin{tikzcd} 
\Tilde{C}_2 & \Tilde{C_1} & \Tilde{C_0}
\arrow["{\Tilde{H}_X^T}", from=1-1, to=1-2]
\arrow["{\Tilde{H}_Z}", from=1-2, to=1-3]
\end{tikzcd}$ be the chain complex of the merged color-surface-BB code. 
Recall that $\mc{Q}_{\mr{CSBB}} = \tilde{Q}(\mc{Q}_{\mr{BB}}, \mc{Q}^{\prime}, \mc{A})$ is constructed by merging $\mc{Q}_{\mr{BB}}$ with $k_b$ copies of the color-surface codes, which we simply denote as $\mc{Q}^{\prime}$, through an adapter system $\mc{A}$, according to Def.~\ref{def:cone_code}.
We re-write the check matrices of $\mc{Q}_{\mr{CSBB}}$ from Eq.~\ref{eq:check_mats_cone_code} in a block form that separates the qubits and checks of the BB code from the rest of the system: $\Tilde{H}_X = \left( \begin{array}{cc}
    H_X & 0 \\
    \alpha & \beta \\
\end{array}\right)$ and $\Tilde{H}_Z = \left( \begin{array}{cc}
    H_Z & \gamma \\
    0 & \delta \\
\end{array}\right)$,
where the first block of rows (resp. columns) are associated with the checks (resp. qubits) of $\mc{Q}_{\mr{BB}}$.
The transversal physical CNOTs between the BB code and the corresponding component of the merged code implement a logical gate if and only if the following diagram commutes~\cite{huang2022homomorphic, xu2024fast}:
\begin{equation}
    \label{eq:quantum_code_homo}
        \begin{tikzcd}
	{\Tilde{C}_2} & {\Tilde{C}_1} & {\Tilde{C}_0} \\
	{C_2} & {C_1} & {C_0}
	\arrow["{\Tilde{H}_X^T}", from=1-1, to=1-2]
	\arrow["{\Tilde{H}_Z}", from=1-2, to=1-3]
        \arrow["{H_X^T}", from=2-1, to=2-2]
	\arrow["{H_Z}", from=2-2, to=2-3]
        \arrow["{\gamma_2}", from=2-1, to=1-1]
	\arrow["{\gamma_1}", from=2-2, to=1-2]
        \arrow["{\gamma_0}", from=2-3, to=1-3]
\end{tikzcd},
    \end{equation}
where the homomorphisms take the form $\gamma_2 = \gamma_1 = \gamma_0 = \left( \begin{array}{c}
     I \\
     0 \\
\end{array}\right)$ (in their corresponding block form). The commutativity can be easily checked:
\begin{equation}
\begin{aligned}
    \Tilde{H}_X^T \gamma_2 & =  \left( \begin{array}{c}
     H_X^T \\
     0 \\
\end{array}\right) = \gamma_1 H_X^T. \\
\Tilde{H}_Z \gamma_1 & =  \left( \begin{array}{c}
     H_Z \\
     0 \\
\end{array}\right) = \gamma_0 H_Z. 
\end{aligned}
\end{equation}
We can further verify that the physical CNOTs implement pair-wise transversal logical CNOTs between the $k$ logical qubits of $\mc{Q}_{\mr{BB}}$ and the corresponding logical qubits of $\mc{Q}_{\mr{CSBB}}$ by checking the propagation of the logical operators. 
\end{proof}

\subsection{Logical translation
\label{sec:logical_translation}}
We implement the logical translation gadget for a BB code by leveraging the translational automorphism of the constructed BB codes. As detailed in the main text, the $k_b$ logical qubits of a $[[n_b, k_b, d_b]]$-$\mc{Q}_{\mr{BB}}$ are distributed into two blocks with indices $\{1, 2,\cdots, k_b/2\}$ and $\{k_b/2 + 1, \cdots, k_b\}$, which can be cyclically permuted within each block using physical permutations. To complete a joint cyclic permutation on $k_b$ logical qubits, we compose the above automorphism gate with an extra swap between the qubits indexed by $1$ and $k_b/2 + 1$. Such an extra swap could be implemented using standard surgery-based approaches~\cite{swaroop2024universal} by, e.g. teleporting the targeted pair of logical qubits to two $[[n_s, 1, d_s \geq d_b]]$ surface codes and then swapping the two surface codes, analogous to how an addressable logical CNOT can be performed on any qLDPC code using a toric-code ancilla~\cite{swaroop2024universal}.
Note that as we only need to perform a \emph{single} swap operation using a single ancilla surgery system, the overheads and the logical error rates of this targeted swap operation will both be much lower than those of the parallel magic-state cultivation procedure, which involves $k_b$ surgery systems.   

\subsection{Clifford fixups
\label{sec:Clifford_fixups}}
\begin{figure}
    \centering
    \includegraphics[width=1\linewidth]{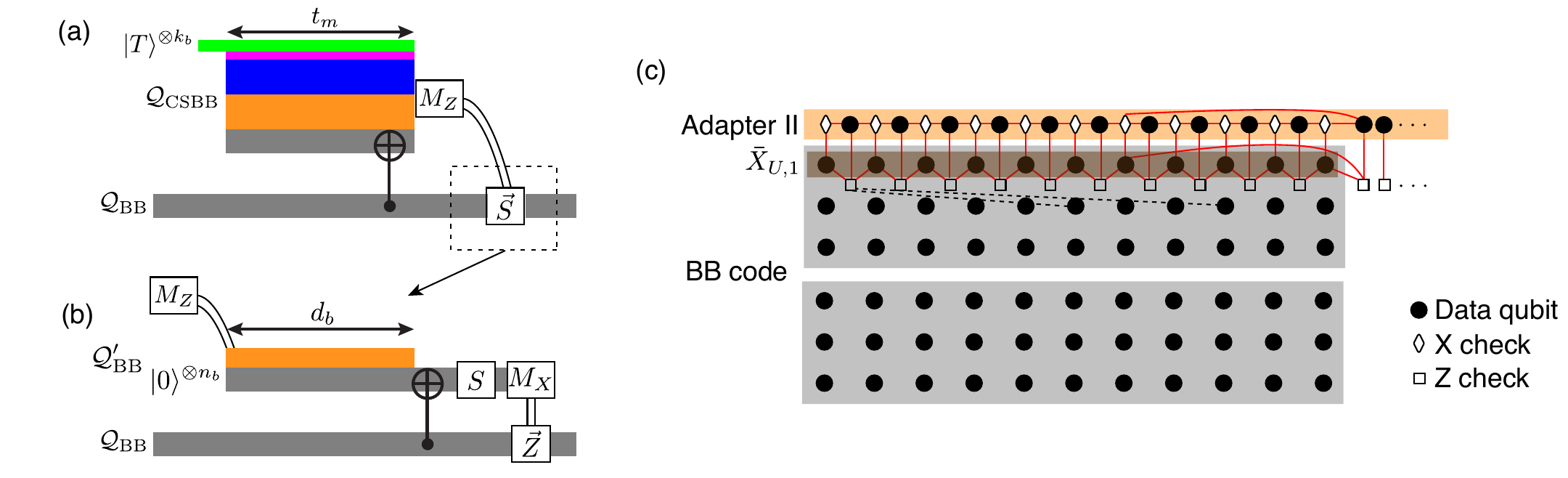}
    \caption{\textbf{Clifford fixups for the BB code}. }
    \label{fig:Clifford_fixups}
\end{figure}

Here, we describe how to implement the Clifford fixups on the data BB code $\mc{Q}_{\mr{BB}}$ when teleporting global $T$ gates in Fig. 6(c) of the main text, which is also copied here in Fig.~\ref{fig:Clifford_fixups}(a). 

Depending on the $Z$ measurement outcomes on the merged color-surface-BB code $\mc{Q}_{\mr{CSBB}}$, we need to apply some addressable $S$ gates, denoted by $\vec{S}$, on $\mc{Q}_{\mr{BB}}$. We implement $\vec{S}$ using the circuit in Fig.~\ref{fig:Clifford_fixups}(b): (1) initialize an ancilla code $\mc{Q}_{\mr{BB}}^{\prime}$ (identical to $\mc{Q}_{\mr{BB}}$) in the $Z$ basis,  (2) resetting a subset (depending on $\vec{S}$) of the $k_b$ logical qubits of $\mc{Q}_{\mr{BB}}^{\prime}$ to $+$ states by doing a set of logical $X$ measurements using surgery systems (indicated by the orange bars in Fig.~\ref{fig:Clifford_fixups}(b)(c)), (3) apply a transversal CNOT between $\mc{Q}_{\mr{BB}}$ and $\mc{Q}_{\mr{BB}}^{\prime}$, (4) apply transversal $S$ gates on $\mc{Q}_{\mr{BB}}^{\prime}$, (5) transversally measure $\mc{Q}_{\mr{BB}}^{\prime}$ in the $X$ basis, (6) apply feedback addressable Pauli $Z$ correction on $\mc{Q}_{\mr{BB}}$. As shown in Fig.~\ref{fig:Clifford_fixups}(c), the surgery system for performing the selective logical $X$ measurements in step (2) is the same as the adapters that are used to merge the BB code with the color codes. For instance, a surgery system for measuring the $X$ logical operator of the first logical qubit of the ``U" block of $\mc{Q}_{\mr{BB}}^{\prime}$ is shown in Fig.~\ref{fig:Clifford_fixups}(c), which is the same as the orange strip in Fig.~\ref{fig:BB_merging}. Since the transversal $Z$ measurements on $\mc{Q}_{\mr{CSBB}}$ in Fig.~\ref{fig:Clifford_fixups}(a) will be random, we only need about $k_b/2$ surgery systems on average. As such, the space-time cost of the $S$ fixups will be much smaller than the global $T$ gates (Fig.~\ref{fig:Clifford_fixups}(b) versus Fig.~\ref{fig:Clifford_fixups}(a)).

\subsection{Resource estimates}
Finally, we estimate the logical error rates and space–time costs of implementing a parallel layer of two-qubit rotations---the key subroutine for simulating the lattice dynamics in Fig. 6(c) of the main text---using the BB-code–based protocols introduced in this section, assuming a physical error rate of $10^{-3}$. 
We benchmark these results against (i) state-of-the-art surface-code protocols and (ii) a BB-code protocol that implements the same task via sequential coupling to a cultivated surface-code magic-state factory~\cite{yoder2025tour}, which we refer to as the low-rate BB protocol.

\begin{figure}
    \centering
    \includegraphics[width=0.8\linewidth]{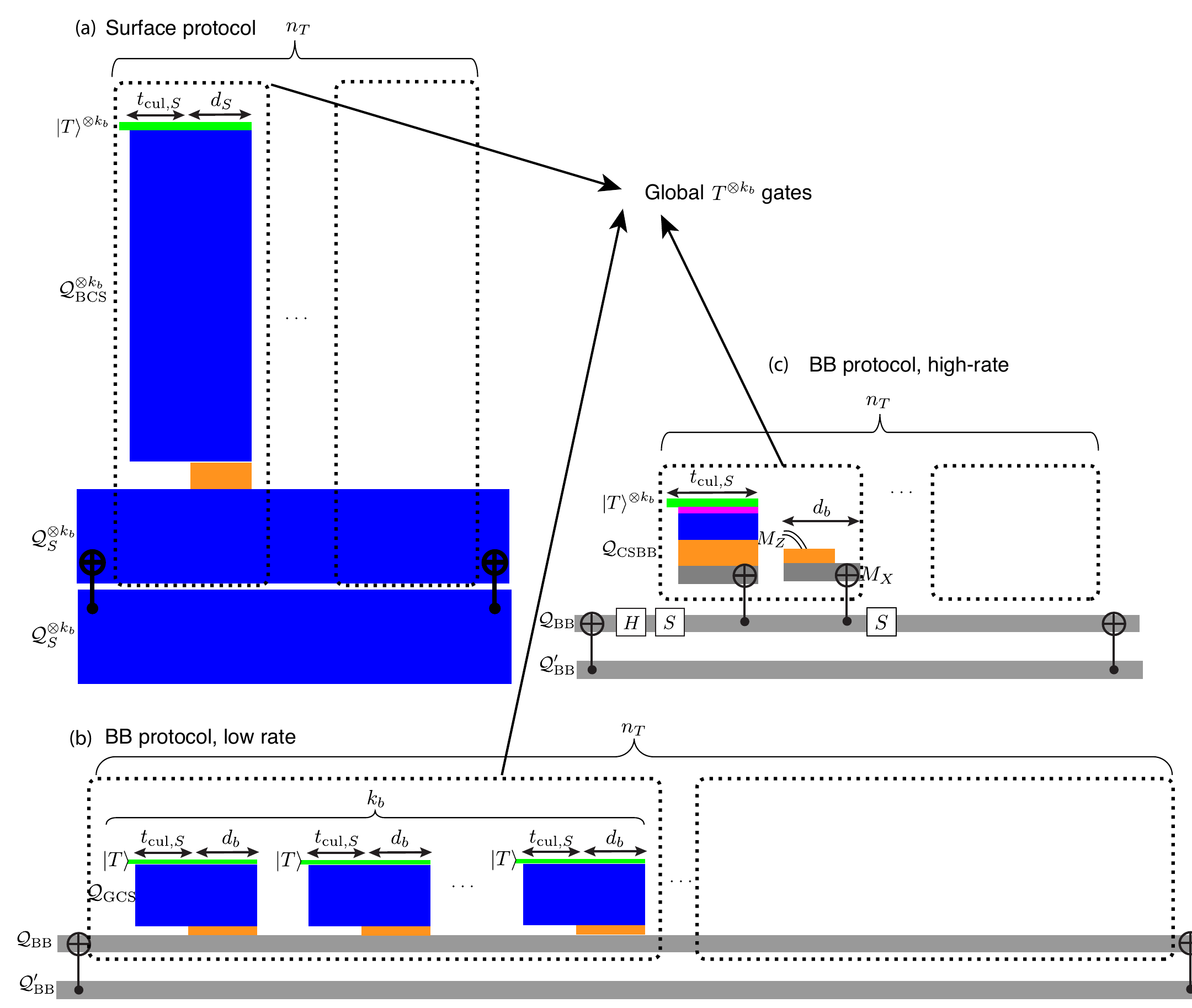}
    \caption{\textbf{Illustration of the space-time costs of (a) Surface code protocol with parallel surface-code cultivation (b) BB protocol with sequential surface-code cultivation (c) BB protocol with parallel BB-cultivation for a layer of two-qubit rotations in Fig. 6(c) of the main text. }}
    \label{fig:resource_estimates}
\end{figure}

The three protocols are illustrated in Fig.~\ref{fig:resource_estimates}. Each implements $k_b$ parallel $ZZ(\theta)$ rotations between two blocks of $k_b$ logical qubits, each encoded in a $k_b$-logical-qubit data code. The operation is realized by sandwiching a global $Z(\theta)^{\otimes k_b}$ rotation on one block between two transversal inter-block CNOTs. The global $Z(\theta)^{\otimes k_b}$ rotations are synthesized from sequences of global $H^{\otimes k_b}$, $S^{\otimes k_b}$, and $T^{\otimes k_b}$ gates. The $T$ gates, in turn, are implemented by coupling the data code to a factory code that hosts cultivated high-fidelity $\ket{T}$ states. Below we detail the implementation for each protocol.

Surface-code protocol.
As shown in Fig.~\ref{fig:resource_estimates}(a), the two data codes are chosen as $k_b$ copies of a rotated $[[81, 1, d_S = 9]]$ surface code $\mc{Q}_S$, and the factory code consists of $k_b$ copies of a distance-$d_{\mr{GCS}} = 15$ grafted color-surface code\cite{gidney2024magic, yoder2025tour}, denoted $\mc{Q}_{\mr{GCS}}$. Using Pauli-based computation~\cite{litinski2019game}, the global $H^{\otimes k_b}$ and $S^{\otimes k_b}$ are absorbed into Clifford transformations, so that the global $T^{\otimes k_b}$ gets transformed to parallel Pauli rotations $P(\pi/8)$ ($P \in {X, Y, Z}$). Each such rotation is implemented by parallel $P \otimes Z$ joint Pauli-product measurements between $\mc{Q}_S$ and $\mc{Q}_{\mr{GCS}}$. These measurements can be carried out in $d_S$ code cycles using $k_b$ copies of a surface-to-surface adapter $\mc{A}_{\mr{S-S}}$ (the orange strip in Fig.\ref{fig:resource_estimates}(a)).
$\mc{A}_{\mr{S-S}}$ simply takes a form of a $d_{\mr{GCS}}$-bit repetition code. 
We denote by $t_{\mr{cul, S}}$ the expected number of code cycles to obtain the cultivated $\ket{T}^{\otimes k_b}$ states in $\mc{Q}{\mr{GCS}}^{\otimes k_b}$. 
We adopt a rough estimate of $t{\mr{cul, S}} = 10$, consistent with the $\sim 4000$ qubit$\times$cycle volume required for a $\ket{T}$ state error rate of $3\times 10^{-5}$~\cite{gidney2024magic}. 
These estimates are admittedly coarse and could be further optimized. 
However, $t_{\mr{cul, S}} = 10$ should be regarded as optimistic, especially given the challenge of performing many (probabilistic) parallel cultivations.

Low-rate BB protocol.
Fig.~\ref{fig:resource_estimates}(b) depicts the analogous construction using BB codes. Each data block is now a $[[66, 6, 8]]$ BB code $\mc{Q}_{\mr{BB}}$ (see Table I in the main text). 
The key difference is that the global $T^{\otimes k_b}$ is implemented sequentially. Specifically, each logical qubit in $\mc{Q}_{\mr{BB}}$ is coupled in turn to a single $\mc{Q}_{\mr{GCS}}$ factory code via a joint Pauli product measurement, using a BB–surface adapter $\mc{A}_{\mr{S-BB}}$ in $d_b$ code cycles. 
For simplicity, we assume $\mc{A}_{\mr{S-BB}}$ is identical to the BB–color-surface adapter $\mc{A}{\mr{CS-BB}}$ defined in Sec.\ref{sec:merged_CSBB_code}, although this assumption is optimistic since $\mc{Q}_{\mr{GCS}}$ is larger than $\mc{Q}_{\mr{CS}}$.
Note that such a sequential BB protocol, adopted in Ref.~\cite{yoder2025tour}, is necessary when, for instance, the data BB code $\mc{Q}_{\mr{BB}}$, e.g. the Gross code in Ref.~\cite{bravyi2024high}, does not have a disjoint logical-operator basis, in which case parallel surgeries are challenging.

High-rate BB protocol.
Finally, Fig.~\ref{fig:resource_estimates}(c) illustrates the fully parallel BB-code protocol introduced in this work. Here the data blocks are again encoded in $\mc{Q}{\mr{BB}} = [[66, 6, 8]]$, but the factory consists of a $[[324, 6, 7]]$ color-surface-BB code $\mc{Q}{\mr{CSBB}}$, which prepares $\ket{T}^{\otimes k_b}$ in an expected $t_{\mr{cul, BB}} \approx 8$ code cycles. These $\ket{T}$ states are then used to teleport the global $T^{\otimes k_b}$ on one $\mc{Q}_{\mr{BB}}$ block via the homomorphic transversal CNOT (Sec.\ref{sec:homo_CNOT}), followed by feedback $S$ gates applied in $d_b$ logical cycles (Sec.\ref{sec:Clifford_fixups}). 
The global $H^{\otimes k_b}$ and $S^{\otimes k_b}$ gates can be applied transversally in $O(1)$ logical cycles using correlated decoding\cite{cain2024correlated, zhou2024algorithmic}. These Clifford operations can be executed concurrently with the factory’s $\ket{T}$ state preparation, incurring no additional code-cycle overhead.

We summarize in Table~\ref{tab:space_cost} the total qubit footprint (including ancillas) of the codes and adapters used in these protocols, and in Table~\ref{tab:time_cost} the numer of code cycles---together with the physical depth of each cycle---accounting for state preparation and measurement---of the most time-consuming surface-code or BB-code cultivation subroutines.
For a small angle such as $\theta = \pi/64$, we can synthesize $Z(\theta)$ using $n_T = 11$ $T/T^{\dagger}$ gates and $13$ $H$ and $S/S^{\dagger}$ gates with a synthesis error $\approx 8\times 10^{-5}$. 
Based on these choices, we estimate the logical error rates and the space- and time-costs for the three protocols below.

\begin{table}[h!]
    \centering
    \begin{tabular}{c|c|c|c|c|c|c}
    \hline
    \hline
                 & $\mc{Q}_{\mr{BB}}$ & $\mc{Q}_S$ & $\mc{Q}_{\mr{CSBB}}$ & $\mc{Q}_{\mr{GCS}}$ &  $\mc{A}_{\mr{S-S}}$ & $\mc{A}_{\mr{S-BB}}$  \\
          \hline
          Physical qubits & $N_{\mr{BB}} = 132$ & $N_S = 161$ & $N_{\mr{CSBB}} = 648$ & $N_{\mr{GCS}} = 454$ & $N_{A, \mr{S-S}} = 29$ & $N_{A, \mr{S-BB}} = 44$\\
    \hline
    \hline
    \end{tabular}
    \caption{\textbf{The number of physical qubits (including ancillas) for different codes and adapters. }}
    \label{tab:space_cost}
\end{table}

\begin{table}[h!]
    \centering
    \begin{tabular}{c|c|c}
    \hline
    \hline
                 & $t_{\mr{cul}, S}$ ($\tau_{\mr{cul}, S}$)& $t_{\mr{cul}, \mr{BB}}$ ($\tau_{\mr{cul}, \mr{BB}}$)  \\
          \hline
          Number of cycles (time steps per cycle) & $10$ ($8$)& $11$ ($14$) \\
    \hline
    \hline
    \end{tabular}
    \caption{\textbf{The number of cycles (the numer of time steps per cycle) for the surface-code cultivation protocol~\cite{gidney2024magic} and the parallel BB-code cultivation protocol presented in this work.} 
    The surface-cultivation cycle depth $\tau_{\mr{cul, S}}$ was extracted from Ref.~\cite{gidney2024magic} and the BB-cultivation cycle depth $\tau_{\mr{cul, BB}}$ assumes a depth-optimal syndrome extraction circuit for the merger color-surface-BB code $\mc{Q}_{\mr{CSBB}}$ in Sec.~\ref{sec:merged_CSBB_code} with a stabilizer weight $12$.
    }
    \label{tab:time_cost}
\end{table}

To simulate a $R$ by $2k_b$ lattice with $K = R \times (2k_b)$ algorithm logical qubits, the total footprint per algorithm qubit for the three protocols are:
\begin{enumerate}
    \item $N_{\mr{surface}}/K = R\times k_b\times (2N_{S} + N_{\mr{GCS}} + N_{A, \mr{S-S}})/K \approx 403$.
    \item $N_{\mr{BB, low-rate}}/K = R\times (2N_{\mr{BB}} + N_{\mr{GCS}} + N_{A, \mr{S-BB}})/K \approx 64$. 
    \item $N_{\mr{BB, high-rate}}/K = R\times (2N_{\mr{BB}} + N_{\mr{CSBB}})/K \approx 76$.
\end{enumerate}
The total number of code cycles are dominated by the number of global $T$ layers $n_T$ multiplied by the code cycles per $T$ layer:
\begin{enumerate}
    \item $t_{\mr{surface}} \approx n_T\times (t_{\mr{cul, S}} + d_S) = 209$.
    \item $t_{\mr{BB, low-rate}} \approx n_T\times k_b \times(t_{\mr{cul}, S} + d_b - 1) = 1122$. 
    \item $t_{\mr{BB, high-rate}} \approx n_T \times (t_{\mr{cul, BB}} + d_b - 1) = 165$.
\end{enumerate}
Note that the extra factor $k_b$ contributing to $t_{\mr{BB, low-rate}}$ arises from the fact that each layer of global $T^{\otimes k_b}$ gates are injected sequentially. 
Here, we have used $d_b - 1$ instead of $d_b$ for $t_{\mr{BB, low-rate}}$ and $t_{\mr{BB, high-rate}}$ as $d_b = 8$ is even for the $[[66, 6, 8]]$ BB code. 

The total physical circuit depth is estimated by multiplying the number of code cycles by the physical depth per cycle---$\tau_{\mr{cul, S}}$ for the surface-code and low-rate BB protocols, and $\tau_{\mr{cul, BB}}$ for the high-rate BB protocol:
\begin{enumerate}
    \item $T_{\mr{surface}} \approx t_{\mr{surface}}\times \tau_{\mr{cul}, S}$ = 1672.
    \item $T_{\mr{BB, low-rate}} \approx t_{\mr{BB, low-rate}}\times \tau_{\mr{cul}, S} = 8976$. 
    \item $T_{\mr{BB, high-rate}} \approx t_{\mr{BB, high-rate}}\times \tau_{\mr{{cul}, BB}} = 2310$.
\end{enumerate}
Note that we have used $\tau_{\mr{cul, BB}} = 14$ by assuming a depth-optimal syndrome extraction circuit for the merged CSBB code with a stabilizer weight $12$~\cite{kang2025quits}. We expect that the logical error rates for a more compact circuit would be comparable to the coloration circuit that we use in the current simulations under the error model that excludes idling errors. 

We then estimate the total space-time volumes per logical qubit by multiplying the footprints per logical qubit by the physical circuit depth:
\begin{enumerate}
    \item $W_{\mr{surface}}/K \approx N_{\mr{surface}}/K\times T_{\mr{surface}} \approx 6.7\times 10^5$.
    \item $W_{\mr{BB, low-rate}}/K \approx N_{\mr{BB, low-rate}}/K\times T_{\mr{BB, low-rate}} \approx 5.7\times 10^5 $.
    \item $W_{\mr{BB, high-rate}}/K \approx N_{\mr{BB, high-rate}}/K\times T_{\mr{BB, low-rate}} = \approx 1.8\times 10^5$.
\end{enumerate}

Finally, we estimate the logical error rate per logical qubit. Since the relevant logical operations across the different codes and protocols all exhibit a comparable error rate per cycle, $p_{L, 0} \approx 4 \times 10^{-6}$, we can approximate the total logical error per qubit for each layer of parallel two-qubit rotations as:
\begin{enumerate}
    \item $p_L(\mr{surface})/K \approx p_{L, 0}\times t_{\mr{surface}} \approx 8\times 10^{-4}$.
    \item $p_L(\mr{\mr{BB, low-rate}})/K \approx p_{L, 0} \times t_{\mr{BB, low-rate}} \approx 4\times 10^{-3}$.
    \item $p_L({\mr{BB, high-rate}})/K \approx p_{L, 0}\times t_{\mr{BB, high-rate}} \approx 7\times 10^{-4}$.
\end{enumerate}

\end{document}